\documentclass[10pt]{article}
\usepackage[left=1in,top=1in,right=1in,bottom=1in,head=.1in,nofoot]{geometry}
\usepackage{mathpazo}
\usepackage{amsfonts}
\usepackage{amsmath}
\usepackage{amssymb}
\usepackage{amsthm}
\usepackage{natbib}
\usepackage{setspace}
\usepackage{courier}
\usepackage{changepage}
\usepackage{enumerate}
\usepackage{mathtools}
\usepackage[capposition=top]{floatrow}
\allowdisplaybreaks
\usepackage[colorlinks,citecolor=blue]{hyperref}
\usepackage{etoolbox} 
\usepackage{multirow} 
\usepackage{lscape} 
\usepackage{tikz} 
\usepackage[flushleft]{threeparttable} 
\usepackage{subcaption} 
\usepackage{graphicx} 
\usepackage[normalem]{ulem} 
\usepackage[toc,title,titletoc,header]{appendix} 
\usepackage{soul} 
\usepackage{enumitem} 

\usepackage{titlesec} 
\titlelabel{\thetitle.\quad} 

\titleformat*{\section}{\bf\Large\center}

\usepackage{booktabs}

\usepackage{wasysym}
\usepackage{colortbl}

\def\bs{\boldsymbol}
\def\R{\textup{R}}
\def\rank{\textup{rank}}
\def\gen{\texttt{g}}
\def\con{\texttt{c}}
\def\mon{\texttt{m}}
\def\wc{\texttt{wc}}

\def\monp{\texttt{mp}}
\def\monn{\texttt{mn}}
\def\mon{\texttt{m}}

\def\I{\mathbbm{1}}
\def\CRE{\text{CRE}}
\def\Pr{\mathbb{P}}

\def\tr{\texttt{t}}
\def\co{\texttt{c}}

\def\qed{\rule{2mm}{2mm}}
\def\independent{\perp \!\!\! \perp}

\newtheorem{theorem}{Theorem}
\newtheorem{lemma}{Lemma}
\newtheorem{proposition}{Proposition}
\theoremstyle{definition}
\newtheorem{example}{Example}
\newtheorem{remark}{Remark}
\newtheorem{assumption}{Assumption}
\newtheorem{algorithm}{Algorithm}
\usepackage{bbm} 

\AtEndEnvironment{remark}{~\qed}
\AtEndEnvironment{example}{~\qed}

\def\sta{\star}
\def\na{\texttt{NA}}
\def\P{\mathbb{P}}
\def\top{\intercal}
\def\S{\mathcal{S}}
\def\vecge{\succcurlyeq}

\usepackage[textsize=tiny, textwidth = 2cm, shadow]{todonotes}

\begin{document}

\onehalfspacing

\title{\bf Randomization Inference with Sample Attrition
}

\author{
Xinran Li, Peizan Sheng, and Zeyang Yu
\footnote{
Xinran Li is Assistant Professor, Department of Statistics, University of Chicago, Chicago, IL 60637, USA (E-mail: \href{mailto:xinranli@uchicago.edu}{xinranli@uchicago.edu}). 
Peizan Sheng is Doctoral Candidate, Harris School of Public Policy, University of Chicago, Chicago, IL 60637, USA (E-mail: \href{mailto:peizan@uchicago.edu}{peizan@uchicago.edu}). 
Zeyang Yu is Assistant Professor, Department of Politics, Princeton University, Princeton, NJ 08544, USA (E-mail: \href{mailto:arthurzeyangyu@princeton.edu}{arthurzeyangyu@princeton.edu}).
}
}

\date{}
\maketitle

\begin{abstract}
\onehalfspacing

Randomization inference is a widely-used and appealing approach for analyzing treatment effects in randomized experiments, as it is finite-sample valid and does not require any distributional assumptions.
However, naive application of randomization inference may suffer from severe size distortion in the presence of sample attrition, where outcome data are missing for some units. 
In this paper, we propose new, computationally efficient methods for randomization inference that remain valid under a broad class of potentially informative missingness mechanisms, allowing a unit’s missingness to depend on its (unobserved) potential outcomes.
Specifically, we construct valid p-values for testing both sharp and bounded null hypotheses on treatment effects via a worst-case consideration of the classical Fisher randomization test.
Leveraging distribution-free test statistics, these worst-case p-values admit closed-form solutions.
Importantly, by incorporating both potential outcomes and potential missingness indicators into the test statistic, our methods can exploit structural assumptions such as monotone missingness, which are commonly adopted in applications due to their plausibility and ability to substantially improve inferential power.
Moreover, our approach connects to a range of partial identification bounds in the literature, which in some sense suggests the sharpness of our tests. 
We illustrate the proposed methods through both simulation studies and an empirical application. An R package implementing the proposed methods is publicly available.


    
\end{abstract}

\setlength{\abovedisplayskip}{2pt}
\setlength{\belowdisplayskip}{2pt}
\setlength{\abovedisplayshortskip}{1pt}
\setlength{\belowdisplayshortskip}{1pt}

\allowdisplaybreaks

\section{Introduction}\label{s1}

Randomization inference, also known as permutation inference, is widely used across the sciences, including fields such as economics, political science, medicine, and genomics \citep{gerber2012field, imbens2015causal, young2019channeling, bates2020causal, zhang2023randomization, chen2024role}. Under a sharp null hypothesis of, say, no treatment effect for any unit, the Fisher randomization test (FRT) provides exact finite-sample inference that relies only on the known treatment assignment mechanism and does not need any distributional assumptions or asymptotic approximations \citep{fisher1935design, lehmann2006testing}. 
However, its implementation requires complete outcome data. 
In practice, many experiments suffer from sample attrition—the failure to observe outcomes for some experimental units \citep{duflo2007using, gerber2012field, Zhaoetal2024}. 
For example, \cite{ghanem2023testing} find that more than 80\% of experimental studies in leading economics journals report missing outcomes. 
Moreover, simply discarding units with missing outcomes and applying standard randomization tests to the remaining observed units may not guarantee validity, particularly when dropouts differ systematically from those who remain. 
This is a well-known issue in the missing data literature \citep{little2019statistical}.

The prevalence of sample attrition in experiments and the widespread use of randomization inference raise several questions. What assumptions, explicit or implicit, do researchers make about missingness mechanisms? How can the size of a randomization test be controlled when testing a null hypothesis on treatment effects in the presence of attrition? 
How can the test be tailored to accommodate different assumptions on missingness mechanisms?
This paper's contribution is to answer these questions, providing theoretical foundations, methodological tools, and empirical guidance for using randomization inference in experiments with sample attrition.

With sample attrition, the usual sharp null hypothesis---such as the assumption of no treatment effect for every unit---is no longer ``sharp'', because the potential outcomes for each unit cannot be fully recovered from the observed data and the null hypothesis. As a result, the classical Fisher randomization test becomes infeasible.
To address this, we propose using the supremum of the p-value from the FRT over all possible configurations of the unknown potential outcomes resulting from sample attrition. While this p-value is valid, it is often computationally intractable due to its complex dependence on potential outcomes that are unobserved and not imputable. 
In particular, both the test statistic and its null distribution may depend on the unknown potential outcomes, and the null distribution often lacks a simple form and requires Monte Carlo approximation. We overcome this challenge by leveraging distribution-free, rank-based test statistics \citep{caughey2021randomization}, which allow us to transform the p-value computation into a worst-case optimization problem over solely the test statistics---one that admits a closed-form solution. This solution naturally aligns with the sharp bounds in \citet{manski1990nonparametric} and \citet{horowitz2000nonparametric} for, e.g., the average treatment effect among all units, where missing outcomes for treated and control units are assumed to lie at the boundary of the outcome support.
Moreover, the use of ranks enhances robustness, enabling our method to yield informative inference even when the potential outcomes are unbounded.

We then consider a series of missingness mechanisms, ranging from general missingness to monotone missingness, and ultimately to sharp missingness. The FRT, when applied with test statistics based solely on potential outcomes, is unable to incorporate information from the missingness mechanism. 
To address this, we propose the use of a composite outcome variable that combines both the original potential outcomes and the potential missingness indicators.
Importantly, this formulation incorporates the implications of the missingness assumptions into the test statistics. For example, under the monotone missingness assumption \citep{zhang2003estimation, lee2009training}, treated units with missing outcomes would also be missing if assigned to control.
We show that the FRT based on these composite outcomes can yield more powerful, yet still valid, tests by leveraging information in the specific missingness mechanism. 

We further improve the power of the test based on the composite outcomes by exploiting the fact that the distribution of missingness types can be (partially) inferred from the observed data. Specifically, we propose a two-step procedure \citep{berger1994p}. In the first step, we construct confidence or prediction sets for the missingness types. In the second step, we impute worst-case composite potential outcomes subject to the constraints implied by the first-step confidence or prediction sets and the missingness assumptions. By combining information from the observed missingness patterns and the assumed missingness mechanisms, this two-step procedure can further enhance the power of the test. 
Moreover, the resulting tests align with the sharp bounds in \citep{zhang2003estimation} and \citet{lee2009training} for the average treatment effect among units who would be observed under both treatment and control, under general and monotone missingness assumptions.

Our paper contributes to several strands of literature. 
First, our work contributes to the literature on statistical inference, particularly randomization-based inference, in the presence of sample attrition.
Existing approaches often invoke assumptions such as missing at random or zero treatment effect on missingness \citep{Kennes2012Choice, ivanova2022randomization, li2021randomization, heng2023design, heussen2024randomization, Zhaoetal2024}.
In contrast, our method accommodates informative missingness, in which missingness can depend on potential outcomes, and provides valid randomization tests under various missingness mechanisms.

Second, this paper contributes to the growing literature on randomization tests beyond sharp null hypotheses. While classical randomization tests are exact under sharp nulls, recent research has extended them to test for weak nulls, such as those involving average treatment effects \citep{chung2013exact, li2017general, bugni2018inference, wu2021randomization, bai2022inference} or quantiles of individual treatment effects \citep{caughey2021randomization, su2022treatment, chen2024enhanced}.
We extend this literature by developing finite-sample valid randomization tests under sample attrition, which renders the classical sharp null hypothesis non-sharp.

Third, it contributes to the literature on the partial identification of treatment effects under sample attrition, where missingness may be informative and may depend on unobserved outcomes.
\citet{manski1990nonparametric} and \citet{horowitz2000nonparametric} provide sharp bounds on treatment effects under minimal assumptions, although these bounds can be wide. Subsequent studies have sought to tighten these bounds by introducing additional assumptions. For instance, \citet{zhang2003estimation} and \citet{lee2009training} impose a monotonicity condition on missingness, and derive bounds on treatment effects among only those units that would be observed under both treatment and control. In contrast, our paper addresses a different but related question: how to conduct randomization inference under sample attrition using a range of missingness assumptions, without imposing any assumptions on the outcome variable? Notably, the worst-case conservative inference procedure developed here directly connects to the bounds in \citet{horowitz2000nonparametric}, \citet{zhang2003estimation}, and \citet{lee2009training}. To our knowledge, this is the first paper to formally and explicitly link robust randomization inference with the framework of partial identification.

\section{Framework and notation}\label{sec:framework}

\subsection{Potential outcomes, potential missingness, and treatment assignment}\label{sec:PO}

Consider a randomized experiment conducted on $n$ units with two treatment arms. We assume that the stable unit treatment value assumption (SUTVA) holds \citep{rubin1980discussion}, and use the potential outcome framework \citep{neyman1923application, rubin1974estimating}. Let $Y_{i}(0)$ and $Y_{i}(1)$ denote the potential outcomes under control and treatment, respectively, for unit $i$ in the full sample. We use $\bs{Y}(0) = (Y_{1}(0), Y_{2}(0), \ldots, Y_{n}(0))^{\intercal}$ and $\bs{Y}(1) = (Y_{1}(1), Y_{2}(1), \ldots, Y_{n}(1))^{\intercal}$ to denote the control and treatment potential outcome vectors for all units in the full sample. Let $\tau_{i} = Y_{i}(1) - Y_{i}(0)$ be the individual treatment effect for unit $i$, and let $\bs{\tau} = (\tau_{1}, \tau_{2}, \ldots, \tau_{n})^{\intercal}$ be the vector of individual treatment effects for the full sample. Let $Z_{i}$ be the treatment assignment for unit $i$, where $Z_{i}$ equals 1 if the unit is assigned to treatment and $0$ otherwise. We use $\bs{Z} = (Z_{1}, Z_{2}, \ldots, Z_{n})^{\intercal}$ to denote the treatment assignment vector for all units. The full sample's outcome $Y_{i}^{\sta}$ for each unit $i$ is realized through $Y_{i}^{\sta} = (1 - Z_{i}) Y_{i}(0)  + Z_{i} Y_{i}(1) $. We use $\bs{Y}^{\sta} = (Y_{1}^{\sta}, Y_{2}^{\sta}, \ldots, Y_{n}^{\sta})^{\intercal}$ to denote the outcome vector for all units in the full sample. Equivalently, the outcome vector is realized through $\bs{Y}^{\sta} = (\bs{1} - \bs{Z}) \circ \bs{Y}(0) + \bs{Z} \circ \bs{Y}(1)$, where $\circ$ denotes element-wise multiplication.

We allow sample attrition in the experiment \citep{duflo2007using, gerber2012field}. Let $M_{i}$ be the missingness indicator for unit $i$. Let $\bs{M} = (M_{1}, M_{2}, \ldots, M_{n})^{\intercal}$ be the vector of missingness indicators. We define the potential missingness indicators $M_i(0)$ and $M_i(1)$ for each units $i$. We use $\bs{M}(0) = (M_{1}(0), M_{2}(0), \ldots, M_{n}(0))^{\intercal}$ and $\bs{M}(1) = (M_{1}(1), M_{2}(1), \ldots, M_{n}(1))^{\intercal}$ to denote the vectors of potential missingness indicators under control and treatment, respectively, for all units. In other words, we assume SUTVA also holds for potential missingness. The observed missingness indicator is realized through $M_{i} = (1 - Z_{i}) M_{i}(0) + Z_{i} M_{i}(1)$, or in vector form, $\bs{M} = (\bs{1} - \bs{Z}) \circ \bs{M}(0) + \bs{Z} \circ \bs{M}(1)$. Note that some of our results hold even without assuming SUTVA for potential missingness; see Remark \ref{rmk:sharp_null_gen_sutva}.

Let $Y_{i}$ denote the observed outcome for unit $i$. Specifically, when $M_{i} = 1$, the realized outcome $Y_{i}^{\sta}$ is observed, and we set $Y_{i} = Y_{i}^{\sta}$. When $M_{i}=0$, the realized outcome $Y_{i}^{\sta}$ is not observed, and we set $Y_i = \na$, where $\na$ indicates ``not available''. 
Define $0 \cdot \na = 0$ and $0 + \na = \na$.
Then the observed outcome $Y_{i}$ has the following equivalent forms:
\begin{align*}
    Y_{i} &= M_{i}Y_{i}^{\sta} + (1-M_i) \na \\
    &= Z_{i} [ M_{i}(1) Y_{i}(1) + \{1-M_i(1)\} \na ] + (1 - Z_{i}) [ M_{i}(0) Y_{i}(0) + \{1-M_i(0)\} \na ].
\end{align*}
Therefore, for units with missing outcomes (i.e., units with $M_{i} = 0$), none of their potential outcomes are observed. For units without missing outcomes (i.e., units with $M_{i} = 1$), one of their potential outcomes is observed. 
For descriptive convenience, 
we further introduce 
\begin{align}\label{eq:n_zm}
    n_{zm} = \sum_{i=1}^n \I\{Z_i = z, M_i = m\}, 
    \qquad (z, m \in \{0,1\}). 
\end{align}

It should be emphasized that sample attrition in this context can encompass a wide range of scenarios. Missing outcomes may result from self-selection \citep{heckman1974shadow, lee2009training}, survey nonresponse \citep{rubin1987multiple, Little01071988, duflo2007using}, or other structural reasons \citep{zhang2003estimation, frangakis_principal_2004}.

\subsection{Design-based inference and treatment assignment mechanism}\label{sec:assign}

In this paper, we conduct design-based inference (also known as randomization-based or finite population inference), treating all potential outcomes (and potential missingness) as fixed constants\footnote{
In the main paper, we treat all potential missingness indicators as fixed constants. We also consider a missing completely at random mechanism, under which these indicators are random; to avoid confusion, we defer this case to the supplementary material.}—or equivalently, as conditioned on—so that randomness solely arises from the experimental design \citep{neyman1923application, fisher1935design, li2017general, 
abadie2020sampling, abadie2023should, roth2023efficient, zhang2023randomization}. 

In design-based inference, the distribution of the treatment assignment $\bs{Z}$, also called the treatment assignment mechanism, is what governs the statistical inference. 
Throughout the paper, we will focus on the \textit{completely randomized experiment} (CRE), under which
\begin{align}\label{eq:CRE}
    \P( \bs{Z} = \bs{z} \mid \bs{Y}(1), \bs{Y}(0), \bs{M}(1), \bs{M}(0) )
    = 
    \begin{cases}
        \frac{1}{{n \choose n_1}}, & \text{if }  \bs{z} \in \{0,1\}^n 
        \text{ and } \sum_{i=1}^n z_i = n_1, \\
        0, & \text{otherwise},
    \end{cases}
\end{align}
where $n_1$ and $n_0$ are two fixed positive integers denoting the numbers of treated and control units, respectively, under the CRE. Importantly, the design in \eqref{eq:CRE} requires that 
the treatment assignment is independent of both the potential outcomes and the potential missingness indicators.

The inference in this paper applies to randomized experiments with exchangeable treatment assignments \citep{caughey2021randomization}, including CREs and Bernoulli randomized experiments (BREs). 
This is because, once we condition on the numbers of treated and control units, the assignment mechanism is equivalent to a CRE, allowing such experiments to be analyzed in the same way.

\section{Missingness mechanisms}\label{sec:miss_mech}

\subsection{General missingness mechanisms}

The missingness mechanism is central to the inference of treatment effects. Random assignment and SUTVA can be ensured by design. However, missingness often arises from self-selection that is neither controlled nor observed by researchers. This may induce dependence between potential missingness and potential outcomes, which we call informative missingness. A canonical case is \textit{threshold missingness}, where outcomes are observed only if they exceed a given threshold. We give two examples of \textit{threshold missingness} below.

\begin{example}\label{eg:threshold_miss_job_train}
Consider an experiment where the outcome is the offered market wage and the treatment is a job training program. Let $Y_i(1)$ and $Y_i(0)$ denote unit $i$’s potential offered wages under treatment and control, and let $c_i$ be the reservation wage. The market wage is observed only if unit $i$ participates in the labor market \citep{heckman1974shadow}:
\begin{align*}
M_i(0)=\I\{Y_i(0) \ge c_i\}, \quad
M_i(1)=\I\{Y_i(1) \ge c_i\}.
\end{align*}
\end{example}

\begin{example}\label{eg:threshold_miss_educ}
Consider an education experiment in which students are randomized to one of two high school programs, and the outcome is the final test score among graduates. 
Let $Y_i(1)$ and $Y_i(0)$ denote unit i’s potential final test scores, and $M_i(1)$ and $M_i(0)$ the corresponding graduation indicators, under treatment and control.
The final test score is observed only if unit $i$ graduates (i.e., does not drop out) \citep{zhang2003estimation}.
\end{example}

Because researchers may have limited information about the missingness mechanism, we develop randomization inference procedures for sharp and bounded nulls that remain valid, though potentially conservative, under a general missingness mechanism allowing arbitrary dependence between potential outcomes and missingness.
The mechanism is formally defined below.

\begin{assumption}[General Missingness]\label{asmp:general}
    $M_i(1)$s and $M_i(0)$s can be any constants in $\{0,1\}$, and their values can depend on the potential outcomes $Y_i(1)$s and $Y_i(0)$s in an arbitrary way. 
\end{assumption}

\subsection{Monotone missingness mechanisms}

We also consider additional assumptions on the missingness mechanism. Importantly, these additional assumptions can improve the power of our inference for treatment effects. This and the following subsections present the assumptions to be studied, along with motivating examples.

The next two assumptions require the treatment effect on the missingness indicator to be monotone, either non-positive or non-negative. We state them formally below.

\begin{assumption}[Monotone Missingness where treatment discourages missingness]\label{asmp:mo_miss_M1_ge_M0}
    $M_i(1)$s and $M_i(0)$s are constants in $\{0,1\}$ such that $M_i(1) \ge  M_i(0)$ for each $1\le i \le n$. 
\end{assumption}

\begin{assumption}[Monotone Missingness where treatment encourages missingness]\label{asmp:mo_miss_M1_le_M0}
    $M_i(1)$s and $M_i(0)$s are constants in $\{0,1\}$ such that $M_i(1) \le  M_i(0)$ for each $1\le i \le n$.  
\end{assumption}

\cite{zhang2003estimation} and \cite{lee2009training} assume Assumption~\ref{asmp:mo_miss_M1_ge_M0}. Consider the job training experiment in Example \ref{eg:threshold_miss_job_train}. The monotone missingness assumption holds when the training program weakly increases each individual's propensity to participate in the labor market. Consider the education experiment in Example \ref{eg:threshold_miss_educ}. The monotone missingness holds when the treatment weakly decreases each student’s propensity to drop out of school.

Note that, under monotone missingness, both Assumptions \ref{asmp:mo_miss_M1_ge_M0} and \ref{asmp:mo_miss_M1_le_M0} can be satisfied through label switching and change of the outcome sign, under which the treatment effects remain unchanged. For example, if the original sample satisfies Assumption \ref{asmp:mo_miss_M1_ge_M0}, then, once we switch the treatment labels and change the outcome signs, Assumption \ref{asmp:mo_miss_M1_le_M0} will hold and the corresponding treatment effects will be the same as those in the original sample. Therefore, when analyzing data, we can proceed with either Assumption \ref{asmp:mo_miss_M1_ge_M0} or \ref{asmp:mo_miss_M1_le_M0}.
However, since switching the treatment labels leads to different test statistics that emphasize slightly different aspects of the treatment effects, we still consider and distinguish between these two monotone missingness mechanisms; see Remark \ref{rmk:test_Y1} for related discussion.

\subsection{Sharp missingness mechanism}

An extreme case of Assumptions \ref{asmp:mo_miss_M1_ge_M0} and \ref{asmp:mo_miss_M1_le_M0} occurs when $M_i(1) = M_i(0)$ for all $i$. When this condition holds, both assumptions are satisfied, implying that Fisher’s sharp null of no treatment effect holds for the missingness indicator \citep{fisher1935design}. We refer to this as the sharp missingness assumption.

\begin{assumption}[Sharp Missingness]\label{asmp:constant_miss}
    $M_i(1)$s and $M_i(0)$s are constants in $\{0,1\}$ such that $M_i(1) =  M_i(0)$ for each $1\le i \le n$. 
\end{assumption}

\cite{heng2023design} show that Assumption \ref{asmp:constant_miss} can still hold even when missingness depends on potential outcomes, provided additionally that (i) Fisher’s null of no treatment effect holds for the outcome and (ii) treatment influences missingness only through the outcome.

\section{Preliminaries: Fisher randomization tests}\label{sec:sharp_null}
 
\subsection{Sharp null hypothesis and Fisher randomization p-value}\label{sec:frt}

A sharp null hypothesis specifies all individual treatment effects \citep{fisher1935design}. It has the following general form:
\begin{align}\label{eq:null_fisher_sharp}
    H_{\bs{\delta}}: \bs{\tau} = \bs{\delta},
\end{align}
where $\bs{\delta}\in \mathbb{R}^n$ is a prespecified constant vector representing the hypothesized treatment effects for all units. Under the null hypothesis $H_{\bs{\delta}}$ and in the absence of missing data, we are able to impute the potential outcomes for all individuals from the observed data:
\begin{align}\label{eq:imputation}
    \bs{Y}(1) = \bs{Y}^{\sta} + (1 - \bs{Z}) \circ \bs{\delta}, 
    \quad 
    \bs{Y}(0) = \bs{Y}^{\sta} - \bs{Z} \circ \bs{\delta}.
\end{align}
Note that the equalities in \eqref{eq:imputation} hold if and only if the sharp null hypothesis $H_{\bs{\delta}}$ is true.

To test the null in \eqref{eq:null_fisher_sharp}, consider a general test statistic of the form $t(\cdot, \cdot): \mathcal{Z} \times \mathbb{R}^{n} \xrightarrow[]{} \mathbb{R}$, which is a generic function with two arguments: a treatment assignment vector $\bs{z} \in \mathcal{Z}$ and an outcome vector $\bs{y} \in \mathbb{R}^{n}$. Following \cite{rosenbaum2002design}, we use test statistics $t(\bs{Z}, \bs{Y}(0))$ that depend on the treatment assignment $\bs{Z}$ and the control potential outcomes in \eqref{eq:imputation} imputed under the sharp null \eqref{eq:null_fisher_sharp}. The tail probability of the randomization distribution of the test statistic $t(\bs{Z}, \bs{Y}(0))$ under the sharp null is
\begin{align}\label{eq:rand_dist}
    G(c) \coloneqq \mathbb{P}( t(\bs{A}, \bs{Y}(0)) \geq c ) = \sum_{\bs{a} \in \mathcal{Z}} \mathbb{P}(\bs{A} = \bs{a}) \I\{ t(\bs{a}, \bs{Y}(0)) \geq c \},
    \quad 
    ( c\in \mathbb{R} )
\end{align}
where $\bs{A}$ is a generic random treatment assignment vector that follows the same distribution as the observed assignment vector $\bs{Z}$. The corresponding randomization p-value under the sharp null is
\begin{align}\label{eq:pval_FRT}
    p_{\bs{Z}} \coloneqq G (t(\bs{Z}, \bs{Y}(0))) = \sum_{\bs{a} \in \mathcal{Z}} \mathbb{P}(\bs{A} = \bs{a}) \I\{ t(\bs{a}, \bs{Y}(0)) \geq t(\bs{Z}, \bs{Y}(0)) \},
\end{align}
which is a valid p-value in the sense that $\P(p_{\bs{Z}} \le \alpha) \le \alpha$ for $\alpha \in (0,1)$. 

\begin{remark}\label{rmk:test_Y1}
Throughout the paper, we use test statistics based on either the control potential outcome or the composite control potential outcome introduced in Section \ref{sec:sharp_null_composite}. As noted by \citet{rosenbaum_effects_2001} and \citet{chen2024enhanced}, such tests primarily capture treatment effects among treated units. By the same logic, the approach can be formulated using test statistics of the form $t(\bs{Z}, \bs{Y}(1))$, based on the treated potential outcome or an analogous composite treated potential outcome; in this case, the test targets treatment effects among control units. This formulation is obtained by switching the treatment and control labels, under which treatment effects remain invariant, although the assumptions on the missingness mechanism change accordingly. For example, Assumption \ref{asmp:mo_miss_M1_ge_M0} becomes Assumption \ref{asmp:mo_miss_M1_le_M0} after label switching, and vice versa.
\end{remark}

\subsection{Distribution-free rank statistics}\label{sec:dist_free}

When sample attrition occurs, the p-value in \eqref{eq:pval_FRT} cannot be computed. In particular, missing outcomes imply that the control potential outcomes of the missing units cannot be imputed under the sharp null. As a result, both the realized test statistic $t(\bs{Z}, \bs{Y}(0))$ and its true randomization distribution \eqref{eq:rand_dist} are unknown.

Consequently, valid randomization-based inference in the presence of attrition requires procedures that maintain size control across all possible configurations of missing outcomes. 
This can be challenging, as attrition affects both the test statistic and its randomization distribution, the latter having no simple form and often requiring Monte Carlo approximation.
We therefore employ the distribution-free, rank-based test statistics proposed by \citet{caughey2021randomization}, specifically the two classes described below.

The first is a class of rank-sum statistics of the following form:
\begin{align}\label{eq:test_stat_R_phi}
    t_{\R, \phi}(\bs{z}, \bs{y})  & = \sum_{i=1}^n z_i \phi( \rank_i(\bs{y}) )
    = \sum_{i=1}^n z_i \phi\left( \sum_{j=1}^n \psi_{i,j}(y_i, y_j) \right)
\end{align}
where $\bs{z}\in \{0,1\}^n$ is the treatment assignment vector, $\bs{y} = (y_1, \ldots, y_n)^\top \in \mathbb{R}^n$ is the outcome vector,
$\rank_i(\bs{y})$ denotes the rank of the $i$-th coordinate of $\bs{y}$, $\phi(\cdot)$ is a prespecified nondecreasing rank transformation, 
and $\psi_{i,j}(\cdot, \cdot)$ is an indicator function for pairwise comparison, which will be defined shortly.
Setting $\phi(r) = r$ reduces \eqref{eq:test_stat_R_phi} to the Wilcoxon rank-sum statistic \citep{wilcoxon1945individual}. Alternatively, with $\phi(r) = \binom{r-1}{s-1} \I(r \ge s)$ for a fixed integer $s > 1$, it becomes the Stephenson rank-sum statistic \citep{robert1985two}.
To break ties, we assume units are randomly ordered and use their indices as a tie-breaker.
For example, $\rank_i(\bs{y}) < \rank_j(\bs{y})$ if and only if (i) $y_i < y_j$ or (ii) $y_i = y_j$ and $i<j$. 
Under this rule, the indicator function $\psi_{i,j}(\cdot, \cdot)$ in \eqref{eq:test_stat_R_phi} is defined for $1\le i, j\le n$ and $y, y'\in \mathbb{R}$ as:
\begin{align}\label{eq:psi_ij} 
    \psi_{i, j}(y, y') = \I\{ y > y' \} + \I\{ y = y' \} \I \{ i \geq j \}.
\end{align}
Thus, $\rank_i(\bs{y}) = \sum_{j=1}^n \psi_{i,j}(y_i, y_j)$ for all $i$, justifying the second equality in \eqref{eq:test_stat_R_phi}. 

The second is a class of generalized Mann-Whitney-type U-statistics:
\begin{align}\label{eq:test_stat_mann_whitney}
t_{\R, \phi}(\bs{z}, \bs{y}) & = \sum_{i=1}^n z_i \phi\left( \sum_{j = 1}^{n} (1 - z_j) \psi_{i, j}(y_i, y_j) \right),
\end{align}
and
\begin{align}\label{eq:test_stat_mann_whitney_negative}
t_{\R, \phi}(\bs{z}, \bs{y}) & = -\sum_{i=1}^n (1 - z_i) \phi\left( \sum_{j = 1}^{n} z_j \psi_{i, j}(y_i, y_j) \right),
\end{align}
where $\bs{z}$, $\bs{y}$, $\phi(\cdot)$, and $\psi_{i, j}(\cdot, \cdot)$ are defined analogously as in \eqref{eq:test_stat_R_phi} and \eqref{eq:psi_ij}. We consider choosing $\phi$ as $\phi(r) = r^{s-1}$, where $s$ is a parameter governing the weights assigned to responses with different ranks. 
When $s = 2$, both \eqref{eq:test_stat_mann_whitney} and \eqref{eq:test_stat_mann_whitney_negative} reduce to the usual Mann-Whitney U-statistic. When $s > 2$, larger responses among the treated (or control) group receive more weight, making the statistic more sensitive to extreme values; this is analogous to rank transformations for Stephenson rank-sum statistics of the form \eqref{eq:test_stat_R_phi}.

We briefly compare the rank-based statistics in \eqref{eq:test_stat_R_phi}, \eqref{eq:test_stat_mann_whitney} and \eqref{eq:test_stat_mann_whitney_negative}. When $\phi$ is the identity, they are equivalent up to a constant shift, reflecting the well-known equivalence between the Wilcoxon rank-sum and Mann–Whitney U statistics. Both are distribution-free and therefore suitable for our proposed randomization inference procedures. Due to these similarities, we use the same notation for the two, specifying the form as needed throughout the paper.
However, for non-identity $\phi$, the statistics differ: \eqref{eq:test_stat_R_phi} is based on full-sample ranks, whereas \eqref{eq:test_stat_mann_whitney} ranks treated units relative to controls and \eqref{eq:test_stat_mann_whitney_negative} ranks control units relative to treated units.
This distinction can be practically important: statistics of the form \eqref{eq:test_stat_mann_whitney} or \eqref{eq:test_stat_mann_whitney_negative} may be easier to optimize when computing valid p-values, especially under monotone missingness.

Importantly, under the CRE and with random tie-breaking, the distribution of $t_{\R, \phi}(\bs{Z},  \bs{y})$ in either \eqref{eq:test_stat_R_phi}, \eqref{eq:test_stat_mann_whitney} or \eqref{eq:test_stat_mann_whitney_negative} does not depend on the value of $ \bs{y}$. Specifically,
$t_{\R, \phi}(\bs{Z},  \bs{y}) \sim t_{\R, \phi}(\bs{Z}, \bs{y}')$ for any constant vectors $ \bs{y},  \bs{y}' \in \mathbb{R}^n$, where $\sim$ denotes equal in distribution \citep{caughey2021randomization}.
Thus, we can define
\begin{equation}\label{eq:G_R_phi}
    G_{\R, \phi}(c) = \mathbb{P}( t_{\R, \phi}(\bs{Z}, \bs{y}) \ge c ), 
    \text{ where } \bs{Z} \text{ is from a CRE and } \bs{y} \text{ can be any fixed vector in } \mathbb{R}^n.
\end{equation}
In other words, the randomization distribution $G_{\R, \phi}(c)$ is determined by the rank transformation $\phi(\cdot)$ and the numbers of treated and control units, $n_1$ and $n_0$.

\section{A Randomization Test Using Only Potential Outcomes}\label{sec:sharp_null_gen}

Throughout the paper, we first focus on the sharp null hypothesis in \eqref{eq:null_fisher_sharp}, and later extend to bounded null hypotheses in Section \ref{sec:validity_p_val_bdd_null}.
We now study randomization tests for the sharp null under the general missingness mechanism of Assumption \ref{asmp:general}, which allows arbitrary dependence between potential missingness and potential outcomes (i.e., arbitrarily informative missingness).

Under sample attrition, the standard randomization p-value for the sharp null in \eqref{eq:null_fisher_sharp} is generally not computable because both the test statistic and its randomization distribution are unknown (see Section \ref{sec:dist_free}). To address this challenge, we use the rank statistic $t_{\R, \phi}(\bs{Z}, \bs{Y}(0))$ introduced in Section \ref{sec:dist_free}, which depends only on control potential outcomes. Due to the distribution-free property of rank statistics, the randomization distribution in \eqref{eq:G_R_phi} does not depend on $\bs{Y}(0)$ and is therefore known exactly. Consequently, the problem of obtaining a valid p-value that maintains size control over all possible configurations of missing outcomes reduces to minimizing the realized test statistic $t_{\R, \phi}(\bs{Z}, \bs{Y}(0))$ subject to the constraints imposed by the observed data and the sharp null hypothesis. In other words, this valid p-value is a worst-case p-value.

This strategy will be used throughout the paper. In later sections, we carefully choose test statistics that incorporate information about potential missingness. This allows us to exploit constraints imposed by the assumptions on the missingness mechanisms and by the distribution of missingness types implied by the observed data, thereby increasing the power of the resulting tests.

\begin{table}[h!]
    \centering
    \scalebox{0.96}{
    \begin{threeparttable}
    \caption{Observed data, potential outcomes and potential missingness, under a sharp null in \eqref{eq:null_fisher_sharp} and different missingness mechanisms.}
    \label{tab:sharp_null}
    \begin{tabular}{ccccccccccc}
    \hline
    $Z_i$ & $M_i$ & $Y_i$ & $M_i(1)$ & $M_i(0)$ & $Y_i(1)$ & $Y_i(0)$ & $Y_i^\wc(0)$ & $\tilde{Y}_{\bs{b}, i}(0)$ & $\tilde{Y}_{\bs{b}, i}^\wc(0)$ & $\tilde{Y}_{\bs{b}, i}^\wc(0)$ \\
    \hline
    \multicolumn{11}{l}{Panel A: General Missingness in Assumption \ref{asmp:general}} \\
    $1$ & $1$ & \checkmark & $1$ & ? & $Y$ & $Y-\delta$ & $Y-\delta$ & $Y-\delta \text{ or } b_{01} $ & $\min\{Y-\delta, b_{01}\}$ & $Y-\delta$ \\
    $1$ & $0$ & ? & $0$ & ? & ? & ?  & $-\infty$ & $b_{00} \text{ or } b_{10}$ & $\min\{b_{00}, b_{10}\}$ & $-\infty$ \\
    $0$ & $1$ & \checkmark & ? & $1$ & $Y+\delta$ & $Y$ & $Y$ & $Y \text{ or } b_{10}$ & $\max\{Y, b_{10}\}$ & $Y$ \\
    $0$ & $0$ & ? & ? & $0$ & ? & ? & $\infty$ & $b_{00} \text{ or } b_{01}$ & $\max\{b_{00}, b_{01}\}$ & $\infty$ \\[1pt]
    \multicolumn{11}{r}{suggested $\bs{b}$: $b_{01} = \infty$ and $b_{10} = -\infty$}\\
    \hline
    \multicolumn{11}{l}{Panel B: Monotone Missingness in Assumption \ref{asmp:mo_miss_M1_ge_M0}, i.e., $M_i(1) \ge M_i(0), \forall i$} \\
    $1$ & $1$ & \checkmark & $1$ & ? & $Y$ & $Y-\delta$ & $Y-\delta$ & $Y-\delta \text{ or } b_{01}$ & $\min\{Y-\delta, b_{01}\}$ & $Y-\delta$ \\
    $1$ & $0$ & ? & $0$ & $0$ & ? & ? & \cellcolor{gray!30}$-\infty$ & $b_{00}$ & $b_{00}$ & \cellcolor{gray!30}$\infty$ \\
    $0$ & $1$ & \checkmark & $1$ & $1$ & $Y+\delta$ & $Y$ & $Y$ & $Y$ & $Y$ & $Y$ \\
    $0$ & $0$ & ? & ? & $0$ & ? & ? & $\infty$ & $b_{00} \text{ or } b_{01}$ & $\max\{b_{00}, b_{01}\}$ & $\infty$ \\[1pt]
    \multicolumn{11}{r}{suggested $\bs{b}$: $b_{00} = b_{01} = \infty$}\\
    \hline
    \multicolumn{11}{l}{Panel C: Monotone Missingness in Assumption \ref{asmp:mo_miss_M1_le_M0}, i.e., $M_i(1)\le M_i(0), \forall i$} \\
    $1$ & $1$ & \checkmark & $1$ & $1$ & $Y$ & $Y-\delta$ & $Y-\delta$ & $Y-\delta$ & $Y-\delta$ & $Y-\delta$ \\
    $1$ & $0$ & ? & $0$ & ? & ? & ? & $-\infty$ & $b_{00} \text{ or } b_{10}$ & $\min\{b_{00}, b_{10}\}$ & $-\infty$ \\
    $0$ & $1$ & \checkmark & ? & $1$ & $Y+\delta$ & $Y$ & $Y$ & $Y \text{ or } b_{10}$ & $\max\{Y, b_{10}\}$ & $Y$ \\
    $0$ & $0$ & ? & $0$ & $0$ & ? & ? & \cellcolor{gray!30}$\infty$ & $b_{00}$ & $b_{00}$ & \cellcolor{gray!30}$-\infty$ \\[1pt]
    \multicolumn{11}{r}{suggested $\bs{b}$: $b_{00} = b_{10} = -\infty$}\\
    \hline
    \multicolumn{11}{l}{Panel D: Sharp Missingness in Assumption \ref{asmp:constant_miss}} \\
    $1$ & $1$ & \checkmark & $1$ & $1$ & $Y$ & $Y-\delta$ & $Y-\delta$ & $Y-\delta$ & $Y-\delta$ & \\
    $1$ & $0$ & ? & $0$ & $0$ & ? & ? & $-\infty$ & $b_{00}$ & $b_{00}$ & \\
    $0$ & $1$ & \checkmark & $1$ & $1$ & $Y+\delta$ & $Y$ & $Y$ & $Y$ & $Y$ & \\
    $0$ & $0$ & ? & $0$ & $0$ & ? & ? & $\infty$ & $b_{00}$ & $b_{00}$ & \\
    \hline
    \end{tabular}
    \begin{tablenotes}
    \item \emph{Note}: In the table, \checkmark means that a quantity is observed, while ? means that a quantity is unknown and can take any value in its support. In the last column, we also give the worst-case imputation with suggested $\bs{b}$. 
    \end{tablenotes}
    \end{threeparttable}}
\end{table}

Panel A in Table \ref{tab:sharp_null} summarizes the available information on potential outcomes and missingness indicators, based on the observed data and the sharp null hypothesis in \eqref{eq:null_fisher_sharp}. From the table, we do not know the value of $Y_{i}(0)$ for units whose outcomes are missing.
Intuitively, $t_{\R, \phi}(\bs{Z}, \bs{Y}(0))$ achieves its minimum value when the potential outcomes $Y_i(0)$ for treated units are as small as possible, while those for control units are as large as possible.
This motivates us to consider the following control potential outcome vector:
$\bs{Y}_{\bs{Z}, \bs{\delta}}^{\gen}(0) = (Y_{\bs{Z}, \bs{\delta}, 1}^{\gen}(0), \ldots, Y_{\bs{Z}, \bs{\delta}, n}^{\gen}(0))^\top,$
where the superscript $\gen$ indicates the general missingness mechanism, the subscript $\bs{\delta}$ represents the sharp null of interest in \eqref{eq:null_fisher_sharp}, and the subscript $\bs{Z}$ emphasizes its dependence on the observed treatment assignment, and
\begin{align}\label{eq:worst_gen}
    Y_{\bs{Z}, \bs{\delta}, i}^{\gen}(0) = \begin{cases}
        Y_{i} - \delta_{i} &\quad \text{if $Z_{i} = 1$ and $M_{i} = 1$}, \\
        -\infty &\quad \text{if $Z_{i} = 1$ and $M_{i} = 0$}, \\
        Y_{i} &\quad \text{if $Z_{i} = 0$ and $M_{i} = 1$}, \\
        \infty &\quad \text{if $Z_{i} = 0$ and $M_{i} = 0$}, \\
    \end{cases}
    \qquad \qquad
    (1\le i \le n). 
\end{align}
Obviously, $\bs{Y}_{\bs{Z}, \bs{\delta}}^{\gen}(0)$ is a ``feasible'' imputed value of $\bs{Y}(0)$ based on the observed data and the sharp null hypothesis. 
The theorem below shows that $\bs{Y}_{\bs{Z}, \bs{\delta}}^{\gen}(0)$ indeed yields the ``worst-case'' configuration of the control potential outcomes, leading to a valid p-value for the sharp null hypothesis $H_{\bs{\delta}}$.

\begin{theorem}\label{thm:sharp_null_gen}
\rm 
Under the CRE and Assumption \ref{asmp:general}, 
a valid p-value for testing the sharp null hypothesis $H_{\bs{\delta}}$ in \eqref{eq:null_fisher_sharp} is 
\begin{align}\label{eq:p_val_sharp_null_orig}
    p_{\bs{Z}, \bs{\delta}}^\gen  \coloneqq G_{\R, \phi}\left( t_{\R, \phi}\left(\bs{Z}, \bs{Y}_{\bs{Z}, \bs{\delta}}^{\gen}(0) \right) \right),
\end{align}
where $t_{\R, \phi}$ is defined in either \eqref{eq:test_stat_R_phi}, \eqref{eq:test_stat_mann_whitney} or \eqref{eq:test_stat_mann_whitney_negative}, $G_{\R, \phi}$ is defined in \eqref{eq:G_R_phi}, and $\bs{Y}_{\bs{Z}, \bs{\delta}}^{\gen}(0)$ is defined  in \eqref{eq:worst_gen}. That is, under $H_{\bs{\delta}}$, $\mathbb{P}\left( p_{\bs{Z}, \bs{\delta}}^\gen \leq \alpha \right) \leq \alpha$ for any $\alpha \in (0, 1)$.
\end{theorem}

\begin{remark}\label{rmk:sharp_null_gen_sutva}
Theorem \ref{thm:sharp_null_gen} remains valid even when the SUTVA is violated for potential missingness.
This is because the p-value in \eqref{eq:p_val_sharp_null_orig} is always an upper bound on the randomization p-value based on the test statistic $t_{\R, \phi}(\bs{Z}, \bs{Y}(0))$, regardless of whether SUTVA for potential missingness holds.
\end{remark}

Our worst-case inferential procedure in Theorem \ref{thm:sharp_null_gen} has a similar flavor with the worst-case identification approach in the partial identification literature \citep{robins1989probability, robins1989estimability, manski1990nonparametric, horowitz2000nonparametric}. This approach assumes that 
(i) the missing values in the treated group are at the lower bound of the support of the outcome and (ii) the missing values in the control group are at the upper bound of the support of the outcome. 
This can then give us a partially identified set for the parameter of interest, such as the average treatment effect. Although we are studying a conceptually different problem, namely, inference for sharp null hypothesis, the worst-case scenario for the p-value nevertheless coincides with the worst-case identification problem.
Moreover, our approach has the advantage that it can still give informative inference results even when the support of the outcome is unbounded.

Note that imposing assumptions on potential missingness (such as Assumptions \ref{asmp:mo_miss_M1_ge_M0}, \ref{asmp:mo_miss_M1_le_M0} or \ref{asmp:constant_miss}) or incorporating information about its distribution, which can be learned as least partially from the observed data, does not change the worst-case imputation of the control potential outcomes.
Consequently, they do not change the worst-case test statistic $t_{\R, \phi}(\bs{Z}, \bs{Y}_{\bs{Z}, \bs{\delta}}^{\gen}(0))$ and the resulting worst-case p-value.
In the next section, we will carefully choose a test statistic that leverages these assumptions to refine worst-case imputation and improve statistical power.

\section{Randomization Tests Using Composite Potential Outcomes}\label{sec:sharp_null_composite}

\subsection{Outcome-missingness composite test statistic}\label{subsec:composite_test_stats}

To use the missingness assumptions in the randomization inference procedure, we modify the test statistic by incorporating potential missingness. Specifically, we introduce the composite control potential outcome for each unit $i$:
$$
\tilde{Y}_{\bs{b},i}(0) \coloneqq Y_i(0) M_i(0) M_i(1) + b_{00} (1 - M_i(0)) (1 - M_i(1)) + b_{01} (1 - M_i(0)) M_i(1) + b_{10} M_i(0) (1 - M(1)),$$ 
where 
$\bs{b} = (b_{00}, b_{01}, b_{10})$ and 
$b_{00}, b_{01}, b_{10} \in\mathbb{R}$. Equivalently, in vector form, $\tilde{\bs{Y}}_{\bs{b}}(0) \coloneqq \bs{Y}(0) \circ \bs{M}(0) \circ \bs{M}(1) + b_{00} \cdot (\bs{1}-\bs{M}(0)) \circ (\bs{1}-\bs{M}(1)) + b_{01} \cdot (\bs{1}-\bs{M}(0)) \circ \bs{M}(1) + b_{10} \cdot \bs{M}(0) \circ (\bs{1}-\bs{M}(1))$.
We use $t(\bs{Z}, \tilde{\bs{Y}}_{\bs{b}}(0))$, which we refer to as the composite test statistic, for randomization inference.

To see the benefit of incorporating potential missingness on inference, we compare the test statistics $t(\bs{Z}, \bs{Y}(0))$ and $t(\bs{Z}, \tilde{\bs{Y}}_{\bs{b}}(0))$. Both statistics contrast the treated and control groups, using either the original or the composite control potential outcomes.
The composite potential outcome coincides with the original potential outcome only for units that would be observed under both treatment and control, and is set to some constant otherwise. Consequently, the statistic $t(\bs{Z}, \tilde{\bs{Y}}_{\bs{b}}(0))$ essentially focuses solely on units with $M_i(1) = M_i(0) = 1$; such a subgroup of units is known as a principal stratum \citep{frangakis_principal_2004}.
This focus is natural. For units outside this principal stratum, at least one potential outcome is unobservable regardless of treatment assignment and may take any value on the real line; such units therefore cannot provide evidence against the null hypothesis. The composite test statistic restricts attention to the principal stratum $\{i: M_i(1) = M_i(0) = 1\}$ rather than the full sample, and, as shown later, can yield sharper and more powerful tests of the null hypothesis.

\subsection{General missingness mechanism}

We now consider testing the sharp null $H_{\bs{\delta}}$ in \eqref{eq:null_fisher_sharp} using the test statistic $t(\bs{Z}, \tilde{\bs{Y}}_{\bs{b}}(0))$ under the general missingness mechanism. 
Similar to Section \ref{sec:sharp_null_gen}, we use a distribution-free statistic in \eqref{eq:test_stat_R_phi}, \eqref{eq:test_stat_mann_whitney} or \eqref{eq:test_stat_mann_whitney_negative}, and we seek to minimize the realized value of the test statistic $t_{\R, \phi}(\bs{Z}, \tilde{\bs{Y}}_{\bs{b}}(0))$. 

Panel A in Table \ref{tab:sharp_null} summarizes the composite potential outcomes under the sharp null in \eqref{eq:null_fisher_sharp}, which depend on treatment assignment, the realized outcome, and counterfactual missingness.
Specifically, for treated units with observed outcomes, the composite potential outcome equals $Y_i - \delta$ if the unit would remain observed under control and $b_{01}$ if it would instead be missing under control; for treated units with missing outcomes, it equals $b_{00}$ if the unit would remain missing under control and $b_{10}$ if it would instead be observed under control.
The corresponding expressions for control units’ composite potential outcomes can be obtained by replacing $Y_i - \delta$ with $Y_i$ and evaluating missingness under treatment.

As in Theorem \ref{thm:sharp_null_gen}, we want $\tilde{Y}_{\bs{b},i}(0)$ to be as small as possible for treated units and as large as possible for control units in order to obtain a valid randomization p-value under the general missingness mechanism. This motivates us to consider the following ``feasible'' values for the composite potential outcome vector:
$\tilde{\bs{Y}}_{\bs{Z},\bs{\delta}, \bs{b}}^{\gen}(0) = (\tilde{Y}_{\bs{Z},\bs{\delta}, \bs{b}, 1}^{\gen}(0), \ldots, \tilde{Y}_{\bs{Z},\bs{\delta}, \bs{b}, n}^{\gen}(0))^\top$ with
\begin{align}\label{eq:worst_gen_YM}
    \tilde{Y}_{\bs{Z},\bs{\delta}, \bs{b}, i}^{\gen}(0)
    & 
    = \begin{cases}
        \min\{Y_{i} - \delta_i, b_{01}\} &\quad \text{if $Z_{i} = 1$ and $M_{i} = 1$}, \\
        \min\{b_{00}, b_{10}\} &\quad \text{if $Z_{i} = 1$ and $M_{i} = 0$}, \\
        \max\{Y_{i}, b_{10}\} &\quad \text{if $Z_{i} = 0$ and $M_{i} = 1$}, \\
        \max\{b_{00}, b_{01}\} &\quad \text{if $Z_{i} = 0$ and $M_{i} = 0$}, \\
    \end{cases}
    \qquad \qquad
    (1\le i \le n). 
\end{align}
The theorem below shows that $ \tilde{\bs{Y}}_{\bs{Z},\bs{\delta}, \bs{b}}^{\gen}(0)$ indeed yields the worst-case configuration of the composite potential outcomes, leading to a valid p-value for testing the sharp null $H_{\bs{\delta}}$ in \eqref{eq:null_fisher_sharp}.

\begin{theorem}\label{thm:sharp_null_gen_comp}
\rm 
Under the CRE and Assumption \ref{asmp:general}, 
a valid p-value for testing the sharp null hypothesis $H_{\bs{\delta}}$ in \eqref{eq:null_fisher_sharp} is 
\begin{align}\label{eq:p_val_sharp_null_comp_gen}
    \tilde{p}_{\bs{Z}, \bs{\delta}, \bs{b}}^\gen  \coloneqq G_{\R, \phi}\left( t_{\R, \phi}\left(\bs{Z}, \tilde{\bs{Y}}_{\bs{Z},\bs{\delta}, \bs{b}}^{\gen}(0)  \right) \right),
\end{align}
where $t_{\R, \phi}$ is defined in either \eqref{eq:test_stat_R_phi}, \eqref{eq:test_stat_mann_whitney} or \eqref{eq:test_stat_mann_whitney_negative}, 
$G_{\R, \phi}$ is defined in
\eqref{eq:G_R_phi}, 
and $\tilde{\bs{Y}}_{\bs{Z},\bs{\delta}, \bs{b}}^{\gen}(0)$ is defined in \eqref{eq:worst_gen_YM}. 
That is, under $H_{\bs{\delta}}$, $\mathbb{P}( \tilde{p}_{\bs{Z}, \bs{\delta}, \bs{b}}^\gen \leq \alpha ) \leq \alpha$ for any $\alpha \in (0, 1)$.
\end{theorem}

The p-value in Theorem \ref{thm:sharp_null_gen_comp} is valid for any pre-specified constants $b_{00}$, $b_{01}$, and $b_{10}$ used to construct the composite potential outcomes. 
We now discuss how the choice of these constants affects the power of the test in Theorem \ref{thm:sharp_null_gen_comp}.
Given the form of the rank-based test statistic, 
the p-value in \eqref{eq:p_val_sharp_null_comp_gen} depends crucially on the relative magnitude of the worst-case configuration $\tilde{\bs{Y}}_{\bs{Z},\bs{\delta}, \bs{b}}^{\gen}(0)$ for treated and control units. 
Regardless of the values of $b_{00}$, $b_{01}$, and $b_{10}$, the composite potential outcomes for the observed treated units are always less than or equal to those for the missing control units, while those for the missing treated units are always less than or equal to those for all the control units. As $b_{01}$ increases and $b_{10}$ decreases, the composite potential outcomes for observed treated units are more likely to exceed those for observed control units.
Therefore, to ``maximize'' the test statistic under the worst-case configuration, or equivalently to ``minimize'' the p-value, we suggest choosing $b_{01}$ as large as possible, i.e., $b_{01} = \infty$, and choosing $b_{10}$ as small as possible, i.e., $b_{10} = -\infty$. 
For descriptive convenience, we define $0 \cdot \infty = 0$ and $0 \cdot -\infty = 0$.

Panel A of Table \ref{tab:sharp_null} summarizes the worst-case configurations of both the control potential outcomes and the composite potential outcomes for any fixed $b_{00}, b_{01}, b_{10} \in \mathbb{R}$, including the suggested choices $b_{01} = \infty$ and $b_{10} = -\infty$.
When using these suggested values of $b_{01}$ and $b_{10}$, the p-value in \eqref{eq:p_val_sharp_null_comp_gen} coincides with that in \eqref{eq:p_val_sharp_null_orig}, because $\tilde{\bs{Y}}_{\bs{Z},\bs{\delta}, \bs{b}}^{\gen}(0)$ reduces to $\bs{Y}_{\bs{Z}, \bs{\delta}}^{\gen}(0)$.
Accordingly, Theorem \ref{thm:sharp_null_gen} is a special case of Theorem \ref{thm:sharp_null_gen_comp} with these suggested values of $b_{01}$ and $b_{10}$, and both yield the same valid p-value for testing the sharp null hypothesis.
Although the test statistic $t(\bs{Z}, \tilde{\bs{Y}}_{\bs{b}}(0))$ offers no advantage over $t(\bs{Z}, \bs{Y}(0))$ under general missingness, it can be beneficial under specific missingness mechanisms, as discussed in the following subsection.
Moreover, it can also offer advantages under general missingness when combined with a two-step approach described in Section \ref{sec:two_step_gen}.

\subsection{Monotone missingness mechanisms}\label{sec:monotone_sharp_null}

We test the sharp null in \eqref{eq:null_fisher_sharp} under missingness mechanisms in which treatment has a monotone effect on the missingness indicator. We consider two cases corresponding to non-negative and non-positive treatment effects, as formalized in Assumptions \ref{asmp:mo_miss_M1_ge_M0} and \ref{asmp:mo_miss_M1_le_M0}, respectively. 
As discussed at the end of Section \ref{sec:sharp_null_gen},
under either Assumption \ref{asmp:mo_miss_M1_ge_M0} or \ref{asmp:mo_miss_M1_le_M0}, 
the worst-case configuration of $\bs{Y}(0)$ remains the same as in \eqref{eq:worst_gen}, and the resulting p-value is the same as that in Theorem \ref{thm:sharp_null_gen}. Thus, the monotone missingness assumptions do not improve the power of tests based on $t_{\R, \phi}(\bs{Z}, \bs{Y}(0))$.

We now focus on the test statistic $t_{\R, \phi}(\bs{Z}, \tilde{\bs{Y}}_{\bs{b}}(0))$. Panels B and C of Table \ref{tab:sharp_null} summarize the available information on the composite potential outcomes implied by the observed data, the sharp null $H_{\bs{\delta}}$, and Assumption \ref{asmp:mo_miss_M1_ge_M0} or \ref{asmp:mo_miss_M1_le_M0}.
Compared with Panel A under Assumption \ref{asmp:general}, the monotone missingness assumptions allow us to impute additional $M_i(0)$ and $M_i(1)$ values.
Specifically, Assumption \ref{asmp:mo_miss_M1_ge_M0} implies $M_i(0)=0$ for missing treated units because $M_i(0)\le M_i(1)=M_i=0$. It also implies $M_i(1)=1$ for observed control units because $M_i(1)\ge M_i(0)=M_i=1$.
Similarly, under Assumption \ref{asmp:mo_miss_M1_le_M0}, $M_i(0)=1$ for observed treated units and $M_i(1)=0$ for missing control units.
The full composite potential outcomes remain unknown, so we consider their worst-case configuration to ensure the validity of the randomization test. As in Section \ref{sec:sharp_null_gen}, we minimize treated units' composite outcomes and maximize those of control units.

We therefore consider the following ``feasible'' configurations of the composite potential outcomes under Assumptions \ref{asmp:mo_miss_M1_ge_M0} and \ref{asmp:mo_miss_M1_le_M0}, as summarized in Panels B and C of Table \ref{tab:sharp_null}.
Under Assumption \ref{asmp:mo_miss_M1_ge_M0}, we define $\tilde{\bs{Y}}^{\monp}_{\bs{Z}, \bs{\delta}, \bs{b}}(0) = (\tilde{Y}^{\monp}_{\bs{Z}, \bs{\delta}, \bs{b}, 1}(0), \ldots, \tilde{Y}^{\monp}_{\bs{Z}, \bs{\delta}, \bs{b}, n}(0))^\top$, with the superscript indicating that the treatment has a monotone positive (more precisely, nonnegative) effect on the missingness indicator: 
\begin{align}\label{eq:worst_YM_mon_p}
    \tilde{Y}^{\monp}_{\bs{Z}, \bs{\delta}, \bs{b}, i}(0)
    & 
    = \begin{cases}
        \min\{Y_{i} - \delta_i, b_{01} \} &\quad \text{if $Z_{i} = 1$ and $M_{i} = 1$}, \\
        b_{00} &\quad \text{if $Z_{i} = 1$ and $M_{i} = 0$}, \\
        Y_{i} &\quad \text{if $Z_{i} = 0$ and $M_{i} = 1$}, \\
        \max\{b_{00}, b_{01} \} &\quad \text{if $Z_{i} = 0$ and $M_{i} = 0$}, \\
    \end{cases}
    \qquad \qquad
    (1\le i \le n). 
\end{align}
Under Assumption \ref{asmp:mo_miss_M1_le_M0}, we define $\tilde{\bs{Y}}^{\monn}_{\bs{Z}, \bs{\delta},\bs{b}}(0) = (\tilde{Y}^{\monn}_{\bs{Z}, \bs{\delta},\bs{b}, 1}(0), \tilde{Y}^{\monn}_{\bs{Z}, \bs{\delta},\bs{b}, 2}(0), \ldots, \tilde{Y}^{\monn}_{\bs{Z}, \bs{\delta},\bs{b}, n}(0))^\top$, with the superscript indicating that the treatment has a monotone negative (more precisely, nonpositive) effect on the missingness indicator:
\begin{align}\label{eq:worst_YM_mon_n}
    \tilde{Y}^{\monn}_{\bs{Z}, \bs{\delta},\bs{b}, i}(0) 
    & 
    = \begin{cases}
        Y_{i} - \delta_i &\quad \text{if $Z_{i} = 1$ and $M_{i} = 1$}, \\
        \min\{ b_{00}, b_{10} \} &\quad \text{if $Z_{i} = 1$ and $M_{i} = 0$}, \\
        \max\{ Y_{i}, b_{10} \} &\quad \text{if $Z_{i} = 0$ and $M_{i} = 1$}, \\
        b_{00} &\quad \text{if $Z_{i} = 0$ and $M_{i} = 0$}, \\
    \end{cases}
    \qquad \qquad
    (1\le i \le n). 
\end{align}
The following theorem shows that \eqref{eq:worst_YM_mon_p} and \eqref{eq:worst_YM_mon_n} indeed give the worst-case configurations under Assumptions \ref{asmp:mo_miss_M1_ge_M0} and \ref{asmp:mo_miss_M1_le_M0}, respectively, and thereby lead to valid p-values for testing the sharp null $H_{\bs{\delta}}$.

\begin{theorem}\label{thm:sharp_null_mon}
\rm 
Under the CRE and Assumption \ref{asmp:mo_miss_M1_ge_M0} or \ref{asmp:mo_miss_M1_le_M0}, 
a valid p-value for testing the sharp null hypothesis $H_{\bs{\delta}}$ in \eqref{eq:null_fisher_sharp} is 
\begin{align}\label{eq:p_val_sharp_null_test_stat_outcome_miss_mono}
    \tilde{p}_{\bs{Z}, \bs{\delta},\bs{b}}^{\mon \square}  \coloneqq G_{\R, \phi}\left( t_{\R, \phi}(\bs{Z}, \tilde{\bs{Y}}^{\mon \square}_{\bs{Z}, \bs{\delta},\bs{b}}(0)  )\right)
    \text{ with } 
    \square = 
    \begin{cases}
        \text{p} & \text{under Assumption \ref{asmp:mo_miss_M1_ge_M0}},\\
        \text{n} & \text{under Assumption \ref{asmp:mo_miss_M1_le_M0}},
    \end{cases}
\end{align}
where $t_{\R, \phi}$ is defined in either \eqref{eq:test_stat_R_phi}, \eqref{eq:test_stat_mann_whitney} or \eqref{eq:test_stat_mann_whitney_negative}, $G_{\R, \phi}$ is defined in \eqref{eq:G_R_phi}, and $\tilde{\bs{Y}}^{\monp}_{\bs{Z}, \bs{\delta},\bs{b}}(0)$ and $\tilde{\bs{Y}}^{\monn}_{\bs{Z}, \bs{\delta},\bs{b}}(0)$ are defined in \eqref{eq:worst_YM_mon_p} and \eqref{eq:worst_YM_mon_n}.
\end{theorem}

We now discuss the choice of $b_{00}$, $b_{01}$, and $b_{10}$, which are important for the power of the test.
Consider first the worst-case configuration of the composite potential outcomes in \eqref{eq:worst_YM_mon_p} under Assumption \ref{asmp:mo_miss_M1_ge_M0}. In this case, the composite outcomes of control units with observed outcomes do not depend on $b_{00}$, $b_{01}$, or $b_{10}$.
Moreover, regardless of the values of $b_{00}$, $b_{01}$, and $b_{10}$, the composite potential outcomes of treated units are always less than or equal to those of control units with missing outcomes (which are $\max\{b_{00}, b_{01}\}$).
As $b_{01}$ increases, observed treated units are more likely to exceed observed control units.
Similarly, as $b_{00}$ increases, missing treated units are also more likely to exceed observed control units.
Therefore, we recommend choosing $b_{00}$ and $b_{01}$ as large as possible for Theorem \ref{thm:sharp_null_mon} with $\square = \text{p}$: that is, using the p-value $\tilde{p}_{\bs{Z}, \bs{\delta},\bs{b}}^{\monp}$ in \eqref{eq:p_val_sharp_null_test_stat_outcome_miss_mono} with $b_{00} = b_{01} = \infty$.

We next consider Assumption \ref{asmp:mo_miss_M1_le_M0}, under which the worst-case configuration of the composite potential outcomes is given in \eqref{eq:worst_YM_mon_n}. 
In this case, the composite potential outcomes of missing treated units are always bounded above by those of the control units.
As $b_{00}$ and $b_{10}$ decrease, the composite potential outcomes of observed treated units are more likely to exceed those of the control units.
Therefore, we recommend choosing $b_{00}$ and $b_{10}$ as small as possible in Theorem \ref{thm:sharp_null_mon} with $\square = \text{n}$; that is, using the p-value $\tilde{p}_{\bs{Z}, \bs{\delta},\bs{b}}^{\monn}$ in \eqref{eq:p_val_sharp_null_test_stat_outcome_miss_mono} with $b_{00} = b_{10} = -\infty$.

With the suggested $\bs{b}$ values, the worst-case configuration under Assumption \ref{asmp:mo_miss_M1_ge_M0} or \ref{asmp:mo_miss_M1_le_M0} matches that under general missingness, except that $-\infty$ values imputed for missing treated units become $\infty$, or $\infty$ values imputed for missing control units become $-\infty$.
Because of the effect increasing property of the rank-sum statistics in \eqref{eq:test_stat_R_phi}, \eqref{eq:test_stat_mann_whitney} and \eqref{eq:test_stat_mann_whitney_negative}, the p-value in \eqref{eq:p_val_sharp_null_test_stat_outcome_miss_mono} from Theorem \ref{thm:sharp_null_mon} under either Assumption \ref{asmp:mo_miss_M1_ge_M0} or \ref{asmp:mo_miss_M1_le_M0} is smaller than or equal to $\tilde{p}_{\bs{Z}, \bs{\delta}, \bs{b}}^\gen$ under the general missing mechanism assumption.
This holds more generally for any prespecified values of $(b_{00}, b_{01}, b_{10})$, and can be seen from the last two columns of Table \ref{tab:sharp_null}.
In other words, by using the monotonicity assumptions in the missingness mechanism, Theorem \ref{thm:sharp_null_mon}, which is based on the composite potential outcomes, yields more powerful tests than Theorem \ref{thm:sharp_null_gen}.

\begin{remark}\label{rmk:equiv_onestep}
Under a sharp null hypothesis of some constant treatment effect $c$ and using the suggested $\bs{b}$ values, the test in Theorem \ref{thm:sharp_null_mon} under Assumption \ref{asmp:mo_miss_M1_ge_M0} is equivalent to the test in Theorem \ref{thm:sharp_null_mon} under Assumption \ref{asmp:mo_miss_M1_le_M0} with switched treatment labels and changed outcome signs. This equivalence may fail, however, under a general sharp null, because the former relies only on the hypothesized effects for treated units, whereas the latter relies on those for control units.
\end{remark}

\subsection{Sharp missingness mechanism}\label{sec:constant_sharp_null}

Testing \eqref{eq:null_fisher_sharp} under the sharp missingness mechanism proceeds as in Sections \ref{sec:sharp_null_gen}, minimizing the original or composite potential outcomes of treated units and maximizing those of control units. For the test statistic $t_{\R, \phi}(\bs{Z}, \bs{Y}(0))$, the worst-case configuration of $Y_i(0)$ again coincides with Panel A of Table \ref{tab:sharp_null}, yielding the same p-value $p_{\bs{Z}, \bs{\delta}}^\gen$ as in \eqref{eq:p_val_sharp_null_orig} under general missingness mechanisms. In contrast, for $t_{\R, \phi}(\bs{Z}, \tilde{\bs{Y}}_{\bs{b}}(0))$, the composite potential outcomes are fully known: for all $1\le i \le n$, $\tilde{Y}_{\bs{b},i}(0) = Y_i(0) M_i(0) M_{i}(1) + b_{00} (1-M_i(0)) (1-M_i(1)) = (Y_i -\delta_i Z_i) M_i + b_{00} (1-M_i)$. The corresponding valid p-value is given in the following proposition.

\begin{proposition}\label{prop:sharp_null_cons}
\rm 
Under the CRE and Assumption \ref{asmp:constant_miss}, 
a valid p-value for testing the sharp null hypothesis $H_{\bs{\delta}}$ in \eqref{eq:null_fisher_sharp} is
\begin{align}\label{eq:sharp_null_cons_compo_p_val}
    \tilde{p}_{\bs{Z}, \bs{\delta}, \bs{b}}^\con = G_{\R, \phi}\left( t_{\R, \phi}(\bs{Z}, (\bs{Y} - \bs{\delta}\circ \bs{Z})\circ \bs{M} +  b_{00}(\bs{1}-\bs{M}) )\right),
\end{align}
where $t_{\R, \phi}$ is defined in either \eqref{eq:test_stat_R_phi}, \eqref{eq:test_stat_mann_whitney} or \eqref{eq:test_stat_mann_whitney_negative}, and $G_{\R, \phi}$ is defined in \eqref{eq:G_R_phi}. 
\end{proposition}

Due to the effect increasing property of the test statistics in \eqref{eq:test_stat_R_phi}, \eqref{eq:test_stat_mann_whitney} and \eqref{eq:test_stat_mann_whitney_negative}, the p-value in Proposition \ref{prop:sharp_null_cons} is smaller than or equal to that in Theorems \ref{thm:sharp_null_gen_comp} and \ref{thm:sharp_null_mon} under the monotone missingness. 
Thus, using composite potential outcomes under sharp missingness yields more powerful tests.

We now discuss the choice of $b_{00}$ in Proposition \ref{prop:sharp_null_cons}, which sets the composite potential outcomes for units with missing outcomes. Under the CRE and Assumption \ref{asmp:constant_miss}, the numbers of missing outcomes are balanced on average between treatment and control groups, so $b_{00}$ may have little effect on test performance.
Below we describe a strategy that avoids specifying $b_{00}$ and is often preferable in practice. Specifically, we propose excluding units with missing outcomes when conducting randomization tests under the sharp missingness mechanism. Specifically, we restrict attention to the subset $\S=\{ i:M_i=1, 1 \leq i \leq n \}$ of units with observed outcomes and view the resulting experiment as a CRE that assigns $n_{\S1}=\sum_{i\in\S} Z_i$ units to treatment and $n_{\S0}=|\S|-n_{\S1}$ to control. Standard randomization tests for sharp null hypotheses then apply, because no outcomes are missing. This procedure is valid because (i) under Assumption \ref{asmp:constant_miss}, $\S=\{i:M_i(1)=M_i(0)=1, 1 \leq i \leq n\}$ is nonrandom, and (ii) the test is conditional on the treatment assignments of units in $\S^{\complement}=\{i:M_i(1)=M_i(0)=0 \}$. Consequently, concerns about missing outcomes and the choice of $b_{00}$ are avoided. We summarize the results in the following theorem.

\begin{theorem}\label{thm:constant_miss}
\rm 
Under the CRE and Assumption \ref{asmp:constant_miss}, 
a valid p-value for testing the sharp null hypothesis $H_{\bs{\delta}}$ in \eqref{eq:null_fisher_sharp} is $p_{\bs{Z}, \bs{\delta}, \S}$, where $p_{\bs{Z}, \bs{\delta}, \S}$ is defined as:
\begin{align}
    p_{\bs{Z}, \bs{\delta}, \S} = G_{\R, \phi, \S}( t_{\R, \phi, \S}(\bs{Z}, \bs{Y} - \bs{\delta}\circ \bs{Z})),
\end{align}
where $t_{\R, \phi, \S}(\cdot)$ and $G_{\R, \phi, \S}(\cdot)$ are defined as:
\begin{align*}
   t_{\R, \phi, \S} (\bs{z}, \bs{y})
   & = 
   t_{\R, \phi} (\bs{z}_{\S}, \bs{y}_{\S}), 
   \ \ 
   \text{and}
   \ \ 
   G_{\R, \phi, \S}(c)
   = \Pr 
   \left( t_{\R, \phi, \S} (\bs{A}, \bs{y}) \ge c
    \right) \text{ for } c \in \mathbb{R},
\end{align*}
with $t_{\R, \phi}$ defined in either \eqref{eq:test_stat_R_phi} or \eqref{eq:test_stat_mann_whitney},  
$G_{\R, \phi, \S}(\cdot)$ defined based on random assignment $\bs{A}_{\S}$ from $\CRE(\S, n_{\S1}, n_{\S0})$, 
and $\bs{z}_{\S}, \bs{y}_{\S}$ and $\bs{A}_{\S}$ being subvectors of $\bs{z}, \bs{y}$ and $\bs{A}$ for units in $\S$.
\end{theorem}

Theorem \ref{thm:constant_miss} essentially tests a “modified” sharp null hypothesis that is restricted to the subset of units whose outcomes are always observed under either treatment arm:
\begin{align}\label{null_fisher_sharp_nonmissing}
H_{\bs{\delta}, \mathcal{S}}:\ \tau_i = \delta_i, \quad i \in \mathcal{S} \equiv \{j : M_j = 1,\ 1 \le j \le n\}.
\end{align}
Moreover, because outcomes are fully observed within $\mathcal{S}$, the procedure in Theorem \ref{thm:constant_miss} applies to generic test statistics, such as the difference-in-means statistic.

\section{Two-Step Randomization Tests for the Sharp Null}\label{sec:two_step_sharp}

The power of the tests in Section \ref{sec:sharp_null_composite} can be further improved by exploiting information about potential missingness. The key intuition is that we can use the observed missingness for treated units to infer the counterfactual missingness for control units, and vice versa. 
Below we present a two-step procedure to improve the power of the tests in Section \ref{sec:sharp_null_composite}.  
As detailed below, we first construct prediction sets for the distribution of potential missingness in the treated or control groups, and then use that to improve the lower bound of the worst-case test statistic values. 
The resulting comparisons between treated and control units in the improved p-values can mirror the corresponding sharp bounds in the  partial identification literature, which in some sense implies the sharpness of our tests. 
In the following, we consider first the monotone missingness under Assumption \ref{asmp:mo_miss_M1_ge_M0} or \ref{asmp:mo_miss_M1_le_M0}, and then the general missingness under Assumption \ref{asmp:general}. 

\subsection{Monotone missingness under Assumption \ref{asmp:mo_miss_M1_ge_M0}}\label{sec:two_step_mp}

Under Assumption \ref{asmp:mo_miss_M1_ge_M0} and 
with the suggested $b_{00}=b_{01} = \infty$ from Section \ref{sec:monotone_sharp_null}, the composite potential outcome for each unit $i$ reduces to 
$\tilde{Y}_{\bs{b},i}(0) = Y_i(0) M_i(0) + \infty (1 - M_i(0))$.
Under the sharp null in \eqref{eq:null_fisher_sharp}, 
with information from the observed data and as summarized in Table \ref{tab:sharp_null} Panel B, the composite potential outcomes are known except for observed treated units. 
Specifically, for an observed treated unit $i$, $\tilde{Y}_{\bs{b},i}(0)$ equals $Y_i - \delta_i$ if the unit would have been observed under control (i.e., $M_i(0)=1$), and equals $\infty$ if the unit would have been missing under control (i.e., $M_i(0)=0$). 
In the worst-case consideration as in \eqref{eq:worst_YM_mon_p} and as summarized in the last column of Panel B in table \ref{tab:sharp_null}, we essentially pretend that all observed treated units would also be observed under control. This, however, may be overly conservative, and thus compromises the power of the test in Theorem \ref{thm:sharp_null_mon}.  

Indeed, the observed missingness in the control group can inform the number of observed treated units who would have been missing under control. 
Specifically, the number of observed control units, $\sum_{i=1}^n (1-Z_i) M_i = \sum_{i=1}^n (1-Z_i) M_i(0)$, follows a Hypergeometric distribution $\text{HG}(N, \sum_{i=1}^n M_i(0), n_0)$\footnote{$\text{HG}(N, m, n)$ denotes the distribution of the number of successes in a sample of size $n$, drawn without replacement from a finite population of size $N$ that contains exactly $m$ successes.}
This can lead to an upper confidence bound for $\sum_{i=1}^n M_i(0)$ and consequently a lower prediction bound for the number of observed treated units who would have been missing under control. Importantly, this implies that at least a certain number of composite potential outcomes for the observed treated units must be $\infty$, rather than all being the imputed control potential outcomes, as in the last column of Panel B in Table \ref{tab:sharp_null}.
Intuitively, this would increase the worst-case value of the test statistic and decrease the p-value, although we still need to take into account the error that may arise in the first step when inferring $\sum_{i=1}^n M_i(0)$. 

Accordingly, we propose a two-step testing procedure. First, we construct an upper confidence bound for $\sum_{i=1}^n M_i(0)$ using, for example, the optimal method in \cite{Wang:2015}, which also implies a lower prediction limit for the number of observed treated units with $M_i(0)=0$. Second, we impose this limit as a constraint when imputing the worst-case configuration of composite control potential outcomes for the observed treated units. 
We apply a Bonferroni-type correction to control the errors that may occur in both steps.
Moreover, to simplify the constrained optimization in the second step, we consider only the Mann--Whitney-type U-statistic in \eqref{eq:test_stat_mann_whitney}, which sums the relative ranks of each treated unit compared with all control units; in particular, it admits a closed-form solution.
We summarize this two-step procedure in Algorithm \ref{alg:mp_two_step_sharp}.
Recall the definition in \eqref{eq:n_zm}. 

\begin{algorithm}\label{alg:mp_two_step_sharp}[Two-step method for constructing a valid p-value under Assumption \ref{asmp:mo_miss_M1_ge_M0}]
\begin{enumerate}[label=(\roman*), topsep=1ex,itemsep=-0.3ex,partopsep=1ex,parsep=1ex]
    \item For a prespecified $\beta \in [0, 1)$, construct a $1-\beta$ upper confidence limit $\hat{M}$ for $\sum_{i = 1}^{n} M_{i}(0)$ using, for example, the method in \cite{Wang:2015}; that is, $\P\left( \sum_{i = 1}^{n} M_{i}(0) \leq \hat{M} \right) \geq 1 - \beta$. 
    Then calculate $\underline{m} \coloneqq \max\{0, n_{11} + n_{01} -  \hat{M}\}$, which is a $1-\beta$ lower prediction limit for $\sum_{i=1}^n Z_i M_i\{1-M_i(0)\}$.  
    \item For each observed treated unit $i$, 
    compute $C_i \coloneqq  \phi(A_{i}) - \phi(B_{i})$, with 
    \begin{align*}
    A_{i} \coloneqq \sum_{j:Z_j=0} \psi_{i,j}( \infty, Y_j M_j +  \infty (1-M_j) ) \quad \text{and} \quad 
    B_i \coloneqq 
        \sum_{j:Z_j=0} \psi_{i,j}( Y_i - 
    \delta_i
    , Y_j M_j +  \infty (1-M_j) ). 
    \end{align*}
    \item Let $\{i: Z_i=M_i=1, 1\le i \le n\} = \{l_1, l_2, \ldots, l_{n_{11}}\}$ such that $C_{l_1} \le C_{l_2} \le \ldots \le C_{l_{n_{11}}}$.     
    \item Impute the composite control potential outcomes by
    $\tilde{\bs{Y}}_{\bs{Z}, \bs{\delta}}^{\monp, \underline{m}}(0)$, with the $i$th element given by: 
    \begin{align}\label{eq:wc_mp_twostep}
    \tilde{Y}_{\bs{Z}, \bs{\delta}, i}^{\monp, \underline{m}}(0)
    & 
    = \begin{cases}
        \infty &\quad \text{if $Z_{i} = 1$, $M_{i} = 1$ and $i \in \{l_1, \ldots, l_{\underline{m}}\}$}, \\
        Y_{i} - \delta_i &\quad \text{if $Z_{i} = 1$, $M_{i} = 1$ and $i \notin \{l_1, \ldots, l_{\underline{m}}\}$}, \\
        \infty &\quad \text{if $Z_{i} = 1$ and $M_{i} = 0$}, \\
        Y_{i} &\quad \text{if $Z_{i} = 0$ and $M_{i} = 1$}, \\
        \infty &\quad \text{if $Z_{i} = 0$ and $M_{i} = 0$}, \\
    \end{cases}
    \quad (1\le i \le n). 
    \end{align}
    \item Construct a p-value by:
    \begin{align}\label{eq:mp_two_step_sharp_p_val}
        \tilde{p}_{\bs{Z}, \bs{\delta}}^{\monp, \text{two-step}} \coloneqq G_{\R, \phi}\left( t_{\R, \phi}\left(\bs{Z}, \tilde{\bs{Y}}^{\monp, \underline{m}}_{\bs{Z}, \bs{\delta}}(0)  \right)\right) + \beta,
    \end{align}
    where $\tilde{\bs{Y}}^{\monp, \underline{m}}_{\bs{Z}, \bs{\delta}, \infty}(0)$ 
    is defined in \eqref{eq:wc_mp_twostep}, 
    $t_{\R, \phi}$ is defined in \eqref{eq:test_stat_mann_whitney}, and $G_{\R, \phi}$ is defined in \eqref{eq:G_R_phi}.
\end{enumerate}
\end{algorithm}

\begin{theorem}\label{thm:sharp_null_mon_two_step}
\rm Under the CRE and Assumption \ref{asmp:mo_miss_M1_ge_M0}, 
for any prespecified $\beta \in [0,1)$, 
Algorithm \ref{alg:mp_two_step_sharp} yields a valid p-value. That is, under $H_{\bs{\delta}}$, $\P( \tilde{p}_{\bs{Z}, \bs{\delta}}^{\monp, \text{two-step}} \leq \alpha ) \leq \alpha$ for any $\alpha \in (0, 1)$.
\end{theorem}

The worst-case imputation in \eqref{eq:wc_mp_twostep} under the two-step procedure is closely connected to the Zhang--Rubin--Lee bounds \citep{zhang2003estimation, lee2009training}. 
Note that when the sample size is large, due to randomization, the proportion of units with $M_i(0) = 1$ would be about the same in the treated and control groups, implying that $(n_{11} - \underline{m})/n_1$ is approximately $n_{01}/n_0$, where  $n_{11}$ and $n_{01}$ are defined in \eqref{eq:n_zm}. 
This implies that the proportion of $\infty$ among treated units in \eqref{eq:wc_mp_twostep} is roughly $1 - (n_{11} - \underline{m})/n_1 \approx 1 - n_{01}/n_0$, the same as the proportion of $\infty$ among control units. 
Thus, in the randomization test and ignoring ties, 
we are essentially comparing the $(n_{11} - \underline{m})/n_{11} \approx (n_{01} / n_0) / (n_{11} / n_1)$ fraction of the observed treated units with the smallest imputed control potential outcomes to all the observed control units. 
Such a comparison exactly mirrors that used in the Zhang--Rubin--Lee lower bound on the average treatment effect 
within the principal stratum of 
units whose outcomes would be observed under both treatment and control; see \citet[Table 6]{zhang2003estimation} and \citet[Proposition 1a]{lee2009training}.

\subsection{Monotone missingness under Assumption \ref{asmp:mo_miss_M1_le_M0}}

Under Assumption \ref{asmp:mo_miss_M1_le_M0} and 
with the suggested $b_{00}=b_{10} = -\infty$ from Section \ref{sec:monotone_sharp_null}, the composite potential outcome for each unit $i$ reduces to 
$\tilde{Y}_{\bs{b},i}(0) = Y_i(0) M_i(1) - \infty (1 - M_i(1))$.
Under the sharp null in \eqref{eq:null_fisher_sharp}, with information from the observed data and as summarized in Table \ref{tab:sharp_null} Panel C, the composite potential outcomes are known except for observed control units, for which they are either the observed outcome $Y_i$ or $-\infty$ depending on their counterfactual missingness had they been assigned to treatment.
The worst-case imputation in Theorem \ref{thm:sharp_null_mon} with the suggested $\bs{b}$ value, as summarized in the last column in Table \ref{tab:sharp_null} Panel C, essentially pretend that all the observed control units would also be observed had they been assigned to treatment, which may be overly conservative. 
Therefore, by the same logic as Section \ref{sec:two_step_mp}, we can use a two-step method to further improve the power of the test in Theorem \ref{thm:sharp_null_mon}. 
Moreover, to facilitate optimization for the worst-case composite potential outcomes, we consider only test statistics of form \eqref{eq:test_stat_mann_whitney_negative}, which admit closed-form solutions. 
We summarize it in the following algorithm.\footnote{We slightly abuse notation: the same symbols may represent different quantities across algorithms.} 

\begin{algorithm}\label{alg:mn_two_step_sharp}[Two-step method for constructing a valid p-value under Assumption \ref{asmp:mo_miss_M1_le_M0}]
\begin{enumerate}[label=(\roman*), topsep=1ex,itemsep=-0.3ex,partopsep=1ex,parsep=1ex]
    \item For $\beta \in [0, 1)$, construct a $1-\beta$ upper confidence limit $\hat{M}$ for $\sum_{i=1}^n M_i(1)$ using, for example, the method in \cite{Wang:2015}; that is, $\P\left( \sum_{i = 1}^{n} M_{i}(1) \leq \hat{M} \right) \geq 1 - \beta$. Then calculate $\underline{m}\coloneqq \max\{0, n_{11} + n_{01} - \hat{M}\}$, 
    which is a $1-\beta$ lower prediction limit for $\sum_{i=1}^n (1-Z_i) M_i\{1-M_i(1)\}$.  
    \item For each observed control unit $i$, compute $C_i \coloneqq \phi(A_{i}) - \phi(B_{i})$, with 
    \begin{align*}
    A_i &\coloneqq \sum_{j:Z_j=1} \psi_{i,j}( Y_i
    , (Y_j-\delta_j) M_j -\infty (1-M_j) ), 
    \\
    B_i & \coloneqq \sum_{j:Z_j=1} \psi_{i,j}( -\infty, (Y_j-\delta_j) M_j -\infty (1-M_j) ). 
    \end{align*}    
    \item Let $\{i: Z_i=0, M_i=1, 1\le i \le n\} = \{l_1, l_2, \ldots, l_{n_{01}}\}$ such that $C_{l_1} \le C_{l_2} \le \ldots \le C_{l_{n_{01}}}$.
    \item Impute the composite control potential outcomes by $\tilde{Y}_{\bs{Z}, \bs{\delta}}^{\monn, \underline{m}}(0)$, with the $i$th element given by:
    \begin{align}\label{eq:wc_mn_twostep}
    \tilde{Y}_{\bs{Z}, \bs{\delta}, i}^{\monn, \underline{m}}(0)
    & 
    = \begin{cases}
        Y_i - \delta_i &\quad \text{if $Z_{i} = 1$, $M_{i} = 1$}, \\
        -\infty &\quad \text{if $Z_{i} = 1$, $M_{i} = 0$}, \\
        Y_i &\quad \text{if $Z_{i} = 0$, $M_{i} = 1$ and $i \notin \{l_1, \ldots, l_{\underline{m}}\}$}, \\
        -\infty &\quad \text{if $Z_{i} = 0$, $M_{i} = 1$ and $i \in \{l_1, \ldots, l_{\underline{m}}\}$}, \\
        -\infty &\quad \text{if $Z_{i} = 0$, $M_{i} = 0$}.\\
    \end{cases}
    \end{align}
    \item Construct a p-value by:
    \begin{align}\label{eq:mn_two_step_sharp_p_val}
        \tilde{p}_{\bs{Z}, \bs{\delta}}^{\monn, \text{two-step}} \coloneqq G_{\R, \phi}\left( t_{\R, \phi}\left(\bs{Z}, \tilde{\bs{Y}}^{\monn, \underline{m}}_{\bs{Z}, \bs{\delta}}(0)  \right)\right) + \beta,
    \end{align}
    where $\tilde{\bs{Y}}^{\monn, \underline{m}}_{\bs{Z}, \bs{\delta}}(0)$
    is defined in \eqref{eq:wc_mn_twostep}, 
    $t_{\R, \phi}$ is defined in \eqref{eq:test_stat_mann_whitney_negative}, and $G_{\R, \phi}$ is defined in \eqref{eq:G_R_phi}.
\end{enumerate}
\end{algorithm}

In Algorithm \ref{alg:mn_two_step_sharp}, we can construct the confidence limit $\hat{M}$ using the method in \citet{Wang:2015} and the fact that $\sum_{i=1}^n Z_i M_i = \sum_{i=1}^n Z_i M_i(1)$ follows $\text{HG}(N, \sum_{i=1}^n M_i(1), n_1)$. 

\begin{theorem}\label{thm:sharp_null_mon_neg_two_step}
\rm Under the CRE and Assumption \ref{asmp:mo_miss_M1_le_M0}, 
for any prespecified $\beta \in [0,1)$, 
Algorithm \ref{alg:mn_two_step_sharp} yields a valid p-value. That is, under $H_{\bs{\delta}}$, $\P( \tilde{p}_{\bs{Z}, \bs{\delta}}^{\monn, \text{two-step}} \leq \alpha ) \leq \alpha$ for any $\alpha \in (0, 1)$.
\end{theorem}

Analogous to the discussion after Theorem \ref{thm:sharp_null_mon_two_step}, 
in the worst-case imputation in \eqref{eq:wc_mn_twostep} and ignoring ties, we are essentially comparing all the observed treated units to approximately $(n_{11}/n_1)/(n_{01}/n_0)$ fraction of observed control units with the largest observed outcomes. This again mirrors the comparison in the Zhang–Rubin–Lee lower bound on the average treatment effect  within the principal stratum of units who can be observed under both treatment and control. 

\begin{remark}
Similar to Remark \ref{rmk:equiv_onestep}, under the sharp null hypothesis of some constant treatment effect,  Algorithm \ref{alg:mp_two_step_sharp} under Assumption \ref{asmp:mo_miss_M1_ge_M0} is equivalent to Algorithm \ref{alg:mn_two_step_sharp} under Assumption \ref{asmp:mo_miss_M1_le_M0} with switched treatment labels and changed outcome signs. Again, this equivalence may fail for general sharp null hypotheses.
\end{remark}

\subsection{General missingness}\label{sec:two_step_gen}

We now consider the general missingness. 
In the worst-case imputation in \eqref{eq:worst_gen_YM} with the suggested $b_{01} = \infty$ and $b_{10} = -\infty$ and as summarized in the last column of Panel A in Table \ref{tab:sharp_null}, 
we pretend that all treated units would have been observed under control, and all control units would have been observed under treatment.  
Similar to the monotone missingness studied in the previous two subsections, we can also use the observed missingness to infer counterfactual missingness, and then use a two-step approach improve the worst-case consideration in the randomization test in Theorems \ref{thm:sharp_null_gen} and \ref{thm:sharp_null_gen_comp}. 
However, both steps become more challenging under the general missingness. 
First, without the monotonicity in Assumption \ref{asmp:mo_miss_M1_ge_M0} or \ref{asmp:mo_miss_M1_le_M0}, the distribution of $(M_i(1), M_i(0))$ is generally not identifiable from the observed data, implying that we need a more refined design of the first step. 
Second, due to the non-identifiability of the joint distribution of the potential missingness, we need to consider a larger search space for the worst-case configuration, implying a more challenging optimization in the second step. 

To address these challenges, in the first step, we construct a two-sided prediction interval for the number of treated units who would have been missing under control, together with an upper prediction bound on the difference between the treated and control groups in the proportions of units with $M_i(1)=M_i(0)=1$ (i.e., units that would be observed under both treatment and control).
As illustrated shortly, these prediction bounds are, in a certain sense, sufficient to ensure the sharpness of our test; in particular, as discussed later, the resulting comparison under the worst-case scenario mirrors that in the sharp identification bound for the average treatment effect within the principle stratum of units that would be observed under both treatment and control. 

In the second step, to ease the optimization, we set $b_{00} = \infty$ and use the test statistic of form \eqref{eq:test_stat_mann_whitney}.\footnote{By a similar logic, we can also set $b_{00} = -\infty$ and use test statistics of form \eqref{eq:test_stat_mann_whitney_negative}; we omit the details for conciseness.}
Moreover, to facilitate the optimization, we derive a closed-form but conservative solution for the minimum value of the test statistic under the constraints imposed by the observed data, the null hypothesis of interest, the prediction bounds on the distribution of potential missingness from the first step, and the number of treated units with $M_i(1)=M_i(0)=1$, with the last quantity requiring further enumeration. 
The conservativeness primarily arises from ties at $-\infty$. In particular, the solution is exact under certain orderings used to break ties, for example, when all missing treated units have smaller indices than all observed control units.

We now present the algorithm for the two-step testing procedure under the general missingness. 
The construction of prediction bounds in the first step is discussed immediately after the algorithm. 
Let $m^{\tr}_{11} = |\{i: Z_i=1, M_i(0)=M_i(1) = 1\}|$, $m^{\tr}_{10} = |\{i: Z_i=1, M_i(0) = 1, M_i(1)=0\}|$, 
and $m^{\co}_{11} = |\{i: Z_i=0, M_i(0) = M_i(1) = 1\}|$. 

\begin{algorithm}\label{alg:gen_two_step_sharp}[Two-step method for constructing a valid p-value under Assumption \ref{asmp:general}]
\begin{enumerate}[label=(\roman*), topsep=1ex,itemsep=-0.3ex,partopsep=1ex,parsep=1ex]
    \item For a prespecified $\beta \in [0, 1)$,
    construct $\underline{m}$, $\overline{m}$ and $\overline{d}$ from the observed data 
    such that 
    $
    \Pr( \underline{m} \le m^{\tr}_{11} + m^{\tr}_{10} \le \overline{m}, \ m^{\co}_{11}/n_0 - m^{\tr}_{11}/n_1  \le \overline{d} ) \ge 1 - \beta. 
    $
    \item Let $\{i: Z_i=0, M_i=1\} = \{l_{1}, l_{2}, \ldots, l_{n_{01}}\}$ such that $\psi_{l_j, l_{j+1}} (Y_{l_j}, Y_{l_{j+1}}) = 0$ for $j$. Define further  
    $\mathcal{F}_J = \{l_{n_{01}-J+1}, l_{n_{01}-J+2}, \ldots, l_{n_{01}}\}$ for $1\le J \le n_{01}$ and $\mathcal{F}_0 = \emptyset$. 
    \item 
    For each treated unit $i$, let
    $A_{i} = n_{01} + \sum_{j:Z_j=0, M_j=0} \I(i>j)$, and, for $0\le J \le n_0$, 
    \begin{align*}
    B_{Ji} &=
    \begin{cases}
        \sum_{j:Z_j=0, M_j=1, j\in \mathcal{F}_J} \psi_{i,j}(Y_i-\delta_i, Y_j)
    + (n_{01}-J), & \text{if } M_i = 1,\\
        \max\{0,\ \sum_{j:Z_j=0, M_j=1} \I(i>j) - J \}, & \text{if } M_i = 0. 
    \end{cases}
    \end{align*}
    \item Compute $C_{Ji} \coloneqq  \phi(A_{i}) - \phi(B_{Ji})$ for each treated unit $i$.    
    Let $C_{J1(1)} \le C_{J 1(2)} \le \ldots \le C_{J1(n_{11})}$ be the sorted value of $\{C_{Ji}: Z_i=1, M_i = 1\}$, and $C_{J0(1)} \le C_{J0(2)} \le \ldots \le C_{J0(n_{10})}$ be the sorted value of $\{C_{Ji}: Z_i=1, M_i = 0\}$. 
    \item 
    Let $\overline{K} \coloneqq \min\{n_{11}, \overline{m}\}$ and 
    $\underline{K} \coloneqq \max\{ 0, \underline{m}-n_{10}\}$.     
    For integer $K \in [\underline{K}, \overline{K}]$, define 
    \begin{align*}
        T_{K} & = \sum_{i=1}^n Z_i \phi(B_{Ji}) + \sum_{j=1}^{n_{11}-K} C_{J1(j)} + \sum_{j=1}^{n_{10}-L} C_{J0(j)}, 
    \end{align*}
    where 
    $J = \min\{ \lfloor n_0 ( \overline{d} + K/n_1 ) \rfloor, n_{01} \}$
    and     
    $L = \min\{ \overline{m} - K, n_{10} \}$. 
    \item Construct a p-value by:
    \begin{align}\label{eq:g_two_step_sharp_p_val}
        \tilde{p}_{\bs{Z}, \bs{\delta}}^{\gen, \text{two-step}} \coloneqq G_{\R, \phi}\left( 
        \inf_{K \in [\underline{K}, \overline{K}]} T_K
        \right) + \beta,
    \end{align}
    where $G_{\R, \phi}$ is defined in \eqref{eq:G_R_phi}, with 
    $t_{\R, \phi}$ defined in \eqref{eq:test_stat_mann_whitney}.
\end{enumerate}
\end{algorithm}

One way to construct the prediction bounds $(\underline{m}, \overline{m}, \overline{d})$ in step (i) of Algorithm \ref{alg:gen_two_step_sharp} is as follows.
Choose $\beta_1, \beta_2 \ge 0$ such that $\beta_1 + \beta_2 = \beta$. 
Let $[\hat{M}_1, \hat{M}_2]$ be a $1-\beta_1$ two-sided confidence interval for $\sum_{i=1}^n M_i(0)$ using the fact that $n_{01} \sim \text{HG}(N, \sum_{i=1}^n M_i(0), n_0)$ and the method in \citet{Wang:2015}. Then set $\underline{m} = \hat{M}_1 - n_{01}$ and $\overline{m} = \hat{M}_2 - n_{01}$. 
Let $q_{\text{HG}}(p; n, m_{11}, n_0)$ denote the $p$ quantile of the Hypergeometric distribution with parameters $(n, m_{11}, n_0)$, and set 
$
\overline{d} = \max_{0\le m_{11} \le n_{11}+n_{01}} \{ n/(n_1n_0) \cdot q_{\text{HG}}(1-\beta_2; n, m_{11}, n_0) - m_{11}/n_1\}. 
$

\begin{theorem}\label{thm:sharp_null_gen_two_step}
\rm Under the CRE and Assumption \ref{asmp:general}, 
for any prespecified $\beta \in [0,1)$, 
Algorithm \ref{alg:gen_two_step_sharp} yields a valid p-value. That is, under $H_{\bs{\delta}}$, $\P( \tilde{p}_{\bs{Z}, \bs{\delta}}^{\gen, \text{two-step}} \leq \alpha ) \leq \alpha$ for any $\alpha \in (0, 1)$.
\end{theorem}

Below we intuitively explain the connection between the randomization test in Theorem \ref{thm:sharp_null_gen_two_step} and the sharp bound in \citet{zhang2003estimation}. 
When the sample size is large, due to randomization, both $\underline{m}$ and $\overline{m}$ are approximately $n_1 \cdot n_{01}/n_0$, while $\overline{d}$ is approximately $0$. 
Consequently, $\underline{K} \approx \max\{0, n_1 n_{01}/n_0 - n_{10}\}$ and $\overline{K} \approx \min\{n_{11}, n_1 n_{01}/n_0\}$. 
In addition, for each $K\in [\underline{K}, \overline{K}]$, 
when computing $T_K$ in step (iv) of Algorithm \ref{alg:gen_two_step_sharp}, we have $J \approx K \cdot n_0/n_1$ and 
$L \approx n_1 n_{01}/n_0 - K$. 

We now explain the test statistic $T_K$ in step (iv) of Algorithm \ref{alg:gen_two_step_sharp}, where $K$ can be interpreted as a candidate value of $m_{11}^\tr$. 
Ignoring ties, $T_K$ corresponds to a comparison between treated and control groups under the following worst-case imputation of the composite potential outcomes. For the treated group, the imputed sample consists of $K$ observed treated units with the smallest imputed control potential outcome, $L \approx n_1 n_{01}/n_0 - K$ negative infinities, and approximately $n_1 - n_1 n_{01}/n_0$ positive infinities.  
For the control group, the imputed sample consists of $J \approx K \cdot n_0/n_1$ observed control units with the largest observed outcomes, $n_{01}-J \approx n_{01} - K \cdot n_0/n_1$ negative infinities, and $n_{00} = n_0 - n_{01}$ positive infinities. 
We can verify that the proportions of positive and minus infinities are approximately the same between the treated and control group. 
Therefore, in the test statistic $T_K$, we are essentially comparing the $K/n_{11}$ fraction of the observed treated units with the smallest imputed outcomes to the $Kn_0/(n_1 n_{01})$ fraction of the observed control units with the largest observed outcomes. 
In step (vi) of Algorithm \ref{alg:gen_two_step_sharp}, we further take a worst case over $K\in [\underline{K}, \overline{K}]$.

To better connect to the bounds in \citet{zhang2003estimation}, we introduce $\pi \coloneqq {n_{01}}/{n_0} - {K}/{n_1}$, which can be interpreted as a candidate value for the proportion of units with $M_i(0)=1$ and $M_i(1) = 0$. Then the above fractions of observed treated and control units being compared in $T_K$ are equivalently 
$n_{01}/n_0/(n_{11}/n_1) - \pi/(n_{11}/n_1)$ and 
$1 - {\pi}/{(n_{01}/n_0)}$, 
with a further worst-case consideration over $\pi \in [ \max\{0, n_{01}/n_0 - n_{11}/n_1\}, \ \min\{n_{01}/n_0, 1 - n_{11}/n_1\} ]$. 
This exactly mirrors the comparison underlying the sharp lower bound of the average treatment effect within the principle stratum of units who would be observed under both treatment and control. 

\section{Validity of the randomization p-values for bounded nulls}\label{sec:validity_p_val_bdd_null}

We now consider the following bounded null hypothesis:\footnote{The bounded null hypothesis of form $H_{\vecge \bs{\delta}}: \bs{\tau} \vecge \bs{\delta}$ can be analogously tested by switching the treatment label or changing the outcome sign.} 
\begin{align}\label{eq:null_bounded}
H_{\preccurlyeq \bs{\delta}} : \bs{\tau} \preccurlyeq \bs{\delta},
\end{align}
where $\bs{\delta} \in \mathbb{R}^{n}$ is a fixed vector. The null hypothesis in \eqref{eq:null_bounded} states that each individual treatment effect is bounded above by the corresponding component of $\bs{ \delta}$. In the special case where $\bs{\delta} = \bs{0}$, this reduces to the null hypothesis that all individual treatment effects are non-positive. 

We give a brief remark on the validity of the randomization p-values developed in Theorems \ref{thm:sharp_null_gen}–\ref{thm:sharp_null_gen_two_step} and Proposition \ref{prop:sharp_null_cons} for testing the bounded null in \eqref{eq:null_bounded}. 
Because the test statistics in \eqref{eq:test_stat_R_phi}, \eqref{eq:test_stat_mann_whitney} and \eqref{eq:test_stat_mann_whitney_negative} are all distribution-free and effect increasing,\footnote{
A statistic $t(\bs{z}, \bs{y})$ is said to be effect increasing if its value weakly increases when the outcomes of treated units (i.e., $y_i$ for $z_i = 1$) are increased and the outcomes of control units (i.e., $y_i$ for $z_i = 0$) are decreased.}
we can verify that these $p$-values are nondecreasing in the hypothesized effects $\delta_i$s. Therefore, under the bounded null in \eqref{eq:null_bounded}, they are no less than their corresponding values using the true treatment effects for imputing potential outcomes. This explains their validity for testing the bounded null hypothesis.  

\section{Simulation studies}\label{sec:sim}
\subsection{A simulation study for testing sharp null hypothesis}\label{sec:sim_sharp_null}

We first conduct a simulation study to evaluate the finite-sample validity of the proposed randomization test under various missingness mechanisms. Specifically, we compare three inferential procedures: (1) the worst-case inferential procedure using the outcome-only test statistic in Section \ref{sec:sharp_null_gen}; (2) the worst-case inferential procedure using the outcome-missingness composite test statistic in Section \ref{sec:sharp_null_composite} with different choices of $\bs{b}$; and (3) the naive procedure that drops units with missing outcomes and applies the standard randomization test to the observed sample. 

We generate potential outcomes as $Y_i(0) = Y_i(1) \overset{\text{i.i.d.}}{\sim} \mathcal{N}(0, 1)$ with $N = 500$, and assign treatment via a CRE allocating half the units to treatment and half to control. We test Fisher's sharp null hypothesis of no treatment effect for any unit, formally stated as $H_0: \bs{\tau}=\bs{0}$, at $10\%$ nominal level. We consider four types of missingness mechanisms, each evaluated under two scenarios in which approximately 5\% and 10\% of units having missing outcomes:
\begin{enumerate}[topsep=1ex,itemsep=-0.3ex,partopsep=1ex,parsep=1ex]
    \item \textsf{(Threshold missingness, regarded as general missingness)}  $M_i(0) = \mathbbm{1}\{ Y_i(0) \leq \text{qnorm}(p) \}$ and $M_i(1) = \mathbbm{1}\{ Y_i(0) \geq \text{qnorm}(q) \}$, with $p = 0.95, q = 0.05$ (5\% missing) or $p = 0.9, q = 0.1$ (10\% missing).

    \item \textsf{(Monotone missingness in Assumption \ref{asmp:mo_miss_M1_ge_M0})}
    $M_{i}(1) = \I\{ Y_{i}(0) \leq \text{qnorm}(p + q) \}$ and $M_{i}(0) = \I\{ Y_{i}(0) \leq \text{qnorm}(p - q) \}$, with $p = 0.95, q = 0.03$ (5\% missing) or $p = 0.9, q = 0.05$ (10\% missing). Note that, by construction,  $M_i(1) \geq M_i(0)$ for all $i$. 

    \item \textsf{(Monotone missingness in Assumption \ref{asmp:mo_miss_M1_le_M0})} $M_{i}(1) = \I\{ Y_{i}(0) \geq \text{qnorm}(p + q) \}$ and $M_{i}(0) = \I\{ Y_{i}(0) \geq \text{qnorm}(p - q) \}$, with $p = 0.05, q = 0.03$ (5\% missing) or $p = 0.1, q = 0.05$ (10\% missing). Note that, by construction,  $M_i(1) \leq M_i(0)$ for all $i$. 

    \item \textsf{(Sharp Missingness)} $M_{i}(1) = M_{i}(0) = \I\{ Y_{i}(0) \leq \text{qnorm}(p) \}$, with $p = 0.95$ (5\% missing) or $p = 0.9$ (10\% missing).
\end{enumerate}

Table~\ref{tab:sim_sharp_null_Type1_Error} reports the empirical Type I error rates of several tests under the missingness mechanisms described above. Column 5 corresponds to the proposed tests based on control potential outcomes.
Columns 6--9 correspond to the proposed tests using composite potential outcomes with different values of $\bs{b}$, where we vary one element of $\bs{b}$ under each missingness mechanism. Column 10 reports the error rates of the ``naive'' method that excludes units with missing outcomes. All tests are based on the Wilcoxon rank-sum statistic.

Table~\ref{tab:sim_sharp_null_Type1_Error} reveals several key findings. First, the proposed worst-case inference procedures control the Type I error rate at or below the nominal level across all missingness mechanisms, confirming their validity. By contrast, the ``naive'' method exhibits substantial Type I error inflation under the threshold and monotone missingness mechanisms, with rejection rates exceeding $75\%$ in several cases, even when the overall proportion of missing outcomes is small. This result highlights an important concern for empirical practice: inference procedures that ignore the missingness mechanism may lead to severe size distortions and misleading conclusions. Under the sharp missingness mechanism, however, the ``naive'' method controls the Type I error rate, consistent with our theoretical results.

Second, the choice of $\bs{b}$ affects the conservativeness of the proposed tests based on composite potential outcomes under the threshold and monotone missingness mechanisms. The optimal choice varies with the missingness mechanism. Specifically, we recommend $b_{01} = \infty, b_{10} = -\infty$ under general missingness (which includes threshold missingness as a special case), $b_{00} = b_{01} = \infty$ under monotone positive missingness, and $b_{00} = b_{10} = -\infty$ under monotone negative missingness. The simulation results align with these recommendations: the empirical rejection probability is closest to the nominal level when the recommended $\bs{b}$ is used. Under the sharp missingness mechanism, the rejection probability remains close to the nominal level for all choices of $b_{00}$, consistent with the discussion in Section~\ref{sec:constant_sharp_null}.

\begin{table}
\centering
\resizebox{\textwidth}{!}{
\begin{threeparttable}
\caption{Simulated Type I Error (\%) Under Different Missingness Patterns}
\label{tab:sim_sharp_null_Type1_Error}
\begin{tabular}{cccccccccc}
    \hline
    \multirow{2}{*}{Missing Mechanism} & \multirow{2}{*}{$b$} & \multirow{2}{*}{Other $b$} & \multirow{2}{*}{$\mathbb{P}(M_i=0)$} & \multirow{2}{*}{$Y_i(0)$} & \multicolumn{4}{c}{$Y_{\bs{b},i}(0)$} & \multirow{2}{*}{Naive} \\
    & & & & & $b=-\infty$ & $b=\text{qnorm}(0.25)$ & $b=\text{qnorm}(0.75)$ & $b=\infty$ & \\ 
    \hline
  threshold & $b_{01}$ & $b_{10} = -\infty, b_{00} = 0$ & 5.10\% & \cellcolor{gray!30}8.82\% & 0.00\% & 0.00\% & 0.98\% & \cellcolor{gray!30}8.82\% & 76.94\% \\ 
  threshold & $b_{01}$ & $b_{10} = -\infty, b_{00} = 0$ & 10.49\% & \cellcolor{gray!30}4.47\% & 0.00\% & 0.00\% & 1.32\% & \cellcolor{gray!30}4.47\% & 99.83\% \\ 
  threshold & $b_{10}$ & $b_{01} = \infty, b_{00} = 0$ & 5.10\% & \cellcolor{gray!30}8.82\% & \cellcolor{gray!30}8.82\% & 1.92\% & 0.00\% & 0.00\% & 76.94\% \\ 
  threshold & $b_{10}$ & $b_{01} = \infty, b_{00} = 0$ & 10.49\% & \cellcolor{gray!30}4.47\% & \cellcolor{gray!30}4.47\% & 2.10\% & 0.00\% & 0.00\% & 99.83\% \\ 
  \hline
  mp & $b_{00}$ & $b_{01} = \infty, b_{10} = 0$ & 6.00\% & 0.76\% & 0.76\% & 1.44\% & 5.30\% & \cellcolor{gray!30}8.44\% & 51.14\% \\ 
  mp & $b_{00}$ & $b_{01} = \infty, b_{10} = 0$ & 10.60\% & 0.00\% & 0.00\% & 0.06\% & 2.49\% & \cellcolor{gray!30}5.83\% & 75.40\% \\ 
  mp & $b_{01}$ & $b_{00} = \infty, b_{10} = 0$ & 6.00\% & 0.76\% & 0.00\% & 0.00\% & 0.98\% & \cellcolor{gray!30}8.44\% & 51.14\% \\ 
  mp & $b_{01}$ & $b_{00} = \infty, b_{10} = 0$ & 10.60\% & 0.00\% & 0.00\% & 0.00\% & 1.64\% & \cellcolor{gray!30}5.83\% & 75.40\% \\ 
  \hline
  mn & $b_{00}$ & $b_{10} = -\infty, b_{01} = 0$ & 5.20\% & 1.21\% & \cellcolor{gray!30}8.71\% & 6.36\% & 2.34\% & 1.21\% & 44.84\% \\ 
  mn & $b_{00}$ & $b_{10} = -\infty, b_{01} = 0$ & 9.60\% & 0.03\% & \cellcolor{gray!30}5.89\% & 3.43\% & 0.24\% & 0.03\% & 75.71\% \\ 
  mn & $b_{10}$ & $b_{00} = -\infty, b_{01} = 0$ & 5.20\% & 1.21\% & \cellcolor{gray!30}8.71\% & 1.87\% & 0.00\% & 0.00\% & 44.84\% \\ 
  mn & $b_{10}$ & $b_{00} = -\infty, b_{01} = 0$  & 9.60\% & 0.03\% & \cellcolor{gray!30}5.89\% & 2.45\% & 0.00\% & 0.00\% & 75.71\% \\ 
  \hline
  sharp & $b_{00}$ & $b_{01} = 0, b_{10} = 0$ & 5.60\% & 0.00\% & 10.42\% & 10.52\% & 10.48\% & 10.46\% & \cellcolor{gray!30}10.05\% \\ 
  sharp & $b_{00}$ & $b_{01} = 0, b_{10} = 0$ & 11.00\% & 0.00\% & 10.32\% & 10.38\% & 10.51\% & 10.47\% & \cellcolor{gray!30}10.25\% \\ 
   \hline
\end{tabular}
\begin{tablenotes}
\footnotesize \item \emph{Note}: The nominal level is set as $\alpha=10\%$. Column 1 displays the missingness mechanism. Column 2 reports the element of $\bs{b}$ that we will vary.
Column 3 reports the values of the other elements of $\bs{b}$ that are kept fixed.
Column 4 reports the proportion of missing units. Column~5 reports the Type~I error rate for the test statistic based only on the imputed control potential outcomes. Columns~6--9 report the Type~I error rates for the test statistics based on the imputed composite control potential outcomes with different values of $\bs{b}$. Column~10 reports the Type~I error rate of the naive approach, which excludes units with missing outcomes. The cells shaded in gray highlight the recommended approach. 
``mp'' and ``mn'' refer to monotone positive and negative missingness in Assumptions \ref{asmp:mo_miss_M1_ge_M0} and \ref{asmp:mo_miss_M1_le_M0}, respectively.
\end{tablenotes}
\end{threeparttable}}
\end{table}

\subsection{A simulation study for testing sharp null hypothesis using two-step method}\label{sec:sim_sharp_null_twostep}

We next conduct a simulation study to compare the power of the two-step methods with that of the one-step methods under various missingness mechanisms. Potential outcomes are generated according to $Y_i(0) \overset{\text{i.i.d.}}{\sim} \mathcal{N}(0,1)$ and $Y_i(1) = Y_i(0) + \delta$, with sample size $N = 500$ and $\delta$ ranging from 0 to 0.8. We again consider the CRE, which assigns half of the units to treatment and the remaining half to control. For each design, we test the sharp null hypothesis $H_0:\bs{\tau}=\bs{0}$ at the $10\%$ nominal level. Missingness indicators are generated as in Section~\ref{sec:sim_sharp_null}, with approximately $20\%$ of units having missing outcomes:
\begin{enumerate}[topsep=1ex,itemsep=-0.3ex,partopsep=1ex,parsep=1ex]
    \item \textsf{(Threshold missingness, regarded as general missingness)}  $M_i(0) = \mathbbm{1}\{ Y_i(0) \leq \text{qnorm}(p) \}$ and $M_i(1) = \mathbbm{1}\{ Y_i(0) \geq \text{qnorm}(q) \}$, with $p = 0.65, q = 0.05$ (20\% missing).
    \item \textsf{(Monotone missingness in Assumption \ref{asmp:mo_miss_M1_ge_M0})}
    $M_{i}(1) = \I\{ Y_{i}(0) \leq \text{qnorm}(p + q) \}
    \ge 
    M_{i}(0) = \I\{ Y_{i}(0) \leq \text{qnorm}(p - q) \}$, with $p = 0.8, q = 0.18$ (20\% missing). 
    \item \textsf{(Monotone missingness in Assumption \ref{asmp:mo_miss_M1_le_M0})} $M_{i}(1) = \I\{ Y_{i}(0) \geq \text{qnorm}(p + q) \}
    \le 
    M_{i}(0) = \I\{ Y_{i}(0) \geq \text{qnorm}(p - q) \}$, with $p = 0.2, q = 0.18$ (20\% missing).
\end{enumerate}

\begin{figure}[h!]
    \centering
    \includegraphics[width=0.8\linewidth]{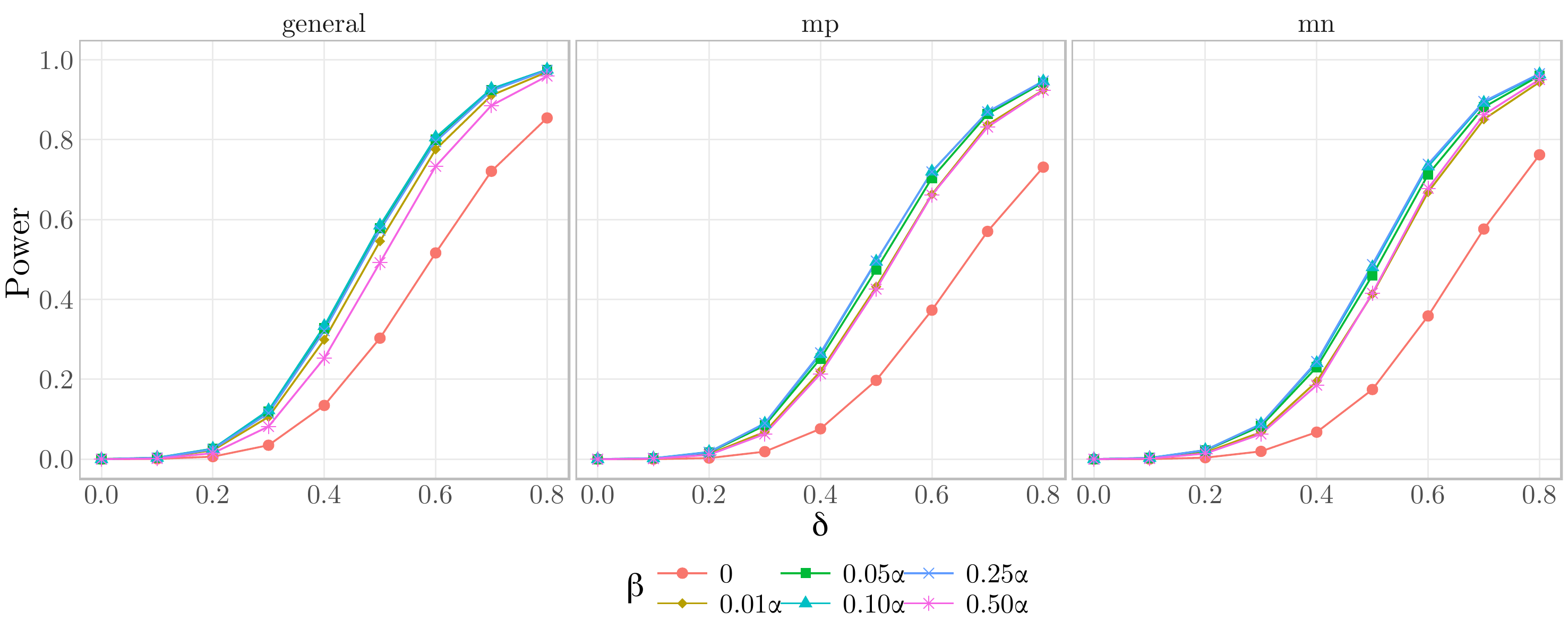}
    \caption{Power Curve of Two-Step Method Under Threshold and Monotone Missingness Patterns}
    \label{fig:sim_sharp_null_power_twostep}
\end{figure}

The results in Figure~\ref{fig:sim_sharp_null_power_twostep} reveal two main findings. First, the two-step method achieves substantially higher power than the one-step method under both threshold and monotone missingness patterns, while maintaining type I error control at the nominal level.
For example, under monotone positive missingness with a treatment effect of 0.5, power increases from 18\% for the one-step method to 49\% for the two-step method with $\beta = 0.1\alpha$. Second, 
the test is relatively robust to a range of positive $\beta$ values. Based on the simulation, we suggest $\beta = 0.1\alpha$ in practice.

\section{Empirical application}

We use data from the National Job Corps Study, a large randomized evaluation funded by the U.S. Department of Labor to assess the Job Corps program’s impact on labor market outcomes, especially wages \citep{lee2009training}. As one of the largest federally funded job training programs, Job Corps serves economically disadvantaged youth aged 16–24, offering residential services, healthcare, and educational and vocational training. Participants typically spend about eight months in the program, receiving roughly 1,100 hours of instruction—equivalent to one academic year of high school.

Even with randomized assignment, estimating the impact of the training program on wages is empirically challenging due to sample selection. Wages are only observed for the employed, but employment itself may be influenced by the program \citep{heckman1974shadow}. As a result, treatment and control groups may not be comparable among those employed. \cite{lee2009training} argues that the missingness mechanism at weeks 90, 135, and 180 is monotone positive.

We begin by testing the sharp null hypothesis that the treatment effect is zero for all individuals. Table \ref{tab:job_corps_p} presents randomization p-values under various missingness assumptions for both one-step and two-step method. We fail to reject the null hypothesis at conventional significance levels under general missingness assumptions and monotone positive/negative missingness assumptions. However, under sharper assumptions—such as sharp missingness or missing at random\footnote{We discuss randomization inference under the missing-at-random mechanism in the supplementary material.}—the sharp null can be rejected at the 5\% level. 
Note that the two-step procedure yields no improvement under either general or monotone missingness: the former is due to high overall missingness, and the latter is due to similar missing proportions between treated and control groups.

Table \ref{tab:job_corps_ci} reports the 95\% confidence intervals derived via Lehmann-style test inversion under the assumption of a constant treatment effect. At week 90 after the treatment, the Job Corps sample includes 5546 treated and 3599 control individuals. Under the general missingness assumption, the 95\% confidence interval for the treatment effect is uninformative: $(-\infty, \infty)$.
However, assuming monotone positive missingness, the interval narrows substantially to $[-0.042, 0.087]$, despite 54\% of wage observations being missing. By the end of the 180-week follow-up period, the interval remains informative, ruling out effects more negative than -8\% and more positive than 12\%. Confidence intervals under stronger assumptions (sharp or random missingness) are even tighter, reflecting the additional statistical power provided by these assumptions.

\begin{table}
    \centering
    \caption{P-Values for Testing the Sharp Null Hypothesis $H_0: \bs{\tau} = \bs{0}$}
    \label{tab:job_corps_p}
    \resizebox{0.8\textwidth}{!}{
    \begin{tabular}{c|ccc|cccc|cc}
    \hline
    \multirow{2}{*}{Week} & \multicolumn{3}{c|}{Missing Proportion} & \multicolumn{6}{c}{Missing Mechanism} \\
    & Total & Treated & Control & General & MP & Sharp & Random & Two-Step General & Two-Step MP \\
    \hline
90 & 0.538 & 0.538 & 0.539 & 1.000 & 0.340 & 0.005 & 0.005 & 1.000 & 0.345 \\
135 & 0.464 & 0.453 & 0.480 & 1.000 & 0.973 & 0.027 & 0.027 & 1.000 & 0.978 \\
180 & 0.431 & 0.417 & 0.452 & 1.000 & 0.973 & 0.015 & 0.015 & 1.000 & 0.974 \\
\hline
    \end{tabular}
    }
\end{table}

\begin{table}
    \caption{95\% CIs for a Constant Treatment Effect}
    \label{tab:job_corps_ci}
    \resizebox{\textwidth}{!}{
    \begin{tabular}{c|ccc|cccc|cc}
    \hline
    \multirow{2}{*}{Week} & \multicolumn{3}{c|}{Missing Proportion} & \multicolumn{6}{c}{Missing Mechanism} \\
    & Total & Treated & Control & General & MP & Sharp & Random & Two-Step General & Two-Step MP \\
    \hline
90 & 0.538 & 0.538 & 0.539 & (-$\infty$, $\infty$) & [-0.042, 0.087] & [0.001, 0.040] & [0.001, 0.040] & (-$\infty$, $\infty$) & [-0.042, 0.089] \\
135 & 0.464 & 0.453 & 0.480 & (-$\infty$, $\infty$) & [-0.085, 0.119] & [0.000, 0.031] & [0.000, 0.031] & (-$\infty$, $\infty$) & [-0.087, 0.121] \\
180 & 0.431 & 0.417 & 0.452 & (-$\infty$, $\infty$) & [-0.080, 0.120] & [0.001, 0.037] & [0.001, 0.037] & (-$\infty$, $\infty$) & [-0.080, 0.121] \\
\hline
    \end{tabular}
    }
\end{table}

\section{Discussion}

Randomization inference has been widely used across scientific disciplines. 
However, naive application in experiments with sample attrition can lead to severe size distortions. We address this problem by developing computationally efficient methods for randomization inference that remain valid under a broad class of potentially informative missingness mechanisms.
These proposed methods are valid for testing both sharp and bounded null hypotheses.

We expect our methods to be practically useful for applied researchers. When missingness mechanisms are largely unknown, our methods under general missingness should be used. If the experiment satisfies monotone or sharp missingness, the corresponding methods should be applied. 
In addition, as discussed in the supplementary material, when missingness is random or believed to be random, randomization inference can be applied directly to the observed sample. An R package implementing the proposed methods is publicly available at \url{https://github.com/peizansheng/riattrition}.

Finally, \cite{kwon2024testing} note that their test of the sharp null of full mediation can also be used to test the sharp null of no treatment effect in our setting. This follows because the sharp null of no treatment effect implies no treatment effect for always-in-sample and never-in-sample units (i.e., $M_i(1)=M_i(0)=1$ and $M_i(1)=M_i(0)=0$, respectively). Their testing procedure differs from ours, and comparing the two approaches is an important direction for future work.

\begin{spacing}{1}

\bibliographystyle{ecta}
\bibliography{rand_inf_joint_po} 

\end{spacing}

\newpage

\setcounter{equation}{0}
\setcounter{section}{0}
\setcounter{figure}{0}
\setcounter{example}{0}
\setcounter{proposition}{0}
\setcounter{corollary}{0}
\setcounter{theorem}{0}
\setcounter{lemma}{0}
\setcounter{table}{0}
\setcounter{condition}{0}
\setcounter{page}{1}

\renewcommand {\theproposition} {A\arabic{proposition}}
\renewcommand {\theexample} {A\arabic{example}}
\renewcommand {\thefigure} {A\arabic{figure}}
\renewcommand {\thetable} {A\arabic{table}}
\renewcommand {\theequation} {A\arabic{equation}}
\renewcommand {\thelemma} {A\arabic{lemma}}
\renewcommand {\thesection} {A\arabic{section}}
\renewcommand {\thetheorem} {A\arabic{theorem}}
\renewcommand {\thecorollary} {A\arabic{corollary}}
\renewcommand {\thecondition} {A\arabic{condition}}

\renewcommand {\thepage} {A\arabic{page}}

\begin{center}
  {\Large \textbf{Supplementary Material for ``Randomization Inference with Sample Attrition''}}
\end{center}

\section{Randomization tests for sharp null under the missing-at-random mechanism}\label{sec:mar_sharp_null}

In this section, we assume non-informative missingness, under which we can infer treatment effects most precisely. Importantly, unlike Assumptions \ref{asmp:general}--\ref{asmp:constant_miss}, the potential missingness indicators are treated as random variables and are not conditioned upon. Intuitively, we assume that the potential missingness indicators are independent of the potential outcomes \citep{little2019statistical, bai2024revisiting}. Since the potential outcomes are already conditioned on in our randomization-based inference and treated as fixed (arbitrary) constants, this assumption effectively requires the potential missingness indicators to be i.i.d. across all units. We formally state this missing at random assumption below, explicitly including the conditioning on potential outcomes to emphasize the non-informativeness of the missingness.

\begin{assumption}[Missing at Random]\label{asmp:mar}
    The potential missingness indicators $(M_i(1),$ $M_i(0))$s are i.i.d.\ across all $1\le i \le n$, 
    and their distribution does not depend on the potential outcomes. 
    Specifically, 
    \begin{align*}
        ((M_1(1), M_1(0)), 
        \ldots, (M_n(1), M_n(0))
        \mid \bs{Y}(1), \bs{Y}(0)
        \overset{\text{i.i.d.}}{\sim} 
        \text{Categorical}(p_{11}, p_{10}, p_{01}, p_{00}),
    \end{align*}
    where $\text{Categorical}(p_{11}, p_{10}, p_{01}, p_{00})$ denotes a distribution with probability mass $p_{mm'}$ at $(m,m')\in \{0,1\}^2$, 
    with $p_{11}+p_{10}+p_{01}+p_{00}=1$, 
    and these probabilities $p_{11}, p_{10}, p_{01}$ and $p_{00}$ are constants that do not depend on the potential outcomes $\bs{Y}(1)$ and $\bs{Y}(0)$. 
\end{assumption}

Assumption \ref{asmp:mar} does not restrict the dependence between $M(1)$ and $M(0)$, so the information on potential missingness and outcomes remains the same as under general missingness mechanisms (see Panel A of Table \ref{tab:sharp_null}). However, Assumption \ref{asmp:mar} implies that, after appropriate conditioning, the experiment can be viewed as a CRE restricted to units with observed outcomes, as in \cite{ghanem2023testing}. Consequently, standard randomization tests can be applied without concern for missing data under this assumption. Specifically, we can show that
\begin{align}\label{eq:imp_mar}
    \bs{Z}_{\S} \mid \bs{Y}(1), \bs{Y}(0), \bs{M}, \bs{Z}_{\S^\complement}, n_{\S1}, n_{\S0} 
    \ \sim \ 
    \CRE(\S, n_{\S1}, n_{\S0}), 
\end{align}
where $\S = \{i: M_i = 1, 1\le i \le n\}$, $\bs{Z}_{\S}$ and $\bs{Z}_{\S^\complement}$ denote the subvectors of $\bs{Z}$ corresponding to units in $\S$ and $\S^\complement$, respectively, $n_{\S1} = \sum_{i\in \S} Z_i$, and $n_{\S0} = |\S|-n_{\S1}$. That is, after proper conditioning, the treatment assignment for units in $\mathcal{S}$ follows the same distribution as that of a CRE, which randomly assigns $n_{\mathcal{S}1}$ units from $\mathcal{S}$ to treatment and the remaining $n_{\mathcal{S}0}$ to control. The following theorem establishes the validity of randomization tests conditional on units with observed outcomes.

\begin{theorem}\label{thm:mar_miss}
\rm 
Under the CRE and Assumption \ref{asmp:mar}, 
a valid p-value for testing the sharp null hypothesis $H_{\bs{\delta}}$ in \eqref{eq:null_fisher_sharp} is 
$
p_{\bs{Z}, \bs{\delta}, \S}
$,
where $p_{\bs{Z}, \bs{\delta}, \S}$ is defined as:
\begin{align}\label{eq:pval_sharpnull_mar}
    p_{\bs{Z}, \bs{\delta}, \S} = G_{\R, \phi, \S}( t_{\R, \phi, \S}(\bs{Z}, \bs{Y} - \bs{\delta}\circ \bs{Z})),
\end{align}
where $t_{\R, \phi, \S}(\cdot)$ and $G_{\R, \phi, \S}(\cdot)$ are defined as:
\begin{align*}
   t_{\R, \phi, \S} (\bs{z}, \bs{y})
   & = 
   t_{\R, \phi} (\bs{z}_{\S}, \bs{y}_{\S}), 
   \ \ 
   \text{and}
   \ \ 
   G_{\R, \phi, \S}(c)
   = \Pr 
   \left( t_{\R, \phi, \S} (\bs{A}, \bs{y}) \ge c
    \right) \text{ for } c \in \mathbb{R},
\end{align*}
with $t_{\R, \phi}$ defined in either \eqref{eq:test_stat_R_phi}, \eqref{eq:test_stat_mann_whitney} or \eqref{eq:test_stat_mann_whitney_negative},  
$G_{\R, \phi, \S}(\cdot)$ defined based on random assignment $\bs{A}_{\S}$ from $\CRE(\S, n_{\S1}, n_{\S0})$, 
and $\bs{z}_{\S}, \bs{y}_{\S}$ and $\bs{A}_{\S}$ being subvectors of $\bs{z}, \bs{y}$ and $\bs{A}$ for units in $\S$.
\end{theorem}

Theorem \ref{thm:mar_miss} is essentially testing the following ``modified'' sharp null hypothesis for units in $\mathcal{S}$ whose outcomes are observed: 
\begin{align}
    H_{\bs{\delta}, \mathcal{S}}: \tau_{i} = \delta_{i}, \quad \text{for $i \in \mathcal{S} \coloneqq \{j: M_j = 1, 1\le j \le n\}$}.
\end{align}
The sharp null $H_{\bs{\delta}}$ implies $H_{\bs{\delta}, \mathcal{S}}$, so testing $H_{\bs{\delta}}$ reduces to conducting a randomization test for $H_{\bs{\delta}, \mathcal{S}}$ using only units with observed outcomes. Note that $\mathcal{S}$ is a random subset of units that generally depends on the treatment assignments. Moreover, the procedure in Theorem \ref{thm:mar_miss} also works when using a generic test statistic, such as the difference-in-means test statistic, since there will be no missing outcomes once we focus only on units in $\S$.

Although the testing procedures in Theorems \ref{thm:constant_miss} and \ref{thm:mar_miss} are the same, the key difference between the two theorems is that $\mathcal{S}$ is random in Theorem \ref{thm:mar_miss}, whereas it is fixed under the sharp missingness mechanism in Theorem \ref{thm:constant_miss}.

Another remark regarding Theorem \ref{thm:mar_miss} concerns the necessity direction. In general, random treatment assignment (such as in the CRE) and $\bs{Z} \independent (\bs{Y}(0), \bs{Y}(1)) \mid \bs{M}$ do not imply $(\bs{M}(0), \bs{M}(1)) \independent (\bs{Y}(0), \bs{Y}(1))$. A counterexample is the case where $\bs{M}(1) = \bs{M}(0) = \bs{Y}^{\sta}(1) = \bs{Y}^{\sta}(0)$, and $\bs{Z}$ is drawn from a CRE.

\section{Useful lemmas}

\begin{lemma}\label{lemma:caughey_et_al_lemma_a7}
\rm Consider any $n \geq 1$, $\bs{z} \in \{ 0, 1 \}^{n}$ and $\bs{y}, \bs{y'} \in \overline{\mathbb{R}}^{n}$, 
where $\overline{\mathbb{R}} := \mathbb{R}\cup \{\pm \infty\}$,
such that:
\begin{align*}
\begin{cases*}
    y_i \leq y_i', \quad \text{if $Z_i = 1$}, \\
    y_i \geq y_i', \quad \text{if $Z_i = 0$}.
\end{cases*}
\end{align*}
Furthermore, construct $\bs{\tilde{y}}, \bs{\tilde{y}'} \in \mathbb{R}$ such that:
\begin{align*}
\tilde{y}_i = 
\begin{cases}
    y_i, \quad &\text{if $y_i \in \mathbb{R}$}, \\
    \gamma, \quad &\text{if $y_i = - \infty$}, \\
    \zeta, \quad &\text{if $y_i = \infty$},
\end{cases} 
\quad 
\tilde{y}_i' = 
\begin{cases}
    y_i', \quad &\text{if $y_i' \in \mathbb{R}$}, \\
    \gamma, \quad &\text{if $y_i' = - \infty$}, \\
    \zeta, \quad &\text{if $y_i' = \infty$},
\end{cases}
\end{align*}
where
\begin{align*}
\gamma = \min\left\{ \{ y_i: y_{i} \in \mathbb{R} \} \cup \{ y_i': y_{i}' \in \mathbb{R} \} \right\} - a, \quad \zeta = \max\left\{ \{ y_i: y_{i} \in \mathbb{R} \} \cup \{ y_i': y_{i}' \in \mathbb{R} \} \right\} + b,
\end{align*}
with $a, b \in \mathbb{R}^{+}$. Then, the following properties hold for any test statistic defined in \eqref{eq:test_stat_R_phi}:
\begin{enumerate}
    \item[(a)] for $\bs{\tilde{y}}, \bs{\tilde{y}'}$:
    \begin{align*}
    \begin{cases*}
    \tilde{y}_i \leq \tilde{y}_i', \quad \text{if $z_i = 1$}, \\
    \tilde{y}_i \geq \tilde{y}_i', \quad \text{if $z_i = 0$};
    \end{cases*}
    \end{align*}
    \item[(b)] for $t_{\R, \phi}\left( \bs{z}, \bs{y} \right)$, $t_{\R, \phi}\left( \bs{z}, \bs{y'} \right)$, $t_{\R, \phi}\left( \bs{z}, \bs{\tilde{y}} \right)$, and $t_{\R, \phi}\left( \bs{z}, \bs{\tilde{y}'} \right)$:
    \begin{align*}
    t_{\R, \phi}\left( \bs{z}, \bs{y} \right) = t_{\R, \phi}\left( \bs{z}, \bs{\tilde{y}} \right), \quad t_{\R, \phi}\left( \bs{z}, \bs{y'} \right) = t_{\R, \phi}\left( \bs{z}, \bs{\tilde{y}'} \right);
    \end{align*}
    \item[(c)] for $t_{\R, \phi}\left( \bs{z}, \bs{y} \right)$, $t_{\R, \phi}\left( \bs{z}, \bs{y'} \right)$:
    \begin{align*}
        t_{\R, \phi}\left( \bs{z}, \bs{y} \right) \leq t_{\R, \phi}\left( \bs{z}, \bs{y'} \right).
    \end{align*}
\end{enumerate}
\end{lemma}

\begin{lemma}\label{lemma:mann_whitney_eff_incres}
\rm Consider any $n \geq 1$, $\bs{z} \in \{ 0, 1 \}^{n}$ and $\bs{y}, \bs{y'} \in \overline{\mathbb{R}}^{n}$ such that:
\begin{align*}
\begin{cases*}
    y_i \leq y_i', \quad \text{if $z_i = 1$}, \\
    y_i \geq y_i', \quad \text{if $z_i = 0$}.
\end{cases*}
\end{align*}
Then, for any test statistic defined in \eqref{eq:test_stat_mann_whitney} and \eqref{eq:test_stat_mann_whitney_negative}:
\begin{align*}
    t_{\R, \phi}\left( \bs{z}, \bs{y} \right) \leq t_{\R, \phi}\left( \bs{z}, \bs{y'} \right).
\end{align*}
\end{lemma}

\begin{lemma}\label{lemma:rand_inf_valid_composite}
\rm Under the CRE and Assumption \ref{asmp:general}, 
the following p-values are valid for testing the sharp null hypothesis $H_{\bs{\delta}}$ in \eqref{eq:null_fisher_sharp}
\begin{align*}
    p_{\bs{Z}, \bs{\delta}} = G_{\R, \phi}\left(t_{\R, \phi}(\bs{Z}, \bs{Y}(0)) \right), \quad p_{\bs{Z}, \bs{\delta}, \bs{b}} = G_{\R, \phi}\left( t_{\R, \phi}\left(\bs{Z}, \tilde{\bs{Y}}_{\bs{b}}(0) \right) \right),
\end{align*}
where $t_{\R, \phi}$ is defined in either \eqref{eq:test_stat_R_phi}, \eqref{eq:test_stat_mann_whitney}, or \eqref{eq:test_stat_mann_whitney_negative}, 
$G_{\R, \phi}$ is defined in
\eqref{eq:G_R_phi}, 
$\bs{Y}(0)$ is defined in \eqref{eq:imputation}, and $\tilde{\bs{Y}}_{\bs{b}}(0)$ is defined in Section \ref{subsec:composite_test_stats}. That is, under $H_{\bs{\delta}}$,
\begin{align*}
    \mathbb{P}\left( p_{\bs{Z}, \bs{\delta}} \leq \alpha \right) \leq \alpha, \quad \mathbb{P}\left( p_{\bs{Z}, \bs{\delta}, \bs{b}} \leq \alpha \right) \leq \alpha,
\end{align*}
for any $\alpha \in (0, 1)$ and $b_{00}, b_{01}, b_{10} \in \mathbb{R}$.
\end{lemma}

\section{Proofs of the main results}\label{appx_sec:proof}

\begin{proof}[Proof of Theorem \ref{thm:sharp_null_gen}] Under $H_{\bs{\delta}}$ in \eqref{eq:null_fisher_sharp}, for $Y_i(0)$ and $Y_{\bs{Z}, \bs{\delta}, i}^{\gen}(0)$, where $1 \leq i \leq n$:
\begin{align*}
Y_i(0) = \begin{cases}
    Y_i - \delta_i \quad &\text{if $Z_i = 1$, $M_i = 1$} \\
    Y_i^{\sta} - \delta_i \quad &\text{if $Z_i = 1$, $M_i = 0$} \\
    Y_i \quad &\text{if $Z_i = 0$, $M_i = 1$} \\
    Y_i^{\sta} \quad &\text{if $Z_i = 0$, $M_i = 0$},
\end{cases} \quad 
Y_{\bs{Z}, \bs{\delta}, i}^{\gen}(0) = \begin{cases}
        Y_{i} - \delta_{i} &\quad \text{if $Z_{i} = 1$, $M_{i} = 1$} \\
        -\infty &\quad \text{if $Z_{i} = 1$, $M_{i} = 0$} \\
        Y_{i} &\quad \text{if $Z_{i} = 0$, $M_{i} = 1$} \\
        \infty &\quad \text{if $Z_{i} = 0$, $M_{i} = 0$}, \\
    \end{cases}
\end{align*}

Therefore, under $H_{\bs{\delta}}$ in \eqref{eq:null_fisher_sharp}, for $Y_{\bs{Z}, \bs{\delta}, i}^{\gen}(0)$ and $Y_i(0)$, where $1 \leq i \leq n$:
\begin{align*}
\begin{cases*}
    Y_{\bs{Z}, \bs{\delta}, i}^{\gen}(0) \leq Y_i(0), \quad \text{if $Z_i = 1$}, \\
    Y_{\bs{Z}, \bs{\delta}, i}^{\gen}(0) \geq Y_i(0), \quad \text{if $Z_i = 0$}.
\end{cases*}
\end{align*}
Lemma \ref{lemma:caughey_et_al_lemma_a7} and Lemma \ref{lemma:mann_whitney_eff_incres} imply that $t_{\R, \phi}\left(\bs{Z}, \bs{Y}_{\bs{Z}, \bs{\delta}}^{\gen}(0) \right) \leq t_{\R, \phi}\left(\bs{Z}, \bs{Y}(0) \right)$.

Because tail probability is nonincreasing, we have 
\begin{align*}
    p_{\bs{Z}, \bs{\delta}} = G_{\R, \phi}(t_{\R, \phi}\left(\bs{Z}, \bs{Y}(0) \right)) \leq G_{\R, \phi}\left( t_{\R, \phi}\left(\bs{Z}, \bs{Y}_{\bs{Z}, \bs{\delta}}^{\gen}(0) \right) \right) = p_{\bs{Z}, \bs{\delta}}^\gen,
\end{align*}
where $p_{\bs{Z}, \bs{\delta}}$ is a valid p-value for testing against the sharp null hypothesis $H_{\bs{\delta}}$ in \eqref{eq:null_fisher_sharp} by Lemma~\ref{lemma:rand_inf_valid_composite}. Now, the desired result follows immediately.
\end{proof}

\begin{proof}[Proof of Theorem \ref{thm:sharp_null_gen_comp}] When $Z_i = 0, M_{i} = 0$, for $\tilde{Y}_{\bs{Z},\bs{\delta}, \bs{b}, i}^{\gen}(0)$ and $\tilde{Y}_{\bs{b},i}(0)$:
\begin{align*}
    \tilde{Y}_{\bs{b},i}(0) = b_{00} \times (1 - M_{i}(1)) + b_{01}  \times M_{i}(1) \leq \max\{ b_{00}, b_{01} \} = \tilde{Y}_{\bs{Z},\bs{\delta}, \bs{b}, i}^{\gen}(0).
\end{align*}

When $Z_i = 0, M_{i} = 1$, for $\tilde{Y}_{\bs{Z},\bs{\delta}, \bs{b}, i}^{\gen}(0)$ and $\tilde{Y}_{\bs{b},i}(0)$:
\begin{align*}
    \tilde{Y}_{\bs{b},i}(0) = Y_i \times M_{i}(1) + b_{10} \times (1 - M_{i}(1)) \leq \max\{ Y_{i}, b_{10} \} = \tilde{Y}_{\bs{Z},\bs{\delta}, \bs{b}, i}^{\gen}(0).
\end{align*}

Under the sharp null hypothesis $H_{\bs{\delta}}$ in \eqref{eq:null_fisher_sharp}, when $Z_i = 1, M_i = 0$, for $\tilde{Y}_{\bs{Z},\bs{\delta}, \bs{b}, i}^{\gen}(0)$ and $\tilde{Y}_{\bs{b},i}(0)$:
\begin{align*}
    \tilde{Y}_{\bs{b},i}(0) = b_{00} \times (1 - M_{i}(0)) + b_{10} \times M_{i}(0) \geq \min\{ b_{00}, b_{10} \} = \tilde{Y}_{\bs{Z},\bs{\delta}, \bs{b}, i}^{\gen}(0).
\end{align*}

Under the sharp null hypothesis $H_{\bs{\delta}}$ in \eqref{eq:null_fisher_sharp}, when $Z_i = 1, M_i = 1$, for $\tilde{Y}_{\bs{Z},\bs{\delta}, \bs{b}, i}^{\gen}(0)$ and $\tilde{Y}_{\bs{b},i}(0)$:
\begin{align*}
    \tilde{Y}_{\bs{b},i}(0) = (Y_i - \delta_{i}) \times M_{i}(0) + b_{01} \times (1 - M_{i}(0)) \geq \min\{ Y_i - \delta_{i}, b_{01} \} = \tilde{Y}_{\bs{Z},\bs{\delta}, \bs{b}, i}^{\gen}(0).
\end{align*}

Therefore, under the sharp null hypothesis $H_{\bs{\delta}}$ in \eqref{eq:null_fisher_sharp}, for $\tilde{\bs{Y}}_{\bs{Z},\bs{\delta}, \bs{b}}^{\gen}(0), \tilde{\bs{Y}}_{\bs{b}}(0)$:
\begin{align*}
\begin{cases*}
    \tilde{Y}_{\bs{Z},\bs{\delta}, \bs{b}, i}^{\gen}(0) \leq \tilde{Y}_{\bs{b},i}(0), \quad \text{if $Z_i = 1$}, \\
    \tilde{Y}_{\bs{Z},\bs{\delta}, \bs{b}, i}^{\gen}(0) \geq \tilde{Y}_{\bs{b},i}(0), \quad \text{if $Z_i = 0$}.
\end{cases*}
\end{align*}
Lemma \ref{lemma:caughey_et_al_lemma_a7} and Lemma \ref{lemma:mann_whitney_eff_incres} imply that $t_{\R, \phi}\left(\bs{Z}, \tilde{\bs{Y}}_{\bs{Z},\bs{\delta}, \bs{b}}^{\gen}(0) \right) \leq t_{\R, \phi}\left(\bs{Z}, \tilde{\bs{Y}}_{\bs{b}}(0) \right)$. Because the tail probability is nonincreasing:
\begin{align*}
    p_{\bs{Z}, \bs{\delta}, \bs{b}} = G_{\R, \phi}\left(t_{\R, \phi}\left(\bs{Z}, \tilde{\bs{Y}}_{\bs{b}}(0) \right)\right) \leq G_{\R, \phi}\left( t_{\R, \phi}\left(\bs{Z}, \tilde{\bs{Y}}_{\bs{Z},\bs{\delta}, \bs{b}}^{\gen}(0) \right) \right) = \tilde{p}_{\bs{Z}, \bs{\delta}, \bs{b}}^\gen,
\end{align*}
where $p_{\bs{Z}, \bs{\delta}, \bs{b}}$ is a valid p-value for testing against the sharp null hypothesis $H_{\bs{\delta}}$ in \eqref{eq:null_fisher_sharp} by Lemma~\ref{lemma:rand_inf_valid_composite}. Now, the desired result follows immediately.
\end{proof}

\begin{proof}[Proof of Theorem \ref{thm:sharp_null_mon}] We first consider the case when Assumption \ref{asmp:mo_miss_M1_ge_M0} holds. Following an identical argument in Theorem \ref{thm:sharp_null_gen_comp}, when $Z_{i} = 0, M_{i} = 0$, for $\tilde{Y}^{\monp}_{\bs{Z}, \bs{\delta}, \bs{b}, i}(0)$ and $\tilde{Y}_{\bs{b},i}(0)$:
\begin{align*}
    \tilde{Y}_{\bs{b},i}(0) \leq \tilde{Y}^{\monp}_{\bs{Z}, \bs{\delta}, \bs{b}, i}(0).
\end{align*}
Similarly, under the sharp null in \eqref{eq:null_fisher_sharp}, when $Z_{i} = 1, M_{i} = 1$, for $\tilde{Y}^{\monp}_{\bs{Z}, \bs{\delta}, \bs{b}, i}(0)$ and $\tilde{Y}_{\bs{b},i}(0)$:
\begin{align*}
    \tilde{Y}_{\bs{b},i}(0) \geq \tilde{Y}^{\monp}_{\bs{Z}, \bs{\delta}, \bs{b}, i}(0).
\end{align*}
Under Assumption \ref{asmp:mo_miss_M1_ge_M0}, $Z_i = 0$ and $M_i = 1$ (i.e., $M_{i}(0) = 1$) implies that $M_{i}(1) = 1$, which further implies:
\begin{align*}
    \tilde{Y}_{\bs{b},i}(0) = Y_i \times 1 \times 1 + b_{00} \times 0 \times 0 + b_{01} \times 0 \times 1 + b_{10} \times 1 \times 0 = Y_i = \tilde{Y}^{\monp}_{\bs{Z}, \bs{\delta}, \bs{b}, i}(0).
\end{align*}
Under the sharp null in \eqref{eq:null_fisher_sharp} and Assumption \ref{asmp:mo_miss_M1_ge_M0}, $Z_i = 1$ and $M_i = 0$ (i.e., $M_{i}(1) = 0$) implies that $M_{i}(0) = 0$, which further implies:
\begin{align*}
    \tilde{Y}_{\bs{b},i}(0) = Y_i(0) \times 0 \times 0 + b_{00} \times 1 \times 1 + b_{01} \times 1 \times 0 + b_{10} \times 0 \times 1 = b_{00} = \tilde{Y}^{\monp}_{\bs{Z}, \bs{\delta}, \bs{b}, i}(0).
\end{align*}

We then consider the case when Assumption \ref{asmp:mo_miss_M1_le_M0} holds. Following an identical argument in Theorem \ref{thm:sharp_null_gen_comp}, when $Z_{i} = 0, M_{i} = 1$, for $\tilde{Y}^{\monn}_{\bs{Z}, \bs{\delta}, \bs{b}, i}(0)$ and $\tilde{Y}_{\bs{b},i}(0)$:
\begin{align*}
    \tilde{Y}_{\bs{b},i}(0) \leq \tilde{Y}^{\monn}_{\bs{Z}, \bs{\delta}, \bs{b}, i}(0).
\end{align*}
Similarly, under the sharp null in \eqref{eq:null_fisher_sharp}, when $Z_{i} = 1, M_{i} = 0$, for $\tilde{Y}^{\monn}_{\bs{Z}, \bs{\delta}, \bs{b}, i}(0)$ and $\tilde{Y}_{\bs{b},i}(0)$:
\begin{align*}
    \tilde{Y}_{\bs{b},i}(0) \geq \tilde{Y}^{\monn}_{\bs{Z}, \bs{\delta}, \bs{b}, i}(0).
\end{align*}
Under Assumption \ref{asmp:mo_miss_M1_le_M0}, $Z_i = 0$ and $M_i = 0$ (i.e., $M_{i}(0) = 0$) implies that $M_{i}(1) = 0$, which further implies:
\begin{align*}
    \tilde{Y}_{\bs{b},i}(0) = Y_i \times 0 \times 0 + b_{00} \times 1 \times 1 + b_{01} \times 1 \times 0 + b_{10} \times 0 \times 1 = b_{00} = \tilde{Y}^{\monn}_{\bs{Z}, \bs{\delta}, \bs{b}, i}(0).
\end{align*}
Under the sharp null in \eqref{eq:null_fisher_sharp} and Assumption \ref{asmp:mo_miss_M1_le_M0}, $Z_i = 1$ and $M_i = 1$ (i.e., $M_{i}(1) = 1$) implies that $M_{i}(0) = 1$, which further implies:
\begin{align*}
    \tilde{Y}_{\bs{b},i}(0) = (Y_i - \delta_i) \times 1 \times 1 + b_{00} \times 0 \times 0 + b_{01} \times 0 \times 1 + b_{10} \times 1 \times 0 = Y_i - \delta_i = \tilde{Y}^{\monn}_{\bs{Z}, \bs{\delta}, \bs{b}, i}(0).
\end{align*}

Therefore, under the sharp null hypothesis $H_{\bs{\delta}}$ in \eqref{eq:null_fisher_sharp}:
\begin{align*}
\begin{cases*}
    \tilde{Y}^{\mon \square}_{\bs{Z}, \bs{\delta}, \bs{b}, i}(0) \leq \tilde{Y}_{\bs{b},i}(0), \quad \text{if $Z_i = 1$}, \\
    \tilde{Y}^{\mon \square}_{\bs{Z}, \bs{\delta}, \bs{b}, i}(0) \geq \tilde{Y}_{\bs{b},i}(0), \quad \text{if $Z_i = 0$},
\end{cases*}
\end{align*}
where $\square = $ p under Assumption \ref{asmp:mo_miss_M1_ge_M0} and $\square =$ n under Assumption \ref{asmp:mo_miss_M1_le_M0}.

Lemma \ref{lemma:caughey_et_al_lemma_a7} and Lemma \ref{lemma:mann_whitney_eff_incres} imply that $t_{\R, \phi}\left(\bs{Z}, \tilde{\bs{Y}}_{\bs{Z},\bs{\delta}, \bs{b}}^{\mon \square}(0) \right) \leq t_{\R, \phi}\left(\bs{Z}, \tilde{\bs{Y}}_{\bs{b}}(0) \right)$. Because the tail probability is nonincreasing,
\begin{align*}
    p_{\bs{Z}, \bs{\delta}, \bs{b}} = G_{\R, \phi}\left(t_{\R, \phi}\left(\bs{Z}, \tilde{\bs{Y}}_{\bs{b}}(0) \right)\right) \leq G_{\R, \phi}\left( t_{\R, \phi}\left(\bs{Z}, \tilde{\bs{Y}}_{\bs{Z},\bs{\delta}, \bs{b}}^{\mon \square}(0) \right) \right) = \tilde{p}_{\bs{Z}, \bs{\delta}, \bs{b}}^{\mon \square},
\end{align*}
where $p_{\bs{Z}, \bs{\delta}, \bs{b}}$ is a valid p-value for testing against the sharp null hypothesis $H_{\bs{\delta}}$ in \eqref{eq:null_fisher_sharp} by Lemma~\ref{lemma:rand_inf_valid_composite}. Now, the desired result follows immediately.
\end{proof}

\begin{proof}[Proof of Proposition \ref{prop:sharp_null_cons}]
We first show that the null hypothesis $H_{\bs{\delta}}$ in \eqref{eq:null_fisher_sharp} and Assumption \ref{asmp:constant_miss} imply that:
\begin{align*}
    \tilde{\bs{Y}}_{\bs{b}}(0) = (\bs{Y} - \bs{\delta}\circ \bs{Z})\circ \bs{M} + b_{00} (\bs{1}-\bs{M}).
\end{align*}
To see this, first consider when $M_i = 1$:
\begin{align*}
    (Y_i - \delta_i Z_i) M_i + b_{00} (1 - M_i) &= (Y_i^\sta - \delta_i Z_i) M_i(0) M_{i}(1) \\
    &= Y_i(0) M_i(0) M_{i}(1) \\
    &= \tilde{Y}_{\bs{b},i}(0),
\end{align*}
where the first equality follows from the fact that Assumption \ref{asmp:constant_miss} implies that $M_i = M_i(0) = M_i(1) = 1$, and the second equality uses the null hypothesis $H_{\bs{\delta}}$ in \eqref{eq:null_fisher_sharp}.

Then, consider the case when $M_i = 0$:
\begin{align*}
    &(Y_i - \delta_i Z_i) M_i + b_{00} (1 - M_i) = b_{00} = b_{00} (1 - M_i(0)) (1 - M_i(1)) = \tilde{Y}_{\bs{b},i}(0),
\end{align*}
where the second equality follows from the fact that Assumption \ref{asmp:constant_miss} implies that $M_i = M_i(0) = M_i(1) = 0$.

Therefore, we conclude that:
\begin{align*}
    \tilde{p}_{\bs{Z}, \bs{\delta},\bs{b}}^\con = G_{\R, \phi}( t_{\R, \phi}(\bs{Z}, (\bs{Y} - \bs{\delta}\circ \bs{Z})\circ \bs{M} + b_{00} (\bs{1}-\bs{M}) )) = G_{\R, \phi}\left( t_{\R, \phi}\left(\bs{Z}, \tilde{\bs{Y}}_{\bs{b}}(0) \right) \right) = p_{\bs{Z}, \bs{\delta}, \bs{b}}.
\end{align*}
The desired result now follows from Lemma~\ref{lemma:rand_inf_valid_composite}.
\end{proof}

\begin{proof}[Proof of Theorem \ref{thm:constant_miss}]
We first show that
\begin{align*}
    \bs{Z}_{\S} \mid \bs{Y}(1), \bs{Y}(0), \bs{M}(1), \bs{M}(0), \bs{Z}_{\S^\complement}, n_{\S1}, n_{\S0} \ \sim \ \CRE(\S, n_{\S1}, n_{\S0}).
\end{align*}
It suffices to show that for all possible $\bs{z} \in \{0, 1\}^{n}$:
\begin{align*}
\mathbb{P}\left(\bs{Z}_{\S}=\bs{z}_{\S} \mid \bs{Y}(1), \bs{Y}(0), \bs{M}(1), \bs{M}(0), \bs{Z}_{\S^\complement}=\bs{z}_{\S^\complement}, n_{\S1} = n_{11}, n_{\S0} = n_{01}\right) = \frac{1}{\binom{n_{11}+n_{01}}{n_{11}}},
\end{align*}
where $n_{11} = \sum_{i=1}^n z_i M_i = \sum_{i=1}^n z_i M_i(0)$ and $n_{01} = \sum_{i=1}^n (1-z_i) M_i = \sum_{i=1}^n (1-z_i) M_i(0)$ because $M_i(0)=M_i(1)=M_i$ under the sharp missingness assumption. Note that under the sharp missingness assumption, $\S$ is fixed and determined by $\bs{M}(0)$.

For $\mathbb{P}\left(\bs{Z}_{\S}=\bs{z}_{\S} \mid \bs{Y}(1), \bs{Y}(0), \bs{M}(1), \bs{M}(0), \bs{Z}_{\S^\complement}=\bs{z}_{\S^\complement}, n_{\S1} = n_{11}, n_{\S0} = n_{01}\right)$:
\begin{align*}
& \mathbb{P}\left(\bs{Z}_{\S}=\bs{z}_{\S} \mid \bs{Y}(1), \bs{Y}(0), \bs{M}(1), \bs{M}(0), \bs{Z}_{\S^\complement}=\bs{z}_{\S^\complement}, n_{\S1} = n_{11}, n_{\S0} = n_{01}\right) \\
=& \frac{\mathbb{P}\left(\bs{Z}_{\S}=\bs{z}_{\S}, \bs{Z}_{\S^\complement}=\bs{z}_{\S^\complement},  n_{\S1} = n_{11}, n_{\S0} = n_{01} \mid \bs{Y}(1), \bs{Y}(0), \bs{M}(1), \bs{M}(0)\right)}{\mathbb{P}\left(  \bs{Z}_{\S^\complement}=\bs{z}_{\S^\complement}, n_{\S1} = n_{11}, n_{\S0} = n_{01} \mid \bs{Y}(1), \bs{Y}(0), \bs{M}(1), \bs{M}(0)\right)} \\
=& \frac{\mathbb{P}\left(\bs{Z}=\bs{z}, n_{\S1} = n_{11}, n_{\S0} = n_{01} \mid \bs{Y}(1), \bs{Y}(0), \bs{M}(1), \bs{M}(0)\right)}{\sum_{\bs{z'}:\bs{z}_{\S^\complement}=\bs{z'}_{\S^\complement}, \sum_{i \in \S} z'_i = n_{11}, \sum_{i \in \S} (1-z'_i) = n_{01}} \mathbb{P}\left(\bs{Z}=\bs{z'}, n_{\S1} = n_{11}, n_{\S0} = n_{01} \mid \bs{Y}(1), \bs{Y}(0), \bs{M}(1), \bs{M}(0)\right)} \\
=&  \frac{\mathbb{P}\left(\bs{Z}=\bs{z}\mid \bs{Y(1)}, \bs{Y(0)}, \bs{M}(1), \bs{M}(0)\right)}{\sum_{\bs{z'}:\bs{z}_{\S^\complement}=\bs{z'}_{\S^\complement}, \sum_{i \in \S} z'_i = n_{11}, \sum_{i \in \S} (1-z'_i) = n_{01}} \mathbb{P}\left(\bs{Z}=\bs{z'} \mid \bs{Y}(1), \bs{Y}(0), \bs{M}(1), \bs{M}(0)\right)} \\
=& \frac{1}{\binom{n_{11}+n_{01}}{n_{11}}},
\end{align*}
where the first equality uses the Bayes' theorem, the second equality uses the law of total probability, the third equality holds because $n_{\S1} = n_{11}$ and $n_{\S0} = n_{01}$ follow directly and are redundant given $\bs{Z} = \bs{z'}$, $\sum_{i \in \S} z'_i = n_{11}$, and $\sum_{i \in \S} (1 - z'_i) = n_{01}$, and the fourth equality holds because the design is a CRE, and the number of terms in the summations in the denominator is equal to ${\binom{n_{11}+n_{01}}{n_{11}}}$.

The desired result follows immediately from the validity of a randomization test in a CRE (see Lemma A4 in \cite{caughey2021randomization}).
\end{proof}

\begin{proof}[Proof of Theorem \ref{thm:sharp_null_mon_two_step}] Under Assumption \ref{asmp:mo_miss_M1_ge_M0}, where suggested $b_{00} = b_{01} = \infty$, we write $\tilde{Y}_{\bs{b},i}(0)$ as $\tilde{Y}_{\infty,i}(0)$ and $\tilde{\bs{Y}}_{\bs{b}}(0)$ as $\tilde{\bs{Y}}_{\infty}(0)$.

First, note that 
if $\sum_{i=1}^n M_i(0)\le \hat{M}$, then 
\begin{align*}
    \sum_{i=1}^n Z_iM_i\{1-M_i(0)\}
    & = n_{11} - \sum_{i:Z_i=1, M_i = 1} M_i(0) \\
    &= n_{11} - \sum_{i:Z_i=1} M_i(0) \\
    &= n_{11} + n_{01} - \sum_{i=1}^n M_i(0) \\
    & \ge n_{11} + n_{01} - \hat{M} \\
    & \coloneqq \underline{m},
\end{align*}
where the second equality uses Assumption \ref{asmp:mo_miss_M1_ge_M0} and the third equality uses the fact that $n_{01} = \sum_{i = 1}^{n}(1 - Z_{i}) M_{i}(0)$. 

We now minimize the test statistic of the form in \eqref{eq:test_stat_mann_whitney} under $H_{\bs{\delta}}$ and Assumption \ref{asmp:mo_miss_M1_ge_M0}. Consider a treated unit $i$ and a control unit $j$, we have:
\begin{align*}
    \psi_{i,j}( \tilde{Y}_{\infty,i}(0), \tilde{Y}_{\infty,j}(0) ) &= \psi_{i,j}( Y_i(0) M_i(0) +  \infty \{1-M_i(0)\}, Y_j M_j +  \infty (1-M_j) ) \\
    & = \begin{cases*}
        A_{ij} & if $M_i = 0$ \\
        A_{ij} & if $M_i = 1$, $M_i(0) = 0$ \\
        B_{ij} & if $M_i = 1$, $M_i(0) = 1$,
    \end{cases*}
\end{align*}
where the second equality uses Assumption \ref{asmp:mo_miss_M1_ge_M0}, $H_{\bs{\delta}}$, and the facts that $A_{ij} \coloneqq  \psi_{i,j}( \infty, Y_j M_j +  \infty (1-M_j) )$ and $B_{ij} \coloneqq \psi_{i,j}( Y_i - \delta_i, Y_j M_j +  \infty (1-M_j) )$.

After some algebraic manipulation, we can verify that:
\begin{align*}
    t_{\R, \phi}\left( \bs{Z}, \tilde{\bs{Y}}_{\infty}(0) \right)  & = \sum_{i:Z_i=1}\phi\left( \sum_{j:Z_j=0} \psi_{i, j}( \tilde{Y}_{\infty,i}(0), \tilde{Y}_{\infty,j}(0) ) \right) 
    \\
    & = \sum_{i:Z_i=1,M_i=0}\phi(A_i) + 
    \sum_{i:Z_i=1, M_i = 1} \phi(B_i) + 
    \sum_{i:Z_i=1, M_i = 1} \{1-M_i(0)\} C_i, 
\end{align*}
where $A_{i}$, $B_{i}$, and $C_{i}$ are defined in Algorithm \ref{alg:mp_two_step_sharp}.

Then, under the constraint that $\sum_{i=1}^n Z_i M_i \{1-M_i(0)\} \ge \underline{m}$, we have:
\begin{align*}
    t_{\R, \phi}\left( \bs{Z}, \tilde{\bs{Y}}_{\infty}(0) \right)  
    & = \sum_{i:Z_i=1,M_i=0}\phi(A_i) + 
    \sum_{i:Z_i=1, M_i = 1} \phi(B_i) + 
    \sum_{i:Z_i=1, M_i = 1} \{1-M_i(0)\} C_i
    \\
    & \ge 
    \sum_{i:Z_i=1,M_i=0}\phi(A_i) + 
    \sum_{i:Z_i=1, M_i = 1} \phi(B_i) + 
    \sum_{i=1}^{\underline{m}} C_{l_i} \\
    &= 
    \sum_{i:Z_i=1,M_i=0}\phi(A_i)  
    + \sum_{i:Z_i=1,M_i=1}\phi(B_i) 
    + \sum_{i:Z_i=1,M_i=1, i \in \{ l_1, \ldots, l_{\underline{m}}\}}(\phi(A_i)-\phi(B_i)) \\
    &= 
    \sum_{i:Z_i=1,M_i=0}\phi(A_i) + 
    \sum_{i:Z_i=1,M_i=1, i \in \{ l_1, \ldots, l_{\underline{m}}\}}\phi(A_i) +
    \sum_{i:Z_i=1,M_i=1, i \notin \{ l_1, \ldots, l_{\underline{m}}\}}\phi(B_i) \\
    &= \sum_{i: Z_i = 1} \phi\left( \sum_{j: Z_j = 0} \psi_{i, j}\left( \tilde{Y}_{\bs{Z}, \bs{\delta}, i}^{\monp, \underline{m}}(0), Y_j M_j + \infty (1 - M_j) \right) \right) \\
    &= t_{\R, \phi}\left( \bs{Z}, \tilde{\bs{Y}}_{\bs{Z}, \bs{\delta}}^{\monp, \underline{m}}(0) \right),
\end{align*}
where the inequality follows from the fact that $C_{l_1}, \ldots, C_{l_{\underline{m}}}$ are the $\underline{m}$ smallest values of $C_i$ among observed treated units.

We now show that $\tilde{p}_{\bs{Z}, \bs{\delta}}^{\monp, \text{two-step}}$ is a valid p-value. 
Note that $\Pr\left( \tilde{p}_{\bs{Z}, \bs{\delta}}^{\monp,\text{two-step}} \le \alpha \right) = 0 \le \alpha$ holds trivially for $\alpha < \beta$. For $\alpha \ge \beta$, we have:
\begin{align*}
     & \Pr\left( \tilde{p}_{\bs{Z}, \bs{\delta}}^{\monp, \text{two-step}} \le \alpha \right) 
    \\
    =& 
    \Pr\left( \tilde{p}_{\bs{Z}, \bs{\delta}}^{\monp, \text{two-step}} \le \alpha, \sum_{i=1}^n M_i(0) \le \hat{M} \right) + \Pr\left( \tilde{p}_{\bs{Z}, \bs{\delta}}^{\monp, \text{two-step}} \le \alpha, \sum_{i=1}^n M_i(0) > \hat{M} \right) \\
    \le & \Pr\left( G_{\R, \phi}\left( t_{\R, \phi}\left(\bs{Z}, \tilde{\bs{Y}}^{\monp, \underline{m}}_{\bs{Z}, \bs{\delta}}(0) \right) \right) \le \alpha - \beta, \sum_{i=1}^n M_i(0) \le \hat{M} \right)  + 
    \Pr\left( \sum_{i=1}^n M_i(0) > \hat{M} \right) \\
    \le & 
    \Pr\left( G_{\R, \phi}\left( t_{\R, \phi}\left(\bs{Z}, \tilde{\bs{Y}}_{\infty}(0) \right) \right) \le \alpha - \beta \right) + 
    \Pr\left( \sum_{i=1}^n M_i(0) > \hat{M} \right) \\
    \le & \alpha - \beta + \beta = \alpha, 
\end{align*}
where the first inequality holds by the construction of Algorithm \ref{alg:mp_two_step_sharp}, the second inequality follows from the fact that $G_{\R, \phi}\left( t_{\R, \phi}\left(\bs{Z}, \tilde{\bs{Y}}_{\infty}(0) \right) \right) \leq G_{\R, \phi}\left( t_{\R, \phi}\left(\bs{Z}, \tilde{\bs{Y}}^{\monp, \underline{m}}_{\bs{Z}, \bs{\delta},\infty}(0) \right) \right)$ when $\sum_{i=1}^n M_i(0)\le \hat{M}$.
\end{proof}

\begin{proof}[Proof of Theorem \ref{thm:sharp_null_mon_neg_two_step}] Under Assumption \ref{asmp:mo_miss_M1_le_M0}, where suggested $b_{00} = b_{10} = - \infty$, we write $\tilde{Y}_{\bs{b},i}(0)$ as $\tilde{Y}_{- \infty,i}(0)$ and $\tilde{\bs{Y}}_{\bs{b}}(0)$ as $\tilde{\bs{Y}}_{- \infty}(0)$.

First, note that if $\sum_{i=1}^n M_i(1) \le \hat{M}$, then 
\begin{align*}
    \sum_{i=1}^n (1 - Z_i) M_i \{1-M_i(1)\}
    & = n_{01} - \sum_{i:Z_i=0, M_i = 1} M_i(1) = n_{01} + n_{11} - \sum_{i=1}^n M_i(1) \\
    & \ge n_{01} + n_{11} - \hat{M} \coloneqq \underline{m},
\end{align*}
where the second equality uses Assumption \ref{asmp:mo_miss_M1_le_M0} and the fact that $n_{01} = \sum_{i = 1}^{n}(1 - Z_{i}) M_{i}(0)$. 

We now minimize the test statistic of the form in \eqref{eq:test_stat_mann_whitney_negative} under $H_{\bs{\delta}}$ and Assumption \ref{asmp:mo_miss_M1_le_M0}. Consider a control unit $i$ and a treated unit $j$, we have:
\begin{align*}
    \psi_{i,j}( \tilde{Y}_{-\infty,i}(0), \tilde{Y}_{-\infty,j}(0) ) &= \psi_{i,j}( Y_i(0) M_i(1) - \infty \{1-M_i(1)\}, (Y_j - \delta_j ) M_j -  \infty (1-M_j) ) \\
    & = \begin{cases*}
        B_{ij} & if $M_i = 0$ \\
        B_{ij} & if $M_i = 1$, $M_i(1) = 0$ \\
        A_{ij} & if $M_i = 1$, $M_i(1) = 1$,
    \end{cases*}
\end{align*}
where the second equality uses Assumption \ref{asmp:mo_miss_M1_le_M0} and the facts that $A_{ij} \coloneqq \psi_{i,j}( Y_i, (Y_j-\delta_j) M_j -\infty (1-M_j) )$ and $B_{ij} \coloneqq \psi_{i,j}( -\infty, (Y_j-\delta_j) M_j -\infty (1-M_j) )$.

After some algebraic manipulation, we can verify that:
\begin{align*}
    t_{\R, \phi}\left( \bs{Z}, \tilde{\bs{Y}}_{-\infty}(0) \right)  & = - \sum_{i:Z_i = 0}\phi\left( \sum_{j:Z_j = 1} \psi_{i, j}( \tilde{Y}_{-\infty,i}(0), \tilde{Y}_{-\infty,j}(0) ) \right) 
    \\
    & = -\sum_{i:Z_i=0,M_i=0}\phi(B_i) - 
    \sum_{i:Z_i=0, M_i = 1} \phi(A_i) + 
    \sum_{i:Z_i=0, M_i = 1} \{1-M_i(1)\} C_i, 
\end{align*}
where $A_{i}$, $B_{i}$, and $C_{i}$ are defined in Algorithm \ref{alg:mn_two_step_sharp}.

Then, under the constraint that $\sum_{i=1}^n (1 - Z_i) M_i \{1-M_i(1)\} \ge \underline{m}$:
\begin{align*}
    t_{\R, \phi}\left( \bs{Z}, \tilde{\bs{Y}}_{- \infty}(0) \right)  
    & = -\sum_{i:Z_i=0,M_i=0}\phi(B_i) - 
    \sum_{i:Z_i=0, M_i = 1} \phi(A_i) + 
    \sum_{i:Z_i=0, M_i = 1} \{1-M_i(1)\} C_i
    \\
    & \ge 
    -\sum_{i:Z_i=0,M_i=0}\phi(B_i) - 
    \sum_{i:Z_i=0, M_i = 1} \phi(A_i) + 
    \sum_{i=1}^{\underline{m}} C_{l_i} \\
    &= 
    - \sum_{i:Z_i=0,M_i=0}\phi(B_i)  
    - \sum_{i:Z_i=0,M_i=1}\phi(A_i) 
    + \sum_{i:Z_i=0,M_i=1, i \in \{ l_1, \ldots, l_{\underline{m}}\}} \left( \phi(A_i) - \phi(B_i) \right) \\
    &= 
    - \sum_{i:Z_i=0,M_i=0}\phi(B_i)  
    - \sum_{i:Z_i=0,M_i=1, i \notin \{ l_1, \ldots, l_{\underline{m}}\}}\phi(A_i) 
    - \sum_{i:Z_i=0,M_i=1, i \in \{ l_1, \ldots, l_{\underline{m}}\}}\phi(B_i) \\
    &= - \sum_{i: Z_i = 0} \phi\left( \sum_{j: Z_j = 1} \psi_{i, j}\left( \tilde{Y}_{\bs{Z}, \bs{\delta}, i}^{\monn, \underline{m}}(0), (Y_j - \delta_j) M_j - \infty (1 - M_j) \right) \right) \\
    &= t_{\R, \phi}\left( \bs{Z}, \tilde{\bs{Y}}_{\bs{Z}, \bs{\delta}}^{\monn, \underline{m}}(0) \right).
\end{align*}
where the inequality follows from the fact that $C_{l_1}, \ldots, C_{l_{\underline{m}}}$ are the $\underline{m}$ smallest values of $C_i$ among observed control units.

We now show that $\tilde{p}_{\bs{Z}, \bs{\delta}}^{\monn, \text{two-step}}$ is a valid p-value. Note that $\Pr\left( \tilde{p}_{\bs{Z}, \bs{\delta}}^{\monn,\text{two-step}} \le \alpha \right) = 0 \le \alpha$ holds trivially for $\alpha < \beta$. For $\alpha \ge \beta$, we have:
\begin{align*}
    & \Pr\left( \tilde{p}_{\bs{Z}, \bs{\delta}}^{\monn, \text{two-step}} \le \alpha \right)
    \\
    = & 
    \Pr\left( \tilde{p}_{\bs{Z}, \bs{\delta}}^{\monn, \text{two-step}} \le \alpha, \sum_{i=1}^n M_i(1) \le \hat{M} \right) + \Pr\left( \tilde{p}_{\bs{Z}, \bs{\delta}}^{\monn, \text{two-step}} \le \alpha, \sum_{i=1}^n M_i(1) > \hat{M} \right) \\
    \le & \Pr\left( G_{\R, \phi}\left( t_{\R, \phi}\left(\bs{Z}, \tilde{\bs{Y}}^{\monn, \underline{m}}_{\bs{Z}, \bs{\delta}}(0) \right) \right) \le \alpha - \beta, \sum_{i=1}^n M_i(1) \le \hat{M} \right)  + 
    \Pr\left( \sum_{i=1}^n M_i(1) > \hat{M} \right) \\
    \le &  
    \Pr\left( G_{\R, \phi}\left( t_{\R, \phi}\left(\bs{Z}, \tilde{\bs{Y}}_{- \infty}(0) \right) \right) \le \alpha - \beta \right) + 
    \Pr\left( \sum_{i=1}^n M_i(1) > \hat{M} \right) \\
    \le & \alpha - \beta + \beta = \alpha, 
\end{align*}
where the first inequality holds by the construction of Algorithm \ref{alg:mn_two_step_sharp}, the second inequality follows from the fact that $G_{\R, \phi}\left( t_{\R, \phi}\left(\bs{Z}, \tilde{\bs{Y}}_{-\infty}(0) \right) \right) \leq G_{\R, \phi}\left( t_{\R, \phi}\left(\bs{Z}, \tilde{\bs{Y}}^{\monn, \underline{m}}_{\bs{Z}, \bs{\delta}}(0) \right) \right)$ when when $\sum_{i=1}^n M_i(1) \le \hat{M}$.
\end{proof}

\begin{proof}[Proof of Theorem \ref{thm:sharp_null_gen_two_step}] Under Assumption \ref{asmp:general}, consider $b_{01} = \infty$, $b_{10} = -\infty$, and $b_{00} = \infty$.

First, we show that, for any prespecified $\beta \in [0,1)$ and $(\overline{m}, \underline{m}, \overline{d})$ constructed in the main paper, we have:
\begin{align*}
    \Pr\left(\underline{m} \leq m_{11}^\tr + m_{10}^\tr \leq \overline{m}, {m_{11}^\co}/{n_0} - {m_{11}^\tr}/{n_1} \leq \overline{d}\right) \geq 1 - \beta.
\end{align*}
Let $m_{11}^\star \coloneqq \sum_{i=1}^n \I \{M_i(1) = M_i(0)=1\} = m_{11}^\tr +  m_{11}^\co$. 
We have 
\begin{align*}
    & \Pr \left(\underline{m} \leq m_{11}^\tr+m_{10}^\tr \leq \bar{m}, \frac{m_{11}^\co}{n_0}-\frac{m_{11}^\tr}{n_1} \leq \bar{d}\right) \\
    \geq &  \Pr\left(\underline{m} \leq m_{11}^\tr+m_{10}^\tr \leq \overline{m}\right) + \Pr\left(\frac{m_{11}^\co}{n_0}-\frac{m_{11}^\tr}{n_1} \leq \bar{d}\right) - 1 \\
    = & \Pr\left(\underline{m}+n_{01} \leq m_{11}^\tr+m_{10}^\tr+n_{01} \leq \overline{m}+n_{01}\right) + \mathbb{P}\left(\frac{m_{11}^\co}{n_0}-\frac{m_{11}^\star-m_{11}^\co}{n_1} \leq \overline{d}\right) - 1 \\
    = & \Pr\left(\hat{M}_1 \leq \sum_{i=1}^n M_{i}(0) \leq \hat{M}_2\right) + \Pr\left(m_{11}^\co \leq \frac{n_1 n_0}{n}\left(\overline{d}+\frac{m_{11}^\star}{n_1}\right)\right) - 1, 
\end{align*}
where the inequality holds by the fact $\Pr(A\cap B)\ge \Pr(A)+\Pr(B)-1$ for any event $A$ and $B$. From \cite{Wang:2015}, we know that 
\[
\Pr\left(\hat{M}_1 \leq \sum_{i=1}^n M_{i}(0) \leq \hat{M}_2\right) \ge 1 - \beta_1. 
\]
By definition, we can verify that $0\le m_{11}^\star \le n_{11} + n_{01}$. 
Recall the construction of $\overline{d}$. We then have
$
\overline{d} \geq n/(n_1n_0) \cdot q_{\text{HG}}(1-\beta_2; n, m_{11}^\star, n_0) - m_{11}^\star/n_1,
$
which further implies that 
$\frac{n_1 n_0}{n}\left(\overline{d}+\frac{m_{11}^\star}{n_1}\right) \ge q_{\text{HG}}(1-\beta_2; n, m_{11}^\star, n_0)$. 
Thus, 
\begin{align*}
    \Pr\left(m_{11}^\co \leq \frac{n_1 n_0}{n}\left(\overline{d}+\frac{m_{11}^\star}{n_1}\right)\right)
    \ge 
    \Pr\left(m_{11}^\co \leq q_{\text{HG}}(1-\beta_2; n, m_{11}^\star, n_0) \right) \ge 1 - \beta_2,
\end{align*}
where the last equality uses the fact that $m_{11}^\co$ follows the Hypergeometric distribution with parameters $(n, m_{11}^\star, n_0)$ and the properties of quantile functions. These then imply that
\begin{align*}
    \Pr \left(\underline{m} \leq m_{11}^\tr+m_{10}^\tr \leq \bar{m}, \frac{m_{11}^\co}{n_0}-\frac{m_{11}^\tr}{n_1} \leq \bar{d}\right)
    \ge 1 - \beta_1 + 1 - \beta_2 - 1 = 1 - \beta. 
\end{align*}

Second, we give a lower bound of the test statistic of form \eqref{eq:test_stat_mann_whitney} under $H_{\bs{\delta}}$, Assumption \ref{asmp:general} and  given values of $(m_{11}^\tr, m_{10}^\tr, m_{11}^\co)$. 
Consider a treated unit $i$ and a control unit $j$ under the sharp null $H_{\bs{\delta}}$, by definition, we have:
\begin{align*}
    \tilde{Y}_{\bs{b},i}(0) = 
    \begin{cases}
        Y_i - \delta_i, & \text{if } M_i(0) = 1, M_i = 1, \\
        \infty, & \text{if } M_i(0) = 0, M_i = 1, \\
        -\infty, & \text{if } M_i(0) = 1, M_i = 0, \\
        \infty, & \text{if } M_i(0) = 0, M_i = 0, 
    \end{cases}
    \qquad 
    \tilde{Y}_{\bs{b},j}(0) = \begin{cases}
        Y_j, & \text{if } M_j = 1, M_j(1) = 1, \\
        -\infty, & \text{if } M_j = 1, M_j(1) = 0, \\
        \infty, & \text{if } M_j = 0. \\
    \end{cases} 
\end{align*}

We consider the following four cases, separately. 
\begin{itemize}
    \item[(a)] Consider a treated unit $i$ with $M_i(0) = 1$ and $M_i = 1$. We have 
    \begin{align*}
        & \quad \ \sum_{j: Z_j = 0} \psi_{i, j}\left(\tilde{Y}_{\bs{b},i}(0), \tilde{Y}_{\bs{b},j}(0)\right)
        \\
        & = 
        \sum_{j: Z_j = 0, M_j = 1, M_j(1) = 1} \psi_{i, j}\left( Y_i - \delta_i, Y_j\right)
        + 
        \sum_{j: Z_j = 0, M_j = 1, M_j(1) = 0} \psi_{i, j}\left( Y_i - \delta_i, -\infty\right)
        \\
        & \quad +
        \sum_{j: Z_j = 0, M_j = 0} \psi_{i, j}\left( Y_i - \delta_i, \infty\right)\\
        & = \sum_{j: Z_j = 0, M_j = 1, M_j(1) = 1} \psi_{i, j}\left( Y_i - \delta_i, Y_j\right) + \big|\{j: Z_j = 0, M_j = 1, M_j(1) = 0\}\big| + 0
        \\
        & = \sum_{j: Z_j = 0, M_j = 1, M_j(1) = 1} \psi_{i, j}\left( Y_i - \delta_i, Y_j\right) +
        n_{01} - m_{11}^\co, 
    \end{align*}
    where the second last equality follows from the fact that $-\infty < Y_i - \delta_i < \infty$, and the last equality follows from the definition of $n_{01}$ and $m_{11}^\co$. 
    By the construction of $(l_1, l_2, \ldots, l_{n_{01}})$ in Step (ii) of the algorithm, we can verify that 
    \begin{align*}
        \psi_{i, l_1 }\left( Y_i - \delta_i, Y_{l_1}\right) \ge \psi_{i, l_2 }\left( Y_i - \delta_i, Y_{l_2}\right) 
        \ge \ldots \ge \psi_{i, l_{n_{01}} }\left( Y_i - \delta_i, Y_{l_{n_{01}}}\right), 
    \end{align*}
    and  
    \begin{align*}
        \sum_{j: Z_j = 0} \psi_{i, j}\left(\tilde{Y}_{\bs{b},i}(0), \tilde{Y}_{\bs{b},j}(0)\right)
        & = \sum_{j: Z_j = 0, M_j = 1, M_j(1) = 1} \psi_{i, j}\left( Y_i - \delta_i, Y_j\right) +
        n_{01} - m_{11}^\co
        \\
        & \ge \sum_{j\in \mathcal{F}_{m_{11}^\co}} \psi_{i, j}\left( Y_i - \delta_i, Y_j\right) +
        n_{01} - m_{11}^\co = B_{m_{11}^\co i},
    \end{align*}
    where $\mathcal{F}_{m_{11}^\co}$ and $B_{m_{11}^\co i}$ are defined as in the algorithm.

    \item[(b)] Consider a treated unit $i$ with $M_i(0) = 0$ and $M_i = 1$. We have 
    \begin{align*}
        & \quad \ \sum_{j: Z_j = 0} \psi_{i, j}\left(\tilde{Y}_{\bs{b},i}(0), \tilde{Y}_{\bs{b},j}(0)\right)
        \\
        & = 
        \sum_{j: Z_j = 0, M_j = 1, M_j(1) = 1} \psi_{i, j}\left( \infty, Y_j\right)
        + 
        \sum_{j: Z_j = 0, M_j = 1, M_j(1) = 0} \psi_{i, j}\left( \infty, -\infty\right)
        +
        \sum_{j: Z_j = 0, M_j = 0} \psi_{i, j}\left( \infty, \infty\right)\\
        & = n_{01} + \sum_{j: Z_j = 0, M_j = 0} \I(i > j) = A_i, 
    \end{align*}
    where the second last equality uses the definition of $n_{01}$ and the fact that $\infty > Y_j$ and $\infty > -\infty$, 
    and the last equality uses the definition of $A_i$. 

    \item[(c)] Consider a treated unit $i$ with $M_i(0) = 0$ and $M_i = 0$. Similar to the case in (b), we have 
    \begin{align*}
        \sum_{j: Z_j = 0} \psi_{i, j}\left(\tilde{Y}_{\bs{b},i}(0), \tilde{Y}_{\bs{b},j}(0)\right)
        & = n_{01} + \sum_{j: Z_j = 0, M_j = 0} \I(i > j) = A_i. 
    \end{align*}

    \item[(d)] Consider a treated unit $i$ with $M_i(0) = 1$ and $M_i = 0$. We have 
    \begin{align*}
        & \quad \ \sum_{j: Z_j = 0} \psi_{i, j}\left(\tilde{Y}_{\bs{b},i}(0), \tilde{Y}_{\bs{b},j}(0)\right)
        \\
        & = 
        \sum_{j: Z_j = 0, M_j = 1, M_j(1) = 1} \psi_{i, j}\left( -\infty, Y_j\right)
        + 
        \sum_{j: Z_j = 0, M_j = 1, M_j(1) = 0} \psi_{i, j}\left( -\infty, -\infty\right)
        +
        \sum_{j: Z_j = 0, M_j = 0} \psi_{i, j}\left( -\infty, \infty\right)\\
        & = 0 + \sum_{j: Z_j = 0, M_j = 1, M_j(1) = 0} \I(i>j) + 0
        = \sum_{j: Z_j = 0, M_j = 1, M_j(1) = 0} \I(i>j),
    \end{align*}
    where the second last equality holds because $-\infty < Y_j$ and $-\infty < \infty$. 
    By the definition of $m_{11}^\co$, we further have 
    \begin{align*}
        \sum_{j: Z_j = 0} \psi_{i, j}\left(\tilde{Y}_{\bs{b},i}(0), \tilde{Y}_{\bs{b},j}(0)\right)
        & = 
        \sum_{j: Z_j = 0, M_j = 1} \I(i>j) - \sum_{j: Z_j = 0, M_j = 1, M_j(1) = 1} \I(i>j)
        \\
        & \ge \max\left\{ 0, \sum_{j: Z_j = 0, M_j = 1} \I(i>j) - m_{11}^\co \right\} = B_{m_{11}^\co i}, 
    \end{align*}
    where the last equality follows from the definition of $B_{m_{11}^\co i}$ in the algorithm. 
\end{itemize}
From the above four cases, we have 
\begin{align}\label{eq:bound_given_m}
     & \quad \ t_{\R, \phi}\left( \bs{Z}, \tilde{\bs{Y}}_{\bs b}(0) \right)  
     \nonumber
     \\
     & = \sum_{i: Z_{i} = 1} \phi\left( \sum_{j: Z_{j} = 0} \psi_{i,j}( \tilde{Y}_{\bs{b},i}(0), \tilde{Y}_{\bs{b},j}(0) ) \right) \nonumber
     \\
     & \ge \sum_{i: Z_{i} = 1, M_i = 1, M_i(0) = 1} \phi( 
     B_{m_{11}^\co i}
     )
     +
     \sum_{i: Z_{i} = 1, M_i = 1, M_i(0) = 0} \phi( 
     A_i
     )
     +
     \sum_{i: Z_{i} = 1, M_i = 0, M_i(0) = 0} \phi( 
     A_i
     )
     \nonumber\\
     & \quad \ +
     \sum_{i: Z_{i} = 1, M_i = 0, M_i(0) = 1} \phi( 
     B_{m_{11}^\co i}
     )\nonumber\\
     & = 
     \sum_{i: Z_{i} = 1, M_i = 1} \phi( 
     B_{m_{11}^\co i}
     )
     + 
     \sum_{i: Z_{i} = 1, M_i = 1, M_i(0) = 0} C_{m_{11}^\co i} 
     + 
     \sum_{i: Z_{i} = 1, M_i = 0} \phi( 
     B_{m_{11}^\co i}
     )
     + 
     \sum_{i: Z_{i} = 1, M_i = 0, M_i(0) = 0} C_{m_{11}^\co i} \nonumber\\
     & = \sum_{i: Z_{i} = 1} \phi( 
     B_{m_{11}^\co i}
     )
     + 
     \sum_{i: Z_{i} = 1, M_i = 1, M_i(0) = 0} C_{m_{11}^\co i} 
     + 
     \sum_{i: Z_{i} = 1, M_i = 0, M_i(0) = 0} C_{m_{11}^\co i} \nonumber\\
     & \ge 
     \sum_{i: Z_{i} = 1} \phi( 
     B_{m_{11}^\co i}
     )
     + 
     \sum_{j=1}^{n_{11} - m_{11}^\tr} C_{m_{11}^\co 1(j)} 
     + 
     \sum_{j=1}^{n_{10} - m_{10}^\tr} C_{m_{11}^\co 0(j)}. 
\end{align}

Third, we prove that the lower bound in \eqref{eq:bound_given_m}, 
\begin{align}\label{eq:bound_given_m_2}
   \sum_{i: Z_{i} = 1} \phi( 
     B_{m_{11}^\co i}
     )
     + 
     \sum_{j=1}^{n_{11} - m_{11}^\tr} C_{m_{11}^\co 1(j)} 
     + 
     \sum_{j=1}^{n_{10} - m_{10}^\tr} C_{m_{11}^\co 0(j)} 
\end{align}
is nonincreasing in $m_{11}^\tr$, $m_{10}^\tr$ and and $m_{11}^\co$. 
By construction, we can verify that, for any treated unit $i$, $C_{Ji}$ is nonnegative. Therefore, \eqref{eq:bound_given_m_2} is nonincreasing in  $m_{11}^\tr$ and $m_{10}^\tr$. 
By the same logic as the last inequality in \eqref{eq:bound_given_m}, for any given $m_{11}^\tr$ and $m_{10}^\tr$, we know that the quantity in \eqref{eq:bound_given_m_2} is the minimum of 
\begin{align}\label{eq:bound_given_m_3}
    \sum_{i: Z_{i} = 1} \phi( 
     B_{m_{11}^\co i}
     )
     + 
     \sum_{i\in \mathcal{T}_{11}} C_{m_{11}^\co i} 
     + 
     \sum_{i\in \mathcal{T}_{10}} C_{m_{11}^\co i}
     = 
     \sum_{i \in \{i:Z_i=1\}\cap \mathcal{T}_{11}^\textrm{c} \cap \mathcal{T}_{10}^\textrm{c}} \phi( 
     B_{m_{11}^\co i}
     )
     + 
     \sum_{i\in \mathcal{T}_{11}} \phi(A_i)
     + 
     \sum_{i\in \mathcal{T}_{10}} \phi(A_i)
\end{align}
over all possible $\mathcal{T}_{11}$ and $\mathcal{T}_{10}$ such that 
\begin{align*}
    \mathcal{T}_{11} \subset \{i: Z_i = 1, M_i = 1\}, \ |\mathcal{T}_{11}| = n_{11} - m_{11}^\tr, \ 
    \mathcal{T}_{10} \subset \{i: Z_i = 1, M_i = 0\}, \  
    |\mathcal{T}_{10}| = n_{10} - m_{10}^\tr. 
\end{align*}
By definition, 
we can verify that, for any treated unit $i$,  $B_{m_{11}^\co i}$ is nonincreasing in $m_{11}^\co$. 
Thus, the quantity in \eqref{eq:bound_given_m_2}, as a minimum of \eqref{eq:bound_given_m_3}, must be nonincreasing in $m_{11}^\co$ as well.

Fourth,
we give a lower bound of the test statistic of form \eqref{eq:test_stat_mann_whitney} under $H_{\bs{\delta}}$, Assumption \ref{asmp:general} and  the constraints that $\underline{m} \leq m_{11}^\tr + m_{10}^\tr \leq \overline{m}$ and ${m_{11}^\co}/{n_0} - {m_{11}^\tr}/{n_1} \leq \overline{d}$. 
Under these constraints, we can verify that 
\begin{align}\label{eq:constr_imp}
    m_{11}^\co \le \min\{ \lfloor n_0 ( \overline{d} + m_{11}^\tr/n_1 ) \rfloor, n_{01} \}, 
    \quad
    m_{10}^\tr \le \min\{ \overline{m} - m_{11}^\tr, n_{10} \}, 
    \quad 
    \underline{K} \le m_{11}^\tr \le \overline{K}. 
\end{align}
From \eqref{eq:bound_given_m}, \eqref{eq:constr_imp} and the above discussion, when $m_{11}^\tr = K$, we have 
\begin{align*}
     t_{\R, \phi}\left( \bs{Z}, \tilde{\bs{Y}}_{\bs b}(0) \right)  
     & \ge 
     \sum_{i: Z_{i} = 1} \phi( 
     B_{m_{11}^\co i}
     )
     + 
     \sum_{j=1}^{n_{11} - m_{11}^\tr} C_{m_{11}^\co 1(j)} 
     + 
     \sum_{j=1}^{n_{10} - m_{10}^\tr} C_{m_{11}^\co 0(j)}\\
     & \ge 
     \sum_{i: Z_{i} = 1} \phi( 
     B_{J i}
     )
     + 
     \sum_{j=1}^{n_{11} - K} C_{J 1(j)} 
     + 
     \sum_{j=1}^{n_{10} - L} C_{J 0(j)}
     = T_K, 
\end{align*}
where $J = \min\{ \lfloor n_0 ( \overline{d} + K/n_1 ) \rfloor, n_{01} \}, L = \min\{ \overline{m} - K, n_{10} \}$, and $T_K$ is defined as in the algorithm. From the range of $m_{11}^\tr$ in \eqref{eq:constr_imp}, we then have 
\begin{align*}
    t_{\R, \phi}\left( \bs{Z}, \tilde{\bs{Y}}_{\bs b}(0) \right) \ge \inf_{K \in [\underline{K}, \overline{K}]}T_K.
\end{align*}

Finally, we show that $\tilde{p}_{\bs{Z}, \bs{\delta}}^{\gen, \text{two-step}}$ is a valid p-value. 
For $\alpha < \beta$, it is obvious that
$\Pr( \tilde{p}_{\bs{Z}, \bs{\delta}}^{\gen, \text{two-step}} \le \alpha ) = 0 \le \alpha$.
For $\alpha \ge \beta$, we have 
\begin{align*}
    & \Pr\left( \tilde{p}_{\bs{Z}, \bs{\delta}}^{\gen, \text{two-step}} \le \alpha \right)
    \\
    = & 
    \Pr\left( \tilde{p}_{\bs{Z}, \bs{\delta}}^{\gen, \text{two-step}} \le \alpha, \underline{m} \leq m_{11}^\tr+m_{10}^\tr \leq \overline{m}, \frac{m_{11}^\co}{n_0}-\frac{m_{11}^\tr}{n_1} \leq \overline{d} \right) \\ 
    & \quad + \Pr\left( \tilde{p}_{\bs{Z}, \bs{\delta}}^{\gen, \text{two-step}} \le \alpha, \left\{\underline{m} \leq m_{11}^\tr+m_{10}^\tr \leq \overline{m}, \frac{m_{11}^\co}{n_0}-\frac{m_{11}^\tr}{n_1} \leq \overline{d} \right\}^{\complement} \right) \\
    \le & \Pr\left( G_{\R, \phi}\left( \inf_{K \in [\underline{K}, \overline{K}]}T_K \right) \le \alpha - \beta, \underline{m} \leq m_{11}^\tr+m_{10}^\tr \leq \overline{m}, \frac{m_{11}^\co}{n_0}-\frac{m_{11}^\tr}{n_1} \leq \overline{d} \right) \\
    & \quad \ + 
    \Pr\left( \left\{ \underline{m} \leq m_{11}^\tr+m_{10}^\tr \leq \overline{m}, \frac{m_{11}^\co}{n_0}-\frac{m_{11}^\tr}{n_1} \leq \overline{d} \right\}^{C} \right) \\
    \le & \Pr\left( G_{\R, \phi}\left( t_{\R, \phi}\left( \bs{Z}, \tilde{\bs{Y}}_{\infty}(0) \right)  \right) \le \alpha - \beta \right) + 
    \Pr\left( \left\{ \underline{m} \leq m_{11}^\tr+m_{10}^\tr \leq \overline{m}, \frac{m_{11}^\co}{n_0}-\frac{m_{11}^\tr}{n_1} \leq \overline{d} \right\}^{C} \right) \\
    \le & \alpha - \beta + \beta = \alpha, 
\end{align*}
where the second inequality holds from the previous discussion.
\end{proof}

\begin{proof}[Proof of Theorem \ref{thm:mar_miss}]
For all possible $\bs{z}, \bs{m} \in \{0, 1\}^{n}$, consider $\mathbb{P}\left(\bs{M}=\bs{m} \mid \bs{Z}=\bs{z}, \bs{Y}(1), \bs{Y}(0)\right)$:
\begin{align*}
&\mathbb{P}\left(\bs{M}=\bs{m} \mid \bs{Z}=\bs{z}, \bs{Y}(1), \bs{Y}(0)\right) \\
=& \Pi_{i = 1}^{n} \mathbb{P}\left(M_i(z_i) = m_i \mid \bs{Z}=\bs{z}, \bs{Y}(1), \bs{Y}(0)\right) \\
=& \Pi_{i = 1}^{n} (p_{11}+p_{10})^{z_i m_i}
(p_{01}+p_{00})^{z_i (1 - m_i)}
(p_{11}+p_{01})^{(1 - z_i) m_i}
(p_{10}+p_{00})^{(1 - z_i) (1 - m_i)} \\
=& (p_{11}+p_{10})^{n_{11}}
(p_{01}+p_{00})^{n_{10}}
(p_{11}+p_{01})^{n_{01}}
(p_{10}+p_{00})^{n_{00}},
\end{align*}
where the first equality uses the SUTVA assumption on potential missingness and the fact that $\{ M_{i}(0), M_{i}(1) \}_{i = 1}^{n}$ is i.i.d. under Assumption \ref{asmp:mar}, the second equality uses the fact that Assumption \ref{asmp:mar} implies that:
\begin{align*}
    \mathbb{P}(M_{i}(z_{i}) = m_i \mid \bs{Y}(1), \bs{Y}(0)) = \sum_{m_i' \in \{0, 1 \}} \mathbb{P}(M_{i}(z_{i}) = m_i, M_i(1 - z_{i}) = m_i' \mid \bs{Y}(1), \bs{Y}(0)),
\end{align*}
and the third equality uses the fact that, for $j, k \in \{0, 1\}$, $n_{jk} = \sum_{i = 1}^{n} \mathbbm{1}\{ z_i = j \}\mathbbm{1}\{ m_i = k \}$.

Then, for all possible $\bs{z}, \bs{m} \in \{0, 1\}^{n}$, consider $\mathbb{P}\left(\bs{Z}=\bs{z}, \bs{M}=\bs{m} \mid \bs{Y(1)}, \bs{Y(0)}\right)$:
\begin{align}
    &\mathbb{P}\left(\bs{Z}=\bs{z}, \bs{M}=\bs{m} \mid \bs{Y(1)}, \bs{Y(0)}\right) \nonumber \\
    =& \mathbb{P}\left(\bs{Z}=\bs{z} \mid \bs{Y(1)}, \bs{Y(0)}\right) \mathbb{P}\left(\bs{M}=\bs{m}\mid \bs{Z}=\bs{z}, \bs{Y(1)}, \bs{Y(0)}\right) \nonumber \\
    =& \frac{1}{\binom{n}{n_1}}(p_{11}+p_{10})^{n_{11}} (p_{01}+p_{00})^{n_{10}} (p_{11}+p_{01})^{n_{01}} (p_{10}+p_{00})^{n_{00}}, \label{eq:p_z_m_mar}
\end{align}
where the second equality holds because the design is a CRE.

To this end, for all possible $\bs{z}, \bs{m} \in \{0, 1\}^{n}$, consider the following conditional distribution of $\bs{Z}_{\S}$:
\begin{align*}
& \mathbb{P}\left(\bs{Z}_{\S}=\bs{z}_{\S} \mid \bs{Y}(1), \bs{Y}(0), \bs{M}=\bs{m}, \bs{Z}_{\S^\complement}=\bs{z}_{\S^\complement}, n_{\S1} = n_{11}, n_{\S0} = n_{01}\right) \\
=& \frac{\mathbb{P}\left(\bs{Z}_{\S}=\bs{z}_{\S}, \bs{M}=\bs{m}, \bs{Z}_{\S^\complement}=\bs{z}_{\S^\complement},  n_{\S1} = n_{11}, n_{\S0} = n_{01} \mid \bs{Y}(1), \bs{Y}(0)\right)}{\mathbb{P}\left(\bs{M}=\bs{m}, \bs{Z}_{\S^\complement}=\bs{z}_{\S^\complement}, n_{\S1} = n_{11}, n_{\S0} = n_{01} \mid \bs{Y}(1), \bs{Y}(0)\right)} \\
=& \frac{\mathbb{P}\left(\bs{Z}=\bs{z}, \bs{M}=\bs{m}, n_{\S1} = n_{11}, n_{\S0} = n_{01} \mid \bs{Y}(1), \bs{Y}(0)\right)}{\sum_{\bs{z'}:\bs{z}_{\S^\complement}=\bs{z'}_{\S^\complement}, \sum m_i z'_i = n_{11}, \sum m_i (1-z'_i) = n_{01}} \mathbb{P}\left(\bs{Z}=\bs{z'}, \bs{M}=\bs{m}, n_{\S1} = n_{11}, n_{\S0} = n_{01} \mid \bs{Y}(1), \bs{Y}(0)\right)} \\
=&  \frac{\mathbb{P}\left(\bs{Z}=\bs{z}, \bs{M}=\bs{m} \mid \bs{Y(1)}, \bs{Y(0)}\right)}{\sum_{\bs{z'}:\bs{z}_{\S^\complement}=\bs{z'}_{\S^\complement}, \sum m_i z'_i = n_{11}, \sum m_i (1-z'_i) = n_{01}} \mathbb{P}\left(\bs{Z}=\bs{z'}, \bs{M}=\bs{m} \mid \bs{Y}(1), \bs{Y}(0)\right)} \\
=& \frac{1}{\binom{n_{11}+n_{01}}{n_{11}}},
\end{align*}
where the first equality uses Bayes' theorem, the second equality uses the law of total probability, the third equality holds because $n_{\S1} = n_{11}$ and $n_{\S0} = n_{01}$ follow directly and are redundant given 
$\bs{Z}=\bs{z'}$, $\sum m_i z'_i = n_{11}$ and $\sum m_i (1-z'_i) = n_{01}$, and the fourth equality follows from equation \eqref{eq:p_z_m_mar} and the number of terms in the summations of the denominator is equal to ${\binom{n_{11}+n_{01}}{n_{11}}}$.

Therefore, we conclude that $\bs{Z}_{\S} \mid \bs{Y}(1), \bs{Y}(0), \bs{M}, \bs{Z}_{\S^\complement}, n_{\S1}, n_{\S0} \sim \CRE(\S, n_{\S1}, n_{\S0})$. The desired result follows immediately from the validity of a randomization test in a CRE (see Lemma A4 in \cite{caughey2021randomization}).
\end{proof}

\section{Proof for the useful lemmas in the Appendix}

\begin{proof}[Proof of Lemma \ref{lemma:caughey_et_al_lemma_a7}]
We first prove (a). For $\bs{\tilde{y}}, \bs{\tilde{y}'}$, first consider the case when $z_i = 1$. If $y_i' = \infty$, by construction, $\tilde{y}_i \leq \zeta = \tilde{y}_i'$. If $y_i' = -\infty$, then by construction, $\tilde{y}_i = \tilde{y}_i' = \gamma$. If $y_i' \in \mathbb{R}$, then either $y_i = - \infty$ holds or $y_i \in \mathbb{R}$ holds. Therefore, by construction, $\tilde{y}_i = \mathbbm{1}\{ y_i = -\infty \} \gamma + \mathbbm{1}\{ y_i \in \mathbb{R} \} y_i \leq y_i' = \tilde{y}_i'$. A similar argument applies when $z_i = 0$. Therefore, we conclude that for $\bs{\tilde{y}}, \bs{\tilde{y}'}$:
\begin{align*}
\begin{cases*}
    \tilde{y}_i \leq \tilde{y}_i', \quad \text{if $z_i = 1$}, \\
    \tilde{y}_i \geq \tilde{y}_i', \quad \text{if $z_i = 0$}.
\end{cases*}
\end{align*}

We then prove (b). By the definition of the rank-sum statistic in \eqref{eq:test_stat_R_phi}, to show $t_{\R, \phi}\left( \bs{z}, \bs{y} \right) = t_{\R, \phi}\left( \bs{z}, \bs{\tilde{y}} \right)$, it suffices to show that $\rank_i(\bs{y}) = \rank_i(\bs{\tilde{y}})$ for all $i \in \{1, \ldots, n \}$. To this end, observe that the following can be verified as true for all $i, j \in \{1, \ldots, n \}$:
\begin{align*}
    \mathbbm{1}\{ y_i > y_j \} = \mathbbm{1}\{ \tilde{y}_i > \tilde{y}_j \}, \quad \mathbbm{1}\{ y_i = y_j \} = \mathbbm{1}\{ \tilde{y}_i = \tilde{y}_j \}.
\end{align*}
Therefore, for $\rank_i(\bs{y})$ and $\rank_i(\bs{\tilde{y}})$:
\begin{align*}
    \rank_i(\bs{y}) &= \sum_{j = 1}^{n} \left( \mathbbm{1}\{ y_i > y_j \} + \mathbbm{1}\{ y_i = y_j \} \mathbbm{1}\{ i \geq j \} \right) \\
    &= \sum_{j = 1}^{n} \left( \mathbbm{1}\{ \tilde{y}_i > \tilde{y}_j \} + \mathbbm{1}\{ \tilde{y}_i = \tilde{y}_j \} \mathbbm{1}\{ i \geq j \} \right) \\
    &= \rank_i(\bs{\tilde{y}}).
\end{align*}
Applying the same logic to $t_{\R, \phi}\left( \bs{z}, \bs{y'} \right)$ and $t_{\R, \phi}\left( \bs{z}, \bs{\tilde{y}'} \right)$, we conclude that $t_{\R, \phi}\left( \bs{z}, \bs{y'} \right) = t_{\R, \phi}\left( \bs{z}, \bs{\tilde{y}'} \right)$.

We now prove (c). Note that for $\bs{\tilde{y}'}$:
\begin{align*}
    \bs{\tilde{y}'} &= \bs{\tilde{y}} + \bs{\tilde{y}'} - \bs{\tilde{y}} \\
    &= \bs{\tilde{y}} + \{ \bs{z} \circ \bs{z} + (\bs{1} - \bs{z}) \circ (\bs{1} - \bs{z}) \} \circ (\bs{\tilde{y}'} - \bs{\tilde{y}}) \\
    &= \bs{\tilde{y}} + \bs{z} \circ \left( \bs{z} \circ (\bs{\tilde{y}}' - \bs{\tilde{y}}) \right) + ( \bs{1} - \bs{z}) \circ \left( (\bs{1} - \bs{z}) \circ (\bs{\tilde{y}}' - \bs{\tilde{y}}) \right).
\end{align*}
Furthermore, by construction:
\begin{align*}
    \bs{z} \circ (\bs{\tilde{y}}' - \bs{\tilde{y}}) \succcurlyeq \bs{0}, \quad (\bs{1} - \bs{z}) \circ (\bs{\tilde{y}}' - \bs{\tilde{y}}) \preccurlyeq \bs{0}.
\end{align*}
Then, for $t_{\R, \phi}\left( \bs{z}, \bs{y} \right)$ and $t_{\R, \phi}\left( \bs{z}, \bs{y'} \right)$:
\begin{align*}
    t_{\R, \phi}\left( \bs{z}, \bs{y'} \right) &= t_{\R, \phi}\left( \bs{z}, \bs{\tilde{y}'} \right) \\
    &= t_{\R, \phi}\left( \bs{z}, \bs{\tilde{y}} + \bs{z} \circ \left( \bs{z} \circ (\bs{\tilde{y}}' - \bs{\tilde{y}}) \right) + ( \bs{1} - \bs{z}) \circ \left( (\bs{1} - \bs{z}) \circ (\bs{\tilde{y}}' - \bs{\tilde{y}}) \right) \right) \\
    &\geq t_{\R, \phi}\left( \bs{z}, \bs{\tilde{y}} \right) = t_{\R, \phi}\left( \bs{z}, \bs{y} \right),
\end{align*}
where the first and the last equalities use part (b) of the lemma, and the inequality uses Proposition 1 in \cite{caughey2021randomization}.
\end{proof}

\begin{proof}[Proof of Lemma \ref{lemma:mann_whitney_eff_incres}]
We first show that for any $y_{i}, y_{j} \in \overline{\mathbb{R}}$ and $i, j \in \{1, \ldots, n\}$, $\psi_{i, j}(y_i, y_j)$ is increasing in $y_{i}$ and decreasing in $y_{j}$. Assume without loss of generality that the ``first'' method is used for breaking the ties. If $i \geq j$, then, $\psi_{i, j}(y_i, y_j) = \mathbbm{1}\{ y_{i} \geq y_{j} \}$. It can be verified that $\mathbbm{1}\{ y_{i} \geq y_{j} \}$ is increasing in $y_{i}$ and decreasing in $y_{j}$. If $i < j$, then, $\psi_{i, j}(y_i, y_j) = \mathbbm{1}\{ y_{i} > y_{j} \}$. It can be verified that $\mathbbm{1}\{ y_{i} > y_{j} \}$ is increasing in $y_{i}$ and decreasing in $y_{j}$.

We first consider the test statistic defined in \eqref{eq:test_stat_mann_whitney}. 
By the monotonicity of $\psi_{i, j}(y_i, y_j)$, for any $i, j \in \{1, \ldots, n\}$ and $y_i, y_j, y_i', y_j' \in \overline{\mathbb{R}}$ that satisfy the following:
\begin{align*}
z_i = 1, z_j = 0, y_i \leq y_i', 
\text{ and } y_j \geq y_j', 
\end{align*}
we have
\begin{align}\label{eq:compare_who_big_monotone}
    \psi_{i, j}(y_i, y_j) \leq \psi_{i, j}\left(y_i, y_j' \right) \leq \psi_{i, j}\left(y_i', y_j' \right).
\end{align}

For $t_{\R, \phi}(\bs{z}, \bs{y})$ and $t_{\R, \phi}(\bs{z}, \bs{y'})$ with $\bs{z}, \bs{y}, \bs{y}'$ satisfying the condition in Lemma \ref{lemma:mann_whitney_eff_incres}:
\begin{align*}
    t_{\R, \phi}(\bs{z}, \bs{y}) = \sum_{i=1}^n z_i \phi\left( \sum_{j = 1}^{n} (1 - z_j) \psi_{i, j}(y_i, y_j) \right) \leq \sum_{i=1}^n z_i \phi\left( \sum_{j = 1}^{n} (1 - z_j) \psi_{i, j}(y_i', y_j') \right) = t_{\R, \phi}(\bs{z}, \bs{y'}),
\end{align*}
where the inequality uses \eqref{eq:compare_who_big_monotone}.

We then consider the test statistic defined in \eqref{eq:test_stat_mann_whitney_negative}.
By the monotonicity of $\psi_{i, j}(y_i, y_j)$, for any $i, j \in \{1, \ldots, n\}$ and $y_i, y_j, y_i', y_j' \in \overline{\mathbb{R}}$ that satisfy the following:
\begin{align*}
z_i = 0, z_j = 1, y_i \geq y_i', 
\text{ and } y_j \leq y_j', 
\end{align*}
we have
\begin{align}\label{eq:compare_who_big_monotone_mwu_neg}
    \psi_{i, j}(y_i', y_j') \leq \psi_{i, j}\left(y_i', y_j \right) \leq \psi_{i, j}\left(y_i, y_j \right).
\end{align}

For $t_{\R, \phi}(\bs{z}, \bs{y})$ and $t_{\R, \phi}(\bs{z}, \bs{y'})$ with $\bs{z}, \bs{y}, \bs{y}'$ satisfying the condition in Lemma \ref{lemma:mann_whitney_eff_incres}:
\begin{align*}
    t_{\R, \phi}(\bs{z}, \bs{y}) &= -\sum_{i=1}^n (1 - z_i) \phi\left( \sum_{j = 1}^{n} z_j \psi_{i, j}(y_i, y_j) \right) \leq - \sum_{i=1}^n (1 - z_i) \phi\left( \sum_{j = 1}^{n} z_j \psi_{i, j}(y_i', y_j') \right) = t_{\R, \phi}(\bs{z}, \bs{y'}),
\end{align*}
where the inequality uses \eqref{eq:compare_who_big_monotone_mwu_neg}.
\end{proof}

\begin{proof}[Proof of Lemma \ref{lemma:rand_inf_valid_composite}]
We first consider $p_{\bs{Z}, \bs{\delta}, \bs{b}}$. Pick arbitrary $b_{00}, b_{01}, b_{10} \in \mathbb{R}$. Recall the definition of $G_{\R, \phi}(c)$:
\begin{align*}
    G_{\R, \phi}(c) = \mathbb{P}\left( t_{\R, \phi}\left(\bs{Z}, \tilde{\bs{Y}}_{\bs{b}}(0) \right) \ge c \right),
\end{align*}
which is the tail probability of the random variable $t_{\R, \phi}\left(\bs{Z}, \tilde{\bs{Y}}_{\bs{b}}(0) \right)$. 
Observe that $t_{\R, \phi}\left(\bs{Z}, \tilde{\bs{Y}}_{\bs{b}}(0) \right)$ is a fixed function of $\bs{Z}$.
Lemma A4 in \cite{caughey2021randomization} implies that $G_{\R, \phi}\left( t_{\R, \phi}\left(\bs{Z}, \tilde{\bs{Y}}_{\bs{b}}(0) \right) \right)$ is stochastically larger than or equal to the uniform distribution on $(0, 1)$. That is, for any $\alpha \in (0, 1)$, $\mathbb{P}\left( G_{\R, \phi}\left( t_{\R, \phi}\left(\bs{Z}, \tilde{\bs{Y}}_{\bs{b}}(0) \right) \right) \leq \alpha \right) \leq \alpha$.

By the same logic, the same conclusion holds for $p_{\bs{Z}, \bs{\delta}}$.
\end{proof}

\end{document}